\pdfoutput=1
\documentclass[twocolumn, tighten]{aastex6}
\usepackage[caption=false]{subfig}
\usepackage{amsmath}
\hypersetup{citecolor=blue,linkcolor = blue,urlcolor = blue}
\begin{document}

\title{An ALMA survey of DCN/H$^{13}$CN and DCO$^+$/H$^{13}$CO$^+$ in protoplanetary disks}
\author{Jane Huang\altaffilmark{1}, Karin I. \"Oberg\altaffilmark{1}, Chunhua Qi\altaffilmark{1}, Yuri Aikawa\altaffilmark{2}, Sean  M. Andrews\altaffilmark{1}, Kenji Furuya\altaffilmark{2}, Viviana V. Guzm\'an\altaffilmark{1}, Ryan A. Loomis\altaffilmark{1}, Ewine F. van Dishoeck\altaffilmark{3}$^,$\altaffilmark{4}, David J. Wilner\altaffilmark{1}}
\altaffiltext{1}{Harvard-Smithsonian Center for Astrophysics, 60 Garden Street, Cambridge, MA 02138, United States of America}
\altaffiltext{2}{Center for Computational Sciences, The University of Tsukuba, 1-1-1, Tennodai, Tsukuba, Ibaraki 305-8577, Japan}
\altaffiltext{3}{Leiden Observatory, Leiden University, P.O. Box 9513, NL-2300 RA Leiden, The Netherlands}
\altaffiltext{4}{Max-Planck Institut f\"ur Extraterrestrische Physik, Giessenbachstr. 1, D-85748 Garching, Germany}
\email{jane.huang@cfa.harvard.edu}
\begin{abstract}
The deuterium enrichment of molecules is sensitive to their formation environment. Constraining patterns of deuterium chemistry in protoplanetary disks is therefore useful for probing how material is inherited or reprocessed throughout the stages of star and planet formation. We present ALMA observations at $\sim0.6''$ resolution of DCO$^+$, H$^{13}$CO$^+$, DCN, and H$^{13}$CN in the full disks around T Tauri stars AS 209 and IM Lup, in the transition disks around T Tauri stars V4046 Sgr and LkCa 15, and in the full disks around Herbig Ae stars MWC 480 and HD 163296. We also present ALMA observations of HCN in the IM Lup disk. DCN, DCO$^+$, and H$^{13}$CO$^+$ are detected in all disks, and H$^{13}$CN in all but the IM Lup disk. We find efficient deuterium fractionation for the sample, with estimates of disk-averaged DCO$^+$/HCO$^+$ and DCN/HCN abundance ratios ranging from $\sim$0.02\textendash0.06 and $\sim$0.005\textendash0.08, respectively, which is comparable to values reported for other interstellar environments. The relative distributions of DCN and DCO$^+$ vary between disks, suggesting that multiple formation pathways may be needed to explain the diverse emission morphologies. In addition, gaps and rings observed in both H$^{13}$CO$^+$ and DCO$^+$ emission provide new evidence that DCO$^+$ bears a complex relationship with the location of the midplane CO snowline. \end{abstract}


\section{Introduction}
Beyond the Solar System, a rich chemistry has been observed in molecular clouds  \citep[e.g.][]{1998ApJS..117..427N, 2001ApJS..132..281S}, around protostars \citep[e.g.][]{2008AA...488..959B, 2014ApJ...788...68O, 2016AA...595A.117J}, and increasingly in protoplanetary disks \citep{2015Natur.520..198O,2016ApJ...823L..10W}. The chemical connection between these interstellar environments and the Solar System is not yet fully understood. Models indicate that some species, such as water, remain intact from the parent molecular cloud throughout the process of planetary system formation, while other molecules, such as methanol, may initially be destroyed at the protoplanetary disk stage and then re-form \citep[e.g.][]{2009AA...495..881V, 2014ApJ...790...97F,2014Sci...345.1590C}. These predictions can now be tested more easily through ALMA observations and measurements from space-based missions. 

Observations of deuterated and non-deuterated isotopologues of molecules are often used to probe the interstellar medium and the Solar System because the sensitivity of their relative abundances to their physical environment facilitates inferences about whether the molecules underwent chemical reprocessing at a given evolutionary stage \citep[e.g.][]{1998Sci...279.1707M, 2007AA...464..245R, 2008ApJ...681.1396Q}. In the interstellar medium (ISM) of the Milky Way, the elemental ratio of deuterium to hydrogen (the D/H ratio) varies from $\sim 0.5\times 10^{-5}$\textendash $2\times10^{-5}$ \citep[e.g.][]{2004ApJ...609..838W,2000ApJ...545..277S, 2010MNRAS.406.1108P}. In contrast, molecular D/H ratios have been observed to be several orders of magnitude greater for a number of species \citep[e.g.][]{1976RvMP...48..513W,1987AA...172..311W,1993AA...271..276J,1995ApJ...447..760V,2001AA...372..998C, 2003AA...403L..37C}. This enhancement, known as deuterium fractionation, is favored at low temperatures because of the lower zero-point energies of the deuterated forms of the molecular ions that initiate the formation pathways of many ISM molecules \citep{1984ApJ...287L..47D}.

Deuterium fractionation starts from a limited number of exchange reactions, such as \citep{1989ApJ...340..906M}: 

\begin{subequations}
\begin{align}
\text{HD} + \text{H}_3^+ & \rightleftharpoons \text{H}_2+\text{H}_2\text{D}^+\label{eq:H2D+}\\
\text{HD} + \text{CH}_3^+ & \rightleftharpoons \text{H}_2+\text{CH}_2\text{D}^+ \label{eq:CH2D+}\\ 
\text{HD} + \text{C}_2\text{H}_2^+ & \rightleftharpoons \text{H}_2+\text{C}_2\text{HD}^+ \label{eq:C2HD+}
\end{align}
\end{subequations}
The zero-point energy of H$_2$D$^+$ is lower than that of H$_3^+$ by $\sim640$ K. \citep{2003AA...406..383R}, while the zero-point energy of CH$_2$D$^+$ is lower than that of CH$_3^+$ by $\sim1100$ to 1200 K \citep{2013JPCA..117.9959R}. (Note that while there is also a zero-point energy difference between HD and H$_2$, the differences for the molecular ions are large enough such that the overall reactions are still exothermic.) H$_2$D$^+$ becomes abundant below $\sim$ 30 K, both because of the increased difficulty of overcoming the energy barrier to re-form H$_3^+$ from H$_2$D$^+$ ($\Delta E\sim$230 K) and because of the freezeout of molecules that can destroy H$_2$D$^+$ \citep[e.g.][]{1992AA...258..479P, 2000AA...361..388R}. Because of the relatively high abundance of H$_3^+$, H$_2$D$^+$ is the dominant deuteron donor below 30 K \citep{2000AA...361..388R}. However, reactions \ref{eq:CH2D+} and \ref{eq:C2HD+} are more exothermic than reaction \ref{eq:H2D+}, with $\Delta E$ ranging from 480 to 660 K for \ref{eq:CH2D+} \citep{2013JPCA..117.9959R} and $\sim550$ K for \ref{eq:C2HD+} \citep{1987ApJ...312..351H}. Hence, CH$_2$D$^+$ and C$_2$HD$^+$ survive more easily in warmer gas compared to H$_2$D$^+$, but the reverse reactions for \ref{eq:CH2D+} and \ref{eq:C2HD+} occur readily above $\sim$80 K, so CH$_2$D$^+$ and C$_2$HD$^+$ dominate over H$_2$D$^+$ between 30 and 80 K \citep{2000AA...361..388R}. 

Because of their relatively high abundances and accessible rotational transitions, DCO$^+$ and DCN are among the primary tracers of deuterium chemistry in the interstellar medium \citep[e.g.][]{2007AA...464..245R, 2012ApJ...749..162O, 2015AA...579A..80G}. Detailed discussions of their formation pathways are provided in \citet{1976RvMP...48..513W}, \citet{1989ApJ...340..906M}, \citet{2001AA...371.1107A}, \citet{2007ApJ...660..441W}, and \citet{2014prpl.conf..859C}. We provide a brief overview of significant reactions below. 

DCO$^+$ is thought to form in the gas phase, primarily via
\begin{equation}
\text{H}_2\text{D}^+ +\text{ CO} \rightarrow \text{H}_2 + \text{DCO}^+. \label{dcop_1}
\end{equation}
This formation pathway is analogous to the primary one for HCO$^{+}$ \citep{1984ApJ...287L..47D}:  
\begin{equation}
{\text{H}_3}^+ + \text{CO} \rightarrow \text{H}_2 + \text{ HCO}^+.
\end{equation}
Like H$_2$D$^+$, DCO$^+$ would only be expected to be abundant below $\sim$30 K \citep{1989ApJ...340..906M}.

Compared to DCO$^+$ and HCO$^+$, DCN and HCN have more complex formation pathways in the interstellar medium. The disk models presented in \citet{2014ApJ...790...97F}, \citet{2015ApJ...807..120A}, and \citet{2015ApJ...810..112O} predict that the major HCN formation mechanisms proceed in the gas phase through small hydrocarbons or their cations:

\begin{subequations}
\begin{align}
\text{CH}_3^+ + \text{e}^- & \rightarrow \text{CH}_2 + \text{H}\\
\text{CH}_2+ \text{N}& \rightarrow \text{HCN}+\text{H}
\end{align}
\end{subequations}
and
\begin{subequations}
\begin{align}
\text{C}+ \text{H}_2 & \rightarrow \text{CH}_2 + h\nu\\
\text{CH}_2+ \text{N}& \rightarrow \text{HCN}+\text{H}
\end{align}
\end{subequations}
and
\begin{subequations}
\begin{align}
\text{CH}_2+ \text{O} & \rightarrow \text{HCO} + \text{H}\\
\text{HCO}+ \text{N}& \rightarrow \text{HCN}+\text{O}
\end{align}
\end{subequations}
The roles of these pathways in interstellar HCN formation were previously described in \citealt{1984ApJS...54...81M} and \citealt{2000MNRAS.311..869H}. Based on laboratory experiments, hydrogenation on grains has also been proposed as an HCN formation pathway in the interstellar medium (\citealt{borget}), but assessing the degree to which such a mechanism would influence HCN abundances in disks would require additional chemical modeling.

The primary DCN formation mechanisms are thought to be \citep[e.g.][]{1989ApJ...340..906M,2001ApJS..136..579T}

\begin{subequations}
\label{DCN_warm_1}
\begin{align}
\text{CH}_2\text{D}^+ + \text{H}_2 & \rightarrow \text{CH}_4\text{D}^+ + h\nu\\
\text{CH}_4\text{D}^+ +\text{ e}^- &\rightarrow \text{CHD}+\text{H}_2+\text{H} \\
\text{CHD}+\text{N}&\rightarrow \text{DCN} + \text{H}
\end{align}
\end{subequations}

and 

\begin{subequations}
\label{DCN_warm_2}
\begin{align}
\text{CH}_2\text{D}^+ +\text{ e}^- &\rightarrow \text{CHD}+\text{H} \\
\text{CHD}+\text{N}&\rightarrow \text{DCN} + \text{H}
\end{align}
\end{subequations}
CHD may also react with O to form DCO, which then undergoes substitution with N to form DCN. Because DCN production is largely initiated by CH$_2$D$^+$ rather than H$_2$D$^+$, DCN is expected to peak in abundance at higher gas temperatures than DCO$^+$ \citep[e.g.][]{1989ApJ...340..906M, 1999AA...351..233A}. 

The models discussed so far suggest that \textit{in situ} disk fractionation chemistry sets the molecular abundance profile, especially for DCO$^+$. Consequently, due to the different temperature dependences of the primary DCN and DCO$^+$ formation pathways, the two molecules are presumed to trace different regions of protoplanetary disks. The vertical structure of a protoplanetary disk consists of a cold midplane in which volatile species freeze out onto dust grains, a warm molecular layer above the midplane, and a surface layer in which high UV radiation leads to photodissociation of molecules \citep{1999AA...351..233A, 2002AA...386..622A}. As a result, disk models typically predict that DCO$^+$ exists mostly near the cold midplane of the outer disk, while DCN peaks in abundance in the warmer interior regions \citep[e.g.][]{1999ApJ...526..314A, 2001AA...371.1107A, 2007ApJ...660..441W}. 

This picture of spatially differentiated DCN and DCO$^+$ distributions in disks is qualitatively consistent with observations of DCN and DCO$^+$ in the TW Hya and HD 163296 disks, the only protoplanetary disks in which sufficiently resolved observations have previously been published for both molecules \citep{2008ApJ...681.1396Q, 2012ApJ...749..162O, 2016ApJ...832..204Y}. Models suggest, though, that such distributions are not necessarily universal in disks. Based on recent upward revisions of the exothermicity for reaction \ref{eq:CH2D+}, which indicate that CH$_2$D$^+$ may be abundant up to gas temperatures of 300 K \citep{2013JPCA..117.9959R}, \citet{2015ApJ...802L..23F} proposed that the primary formation mechanisms for DCO$^+$ in disks could instead be 
\begin{equation}
\text{CH}_2\text{D}^+ + \text{O} \rightarrow \text{DCO}^+ + \text{H}_2
\end{equation}
and
\begin{equation}
\text{CH}_4\text{D}^+ +\text{ CO} \rightarrow  \text{DCO}^+ + \text{CH}_4, 
\end{equation}
where CH$_4$D$^+$ is derived from CH$_2$D$^+$. DCO$^+$ would then no longer be limited to the cold outer disk midplane, and could instead be abundant in the inner disk. Another reaction that could produce DCO$^+$ in the warm inner disk is \citep{1985ApJ...294L..63A}:
\begin{equation}
\text{HCO}^+ + \text{D} \rightarrow \text{DCO}^+ + \text{H}. 
\end{equation}

Furthermore, the following two pathways have been proposed for DCN production in cold gas \citep{2007ApJ...660..441W}:   

\begin{subequations}
\label{DCN_cold_1}
\begin{align}
\text{H}_2\text{D}^+ +\text{ CO} & \rightarrow \text{H}_2 + \text{DCO}^+\\
\text{DCO}^+ + \text{HNC} & \rightarrow \text{DCNH}^+ + \text{CO}\\
\text{DCNH}^+ + e^{-} & \rightarrow \text{DCN} + \textup{H}
\end{align}
\end{subequations}
and 
\begin{subequations}
\label{DCN_cold_2}
\begin{align}
\text{D}_3^+ + \text{HNC} & \rightarrow \text{DCNH}^+ + \text{D}_2\\
\text{DCNH}^+ + \text{e}^- & \rightarrow \text{DCN} + \text{H},
\end{align}
\end{subequations}
where D$_3^+$ formation is initiated by H$_2$D$^+$. If these alternative pathways are active in disks, the distributions of DCN and DCO$^+$ may not always show the differentiation that has been observed in the TW Hya and HD 163296 disks. 

Another motivation for examining deuterium fractionation in a range of disk types is that most theoretical studies of protoplanetary disk chemistry have been based on models of full T Tauri disks \citep[e.g.][]{1999ApJ...526..314A, 2007ApJ...660..441W, 2010ApJ...722.1607W, 2014ApJ...790...97F}. More theoretical attention has recently been paid to the chemistry of transition disks and disks around Herbig Ae stars, but has not yet directly addressed deuterium fractionation \citep{2007AA...463..203J, 2011ApJ...743L...2C,2013AA...559A..46B, 2015AA...582A..88W}. The radiation fields and temperature structures of transition disks and Herbig Ae disks may lead to substantially different chemistry; the presence of a gap in a transition disk increases the exposure of the outer disk midplane to radiation from the central star \citep[e.g.][]{2011ApJ...743L...2C}, while Herbig Ae stars have higher UV fluxes, lower X-ray fluxes, and warmer disks than T Tauri stars \citep[e.g.][]{2008AA...491..821S, 2009AA...508..707P}. Herbig Ae and T Tauri stars may also have differing wind levels, which can affect cosmic-ray ionization and accretion in disks \citep[e.g.][]{2013ApJ...769...76B, 2014ApJ...797..112C,2014ApJ...794..123C}. Thus, it is useful to examine the generalizability of existing disk models by seeking observational constraints on how the abundance profiles of deuterated molecules may be linked to spectral type and overall disk morphology. 

To explore the distribution patterns of DCN and DCO$^+$ in protoplanetary disks, we used the Atacama Large Millimeter/submillimeter Array to observe the $J = 3-2$ lines of DCO$^+$, DCN, H$^{13}$CO$^+$, and H$^{13}$CN  at sub-arcsecond spatial resolution in a diverse sample of six disks. We describe the sample selection in Section \ref{sec:sampleselection} and the observations and data reduction in Section \ref{sec:observations}. Results are presented in Section \ref{sec:results}. The chemical and physical implications of the observations are discussed in Section \ref{sec:discussion}. A summary is presented in Section \ref{sec:summary}. 

\begin{deluxetable*}{ccccccccccc}
\tablecaption{Stellar Properties\label{tab:stellarproperties}}
\tablehead{
\colhead{Source} & \colhead{R.A.} &\colhead{ Decl.} &\colhead{ Sp.}&\colhead{ $L_*$} &\colhead{Age}&\colhead{$M_*$} &\colhead{$\dot M$}&\colhead{$T_\text{eff}$}&\colhead{Dist.}&\colhead{Ref.}\\
\colhead{}& \colhead{ (J2000)} & \colhead{(J2000)} &\colhead{Type}& \colhead{($L_\odot$)} &\colhead{(Myr)}& \colhead{($M_\odot$)}&\colhead{($10^{-9}$ $M_\odot$ yr$^{-1}$)} &\colhead{(K)}&\colhead{(pc)}}
\colnumbers
\startdata
AS 209 &16 49 15.30 & -14 22 08.6 & K5 &1.5&1.6&0.9&51&4250&126&1\textendash4\\
IM Lup& 15 56 09.18 & -37 56 06.1 &M0&0.93& 1 &1.0 &0.01 &3850& 161\tablenotemark{a}&4\textendash7\\
V4046 Sgr&18 14 10.47 & -32 47 34.5&K5, K7&0.49, 0.33 &24& 1.75\tablenotemark{b}&0.5&4370& 72 & 8\textendash12\\
LkCa 15 &04 39 17.80 & +22 21 03.5& K3 &0.74&3\textendash 5&0.97&1.3& 4730&140&3, 4, 13, 14\\
MWC 480 & 04 58 46.27 & +29 50 37.0 & A4 &11.5& 7&1.65&126&8250&142& 14\textendash17\\
HD 163296 & 17 56 21.29 & -21 57 21.9&A1&30&4&2.25&69&9250&122&16\textendash18\\
\enddata
\tablenotetext{a} {IM Lup values for $L_*$ and age \citep{2015AA...580A..26G}, $M_*$ \citep{2008AA...489..633P}, and $\dot M$ \citep{2010AA...519A..97G} were derived using an older distance estimate of 190 pc.}
\tablenotetext{b} {Stellar mass listed is the sum of the binary masses}
\tablerefs{(1) \citet{1988cels.book.....H}, (2) \citet{2009ApJ...700.1502A}, (3) \citet{2015MNRAS.450.3559N}, (4) \citet{2016AA...595A...2G}, (5) \citet{2015AA...580A..26G}, (6) \citet{2008AA...489..633P}, (7) \citet{2010AA...519A..97G}, (8) \citet{2000IAUS..200P..28Q}, (9) \citet{2012ApJ...759..119R}, (10) \citet{2015MNRAS.454..593B}, (11) \citet{2011MNRAS.417.1747D}, (12) \citet{2006AA...460..695T}, (13) \citet{2010ApJ...717..441E}, (14) \citet{2000ApJ...545.1034S}, (15) \citet{1994AJ....108.1872K}, (16) \citet{2011AJ....141...46D}, (17) \citet{2009AA...495..901M}, (18) \citet{1998AA...330..145V}}
\end{deluxetable*}

\begin{deluxetable*}{cccccccc}
\tablecaption{Disk Properties\label{tab:diskproperties}}
\tablehead{
\colhead{Source} &\colhead{Disk Type}&\colhead{Disk Incl.\tablenotemark{a}} &\colhead{ Disk P.A.\tablenotemark{a}} &\colhead{Disk Mass\tablenotemark{b}}&\colhead{$R_{\mathrm{CO}}$}&\colhead{Systemic vel.}&\colhead{Ref.}\\
\colhead{} & \colhead{}& \colhead{(deg.) }&\colhead{(deg.)}&\colhead{($M_\odot$)}&\colhead{(AU)}&\colhead{(km s$^{-1}$)}}
\colnumbers
\startdata
AS 209&T Tauri full&38&86&0.015&340&4.6&1, 2, 3, 4 \\
IM Lup&T Tauri full&50&144.5&0.17&970&4.4& 5, 6\\
V4046 Sgr&T Tauri transition&33.5&76&0.028&365&2.9&7 \\
LkCa 15&T Tauri transition&52&60\tablenotemark{c}&0.05\textendash0.1&900&6.3&8, 9\\
MWC 480&Herbig Ae full&37&148\tablenotemark{c}&0.11&460&5.1&3, 8, 10\\
HD 163296&Herbig Ae full&48.5&132 &0.17&540&5.8&11, 12\\
\enddata
\tablenotetext{a}{Disk geometry derived from modeling millimeter/sub-millimeter CO emission}
\tablenotetext{b}{Disk masses derived from dust continuum observations except for V4046 Sgr, which is derived from $^{12}$CO observations. All quoted estimates assume the ISM gas-to-dust ratio of 100:1.}
\tablenotetext{c} {The tabulated position angle (P.A.) values differ from those given in \citealt{2007AA...467..163P}, which defines P.A. in terms of the rotation axis. We converted these values using the convention that P.A. is the angle east of north made by the disk semi-major axis.}

\tablerefs{(1) \citet{2009ApJ...700.1502A}, (2) \citet{2016AA...588A..53T}, (3) \citet{2011ApJ...734...98O}, (4) \citet{2016ApJ...823L..18H}, (5) \citet{2016ApJ...832..110C}, (6) \citet{2009AA...501..269P}, (7) \citet{2012ApJ...759..119R}, (8) \citet{2007AA...467..163P}, (9) \citet{2012ApJ...747..136I}, (10) \citet{2001ApJ...547.1077C}, (11) \citet{2015ApJ...813...99F}, (12) \citet{2007AA...469..213I}}
\end{deluxetable*}

\section{Sample Selection and Characteristics} \label{sec:sampleselection}
The sample, chosen to span a range of spectral types and disk morphologies, consists of two T Tauri stars with full disks (AS 209 and IM Lup), two T Tauri stars with transition disks (V4046 Sgr and LkCa 15), and two Herbig Ae stars with full disks (MWC 480 and HD 163296). To enable comparisons between the distributions of DCN and DCO$^+$, the targets were mainly chosen from disks in which DCO$^+$ had previously been observed in DISCS: The Disk Imaging Survey of Chemistry with the Submillimeter Array \citep{2010ApJ...720..480O, 2011ApJ...734...98O}. The typical spatial resolution of the previous observations was on the order of several arcseconds, compared to $\sim0.6''$ in this work. The angular resolution of $\sim0.6''$ translates to physical scales of $\sim70$\textendash$80$ AU for most of the targets, which corresponds to the scales at which previous observations and disk chemistry models indicate that their emission substructures could be resolved \citep[e.g.][]{2002AA...386..622A, 2007ApJ...660..441W, 2008ApJ...681.1396Q,2012ApJ...749..162O,2013AA...557A.132M}.  DCO$^+$ was not detected in the MWC 480 disk in DISCS, but this disk was selected as a target for the ALMA survey because it is considered a ``benchmark" Herbig Ae disk \citep[e.g.][]{1997ApJ...490..792M, 2007AA...467..163P, 2010ApJ...719.1565G}. The HD 163296 disk was also included in the survey as the only Herbig Ae disk in which DCO$^+$ had previously been detected \citep{2013AA...557A.132M}. Of the sources in the survey, DCN has only previously been detected in the LkCa 15 and HD 163296 disks \citep{2001PhDT.........1Q, 2016ApJ...832..204Y}. \citet{2010ApJ...720..480O} and \citet{2011ApJ...734...98O} reported non-detections of DCN for the other targets.

\section{Observations and Data Reduction} \label{sec:observations}
\subsection{Observational Setup}

\begin{deluxetable}{cccc}
\tablecaption{Targeted Lines\label{tab:targetedlines}}
\tablehead{
\colhead{Transition} &\colhead{Frequency} &\colhead{$E_u$} &\colhead{$S_{ij}\mu^2$}\\
\colhead{} & \colhead{(GHz)}&\colhead{(K)}&\colhead{(D$^2$)}}
\startdata
DCO$^+$ $J = 3-2$ &216.11258&20.74&45.62\\
DCN $J=3-2$&217.23853&20.85&80.50\\
H$^{13}$CN $J = 3-2$ &259.01180&24.86&80.20\\
H$^{13}$CO$^+$ $J =3-2$&260.25534&24.98&45.62\\
HCN $J = 3-2$&265.88643 &25.52&80.20
\enddata
\end{deluxetable}

We observed the $J = 3-2$ lines of DCO$^+$, DCN, H$^{13}$CO$^+$, and H$^{13}$CN, which have upper energy levels of $\sim20$ to 25 K, comparable to the gas temperatures at which DCN and DCO$^+$ are expected to be abundant. H$^{13}$CN and H$^{13}$CO$^+$ were targeted rather than the main isotopologues because the corresponding transitions of the latter are expected to be optically thick in disks \citep{2004AA...425..955T}. The properties of the targeted transitions, taken from the Cologne Database for Molecular Spectroscopy \citep{2001AA...370L..49M,2005JMoSt.742..215M}, are summarized in Table \ref{tab:targetedlines}.

The six disks were observed during ALMA Cycle 2 (project code \dataset[ADS/JAO.ALMA\#2013.1.00226.S]{https://almascience.nrao.edu/aq/?project\_code=2013.1.00226.S}) in two Band 6 spectral settings. The array configuration and calibrators for each observation are described in Table \ref{tab:settings}. The 1.1 mm setting, which targeted H$^{13}$CN and H$^{13}$CO$^+$ $J = 3-2$, was configured with 14 spectral windows (SPWs). The 1.4 mm setting, which targeted DCN and DCO$^+$ $J = 3-2$, was configured with 13 SPWs. Details of the spectral setups are listed in Tables \ref{tab:1.1 mm} and \ref{tab:1.4 mm}. Because narrow spectral windows were needed to achieve high spectral resolution for the targeted lines, each observation used the bandwidth-switching mode, in which calibrators were observed in wider spectral windows in order to improve the signal-to-noise ratio. For observations spanning multiple execution blocks, the individual blocks were first imaged separately to check for consistency. \footnote{The DCO$^+$ and DCN lines for the MWC 480 and LkCa 15 disks were also observed for three minutes each on 2015 July 25, but these observations were not combined with the other two 12 minute execution blocks because of the poorer signal-to-noise ratio and $uv$ coverage. An additional execution block was also originally delivered for the 1.4 mm setting of HD 163296, but the ALMA helpdesk subsequently advised excluding it from analysis due to poor weather.}

\begin{deluxetable*}{ccccccccc}
\tablecaption{ALMA Observation Details\label{tab:settings}}
\tablehead{
\colhead{Source} & \colhead{Date} &\colhead{ Setting} &\colhead{Antennas}&\colhead{Baselines} &\colhead{On source}&\colhead{Bandpass} &\colhead{Phase}&\colhead{Flux}\\
\colhead{}& \colhead{} & \colhead{} &\colhead{}& \colhead{(m)} &\colhead{int. (min)}& \colhead{Cal.}&\colhead{Cal.}&\colhead{Cal.}}
\colnumbers
\startdata
\multicolumn{9}{c}{Cycle 2 Observations} \\
\hline
AS 209 &2014 July 17&1.1 mm & 32 &20\textendash650&21& J1733-1304 & J1733-1304 & Titan\\
& 2014 July 2 & 1.4 mm & 34 & 20\textendash650 & 21 & J1733-1304 & J1733-1304 & Titan\\
IM Lup &2014 July 17 & 1.1 mm & 32 &20\textendash650  & 21 &J1427-4206&J1534-3526&Titan\\
&2014 July 6&1.4 mm &31 &20\textendash650 &21&J1427-4206&J1534-3526&Titan\\
V4046 Sgr & 2015 May 13 &1.1 mm & 36&21\textendash558 & 21 &J1924-2914&J1826-2924&Titan\\
&2014 June 9 &1.4 mm & 33 &20\textendash646& 21 &  J1924-2914 & J1802-3940& J1924-2914\\
LkCa 15/MWC 480\tablenotemark{a}& 2014 June 15 &1.1 mm & 33 &20\textendash650 & 21 &J0510+1800&J0510+1800&J0510+1800\\
&2014 July 29 & 1.4 mm & 31 & 24\textendash820& 12 & J0510+1800 & J0510+1800&J0510+1800\\
&2015 June 6&1.4 mm & 37&21\textendash 784&12 &J0510+1800&J0510+1800&J0510+1800\\
HD 163296 &2014 July 16&1.1 mm & 32&34\textendash 650 & 13 &J1733-1304&J1733-1304&J1733-1304\\
&2015 May 13&1.1 mm &36 &21\textendash558&21&J1733-1304&J1733-1304&Titan\\
&2014 July 2 & 1.4 mm & 34 &20\textendash650& 21 &J1733-1304&J1733-1304 & J1733-1304\\
&2015 May 13&1.4 mm &36 &21\textendash558&13&J1733-1304&J1733-1304&Titan\\
\hline
\multicolumn{9}{c}{Cycle 3 Observations} \\
\hline
IM Lup &2016 May 1 & 1.1 mm\tablenotemark{b} & 41 &15\textendash630  & 12 &J1517-2422&J1610-3958&Titan\\
\enddata
\tablenotetext{a}{LkCa 15 and MWC 480 were observed in the same scheduling blocks.}
\tablenotetext{b}{These observations were also at 1.1 mm, but the spectral setup differs from the Cycle 2 observations and is described in more detail in Appendix \ref{sec:setups}.}
\tablehead{
\colhead{Source} & \colhead{Date} &\colhead{ Setting} &\colhead{Antennae}&\colhead{Baselines} &\colhead{On source}&\colhead{Bandpass} &\colhead{Phase}&\colhead{Flux}\\
\colhead{}& \colhead{} & \colhead{} &\colhead{}& \colhead{(m)} &\colhead{int. (min)}& \colhead{Cal.}&\colhead{Cal.}&\colhead{Cal.}}

\end{deluxetable*}

Because H$^{13}$CN was not detected in the IM Lup disk, we also present observations of HCN $J = 3-2$ from ALMA Cycle 3 (project code \dataset[ADS/JAO.ALMA\#2015.1.00964.S. ]{https://almascience.nrao.edu/aq/?project\_code=2015.1.00964.S}). Details of the array configurations and spectral setup are provided in Tables \ref{tab:settings} and \ref{tab:HCNsetting}.

Images from an earlier reduction of the IM Lup DCO$^+$ and H$^{13}$CO$^+$ observations were first published in \citet{2015ApJ...810..112O} as part of their analysis of CO ice desorption in the outer disk. Earlier reductions of the H$^{13}$CN data for MWC 480 were first presented in \citet{2015Natur.520..198O} in an analysis of cyanide abundances. These data are presented again in this work in conjunction with new data for a consistent analysis of deuterium fractionation for the full sample. The H$^{13}$CN data for all six disks are also discussed further in an analysis of $^{14}$N/$^{15}$N fractionation in \citet{Guzmaninpress}. 

The spectral setups with multiple narrow windows were configured to search for other lines in addition to the four targeted for the deuterium survey. Because those searches are outside the scope of the present survey, they will be discussed in separate publications. In addition to the papers mentioned above, results from the secondary line searches have also been presented in \citet{2015ApJ...809L..26H}, \citet{2016ApJ...823L..18H}, and \citet{2016ApJ...832..110C}.

\subsection{Data Reduction}

Initial flux, phase, and bandpass calibrations of the data were performed by ALMA/NAASC staff. The Butler-JPL-Horizons 2012 model was used for Titan, the flux calibrator for about half the observations. The quasar flux calibration models for the other observations, as well as modifications of the models specified in the scripts provided by ALMA/NAASC, are described in Table \ref{tab:fluxcal}. In addition, the calibration scripts were modified for the 1.1 and 1.4 mm settings of AS 209, 1.4 mm setting of V4046 Sgr, 1.1 mm setting of IM Lup, and 1.1 and 1.4 mm settings of HD 163296 to scale visibility weights properly\footnote{See \url{https://casaguides.nrao.edu/index.php/DataWeightsAndCombination}} and executed in CASA 4.4.0 to re-calibrate the visibilities. 

Subsequent data reduction and imaging were also completed with CASA 4.4.0 for the Cycle 2 data and CASA 4.5.3 for the Cycle 3 data. For each disk, the 258 GHz continuum was phase self-calibrated by obtaining solutions from combining the line-free portions of the spectral windows in the upper sideband of the 1.1 mm setting. The continuum was then imaged by CLEANing with a Briggs robust weighting parameter of 0.5. The self-calibration tables produced from the 258 GHz continuum data were then applied to the SPWs covering the H$^{13}$CN and H$^{13}$CO$^+$ lines. Likewise, phase self-calibration of the SPWs covering the DCN and DCO$^+$  lines was performed using solutions obtained from combining the line-free portions of the SPWs in the lower sideband of the 1.4 mm setting and phase self-calibration of the IM Lup HCN data was performed using solutions from combining the line-free portions of the upper sideband SPWs.

\begin{deluxetable*}{ccccc}
\tablecaption{Properties of the 258 GHz Dust Continuum\label{tab:continuum}}
\tablehead{
\colhead{Source} &\colhead{$F_\text{cont}$\tablenotemark{a}} &\colhead{Peak flux density\tablenotemark{a}}&\colhead{Beam (PA)}&\colhead{Detected Radius}\\
\colhead{} & \colhead{(mJy)}&\colhead{(mJy beam$^{-1}$)}&&\colhead{(AU)}}
\startdata
AS 209&350.7$\pm$1.4&87.8$\pm$0.2&$0\farcs43\times0\farcs42$ $(67\fdg9)$&190\\
IM Lup&276$\pm$2&72.7$\pm$0.2&$0\farcs44\times0\farcs37$ $(70\fdg8)$&310\\
V4046 Sgr&422.6$\pm$1.2&64.8$\pm$0.2&0$\farcs52\times0\farcs43$ $(-80\fdg4)$&110\\
LkCa 15&$204.4\pm0.7$&39.0$\pm$0.2&$0\farcs58\times0\farcs42$ $(-11\fdg40)$&200\\
MWC 480&$373.1\pm1.1$&195.5$\pm$0.3&$0\farcs67\times0\farcs39$ $(-6\fdg3)$&200\\
HD 163296&$857\pm5$& 203.5$\pm$0.3 & $0\farcs52\times0\farcs40$ $(-89\fdg99)$&260
\enddata
\tablenotetext{a}{Uncertainties do not include 15\% systematic flux uncertainties}
\end{deluxetable*}

As a preliminary step for choosing CLEAN and spectral extraction masks for the lines, the radius of the millimeter dust disk was estimated by using the Python package \textsc{scikit-image} \citep{scikit-image} to fit ellipses to the 3$\sigma$ contours of the 258 GHz continuum images, where $\sigma$ is the rms measured from a signal-free portion of the continuum map. Since the radii, listed in Table \ref{tab:continuum}, are based on the detected extent of the millimeter dust emission, they are not necessarily similar to the often-derived characteristic radius values, which describe the radius at which the surface density transitions between a power-law profile in the inner disk and an exponentially declining profile in the outer disk \citep{2008ApJ...678.1119H}.

The 258 GHz continuum flux densities were obtained by integrating interior to the $3\sigma$ contour. Continuum fluxes and rms values are listed in Table \ref{tab:continuum}. The statistical uncertainty of the continuum flux for each source was estimated by using the elliptical fit to the $3\sigma$ contour to measure flux densities at 500 random signal-free positions in a $25''\times25''$ continuum image, then taking the standard deviation of these measurements.

After self-calibration, the continuum was subtracted in the $uv$-plane from the SPWs containing the targeted lines. Most lines were imaged and CLEANed down to a 3$\sigma$ threshold at a binned resolution of 0.5 km s$^{-1}$ using a Briggs parameter of 1.0 and CLEAN masks based on the emission in each channel. Signal-free channels from the image cubes were used to calculate $\sigma$ values. SPWs without strong, well-resolved line emission (targeting H$^{13}$CN and DCN in the MWC 480 and HD 163296 disks, and H$^{13}$CN in the LkCa 15 disks) were CLEANed with a Briggs parameter of 2.0 for better sensitivity. Since their emitting region is more ambiguous, we used elliptical CLEAN masks based on the position angles and inclinations listed in Table \ref{tab:diskproperties} and the millimeter dust disk radii estimates listed in Table \ref{tab:continuum}. For the IM Lup disk, we used elliptical masks based on the extent of the HCN emission.

 \begin{deluxetable*}{cccccccc}
\tablecaption{Line observations\label{tab:lineobservations}}
\tablehead{
\colhead{Source} &\colhead{Line} &\colhead{Int.}&\colhead{Moment}&\colhead{Chan.}&\colhead{Mask}&\colhead{Int.}&\colhead{Beam (PA)}\\
&&\colhead{range}&\colhead{zero rms}&\colhead{rms}&\colhead{axis}&\colhead{Flux}&}
\colnumbers
\startdata
AS 209&H$^{13}$CO$^+$  &1.5\textendash7.5&8.2&4.2&4$''$&$230\pm30$&$0\farcs49\times0\farcs47$ $(-174\fdg5)$\\
&DCO$^+$ &1.5\textendash7.5&10.&5.1&4$''$&$480\pm30$&$0\farcs63\times0\farcs60$ $(-75\fdg6)$\\
&H$^{13}$CN &1.5\textendash7.5&8.7&4.0&3\farcs5&$210\pm30$&$0\farcs51\times0\farcs49$ $(-11\fdg0)$\\
&DCN&1.5\textendash7.5&11.&5.4&3\farcs5&340$\pm30$&$0\farcs63\times0\farcs60$ $(-78\fdg1)$\\
IM Lup&H$^{13}$CO$^+$ &$2.0\textendash$7.0&6.7&3.6&7\farcs5&$390\pm40$&$0\farcs47\times0\farcs41$ $(+77\fdg6)$\\
&DCO$^+$ &2.0\textendash7.0&5.5&2.8&7\farcs5&$490\pm20$&$0\farcs65\times0\farcs48$ $(-82\fdg4)$\\
&H$^{13}$CN &1.5\textendash7.5&7.1&3.6&7\farcs5&\textless 60\tablenotemark{a}&$0\farcs49\times0\farcs42$ $(+80\fdg0)$\\
&DCN&1.5\textendash7.5&6.1&2.9&7\farcs5&90$\pm$13\tablenotemark{b}&0$\farcs67\times0\farcs49$ $(-74\fdg3)$\\
&HCN&1.5\textendash7.5&11.&4.8&7\farcs5&3300$\pm$70&$0\farcs55\times0\farcs54$ $(11\fdg0)$\\
V4046 Sgr&H$^{13}$CO$^+$&-1.0\textendash7.0&9.1&3.4&8$''$&$820\pm50$&0$\farcs58\times0\farcs47$ $(+86\fdg1)$\\
&DCO$^+$ &-1.0\textendash7.0&6.3&2.7&8$''$&$650\pm30$&$0\farcs81\times0\farcs53$ $(-88\fdg2)$\\
&H$^{13}$CN &-3.0\textendash9.0&11.&3.3&$5''$&$930\pm50$&0$\farcs58\times0\farcs47$ $(+86\fdg0)$\\
&DCN &-3.0\textendash9.0&8.5&2.9&5$''$&$190\pm20$&$0\farcs80\times0\farcs53$ $(-88\fdg5)$\\
LkCa 15&H$^{13}$CO$^+$&3.0\textendash9.0&7.8&3.7&6\farcs5&$380\pm30$&$0\farcs64\times0\farcs48$ $(-13\fdg5)$\\
&DCO$^+$&3.0\textendash9.0&5.9&2.7&6\farcs5&$400\pm30$&$0\farcs64\times0\farcs5$ $(+15\fdg9)$\\
&H$^{13}$CN&3.0\textendash9.0&7.6&3.7&5\farcs5&100$\pm$30&$0\farcs67\times0\farcs51$ $(-14\fdg2)$\\
&DCN&3.0\textendash9.0&5.9&3.0&5\farcs5&$280\pm30$&$0\farcs63\times0\farcs49$ $(+15\fdg3)$\\
MWC 480&H$^{13}$CO$^+$ &$1.0\textendash9.0$&8.7&3.6&5$''$&360$\pm$30&$0\farcs72\times0\farcs46$ $(-6\fdg8)$\\
&DCO$^+$&1.0\textendash9.0&6.6&2.7&5$''$&$420\pm30$&$0\farcs73\times0\farcs48$ $(+14\fdg4)$\\
&H$^{13}$CN&1.0\textendash9.0&8.7&3.6&3\tablenotemark{c}&$150\pm20$&$0\farcs76\times0\farcs48$ $(-7\fdg0)$\\
&DCN&1.0\textendash9.0&6.5&2.8&3\tablenotemark{c}&$70\pm20$&$0\farcs76\times0\farcs50$ $(+14\fdg0)$\\
HD 163296&H$^{13}$CO$^+$&1.0$\textendash$11.0&7.4&2.6&8\farcs5&$620\pm40$&$0\farcs57\times0\farcs45$ $(-88\fdg2)$\\
&DCO$^+$&1.0\textendash11.0&8.3&2.8&8\farcs5&$1290\pm40$&$0\farcs66\times0\farcs57$ $(+57\fdg1)$\\
&H$^{13}$CN&1.0\textendash11.0&7.2&2.8&4\tablenotemark{c}&$170\pm30$&$0\farcs59\times0\farcs46$ $(-87\fdg9)$\\
&DCN&1.0\textendash11.0&8.2&3.0&4\tablenotemark{c}&$120\pm20$&$0\farcs69\times0\farcs59$ $(+64\fdg4)$\\
\enddata
\tablecomments{Column descriptions: (1) Source name. (2) Molecule corresponding to targeted $J=3-2$ transition. (3) Velocity range (km s$^{-1}$) integrated across to measure flux and produce integrated intensity maps. (4) Moment zero map rms (mJy beam$^{-1}$ km s$^{-1}$). (5) Channel rms for bin sizes of 0.5 km s$^{-1}$. (6) Major axes of elliptical spectral extraction masks (7) Integrated flux (mJy km s$^{-1}$). Uncertainties do not include systematic flux uncertainties. (8) Synthesized beam dimensions.}
\tablenotetext{a}{Upper limit quoted as 3$\times$flux density uncertainties}
\tablenotetext{b}{Flux estimated based on HCN emitting region}
\tablenotetext{c}{Spectral extraction mask based on extent of millimeter dust emission}
\end{deluxetable*}

\begin{figure*}[htp]
\epsscale{1.2}
\plotone{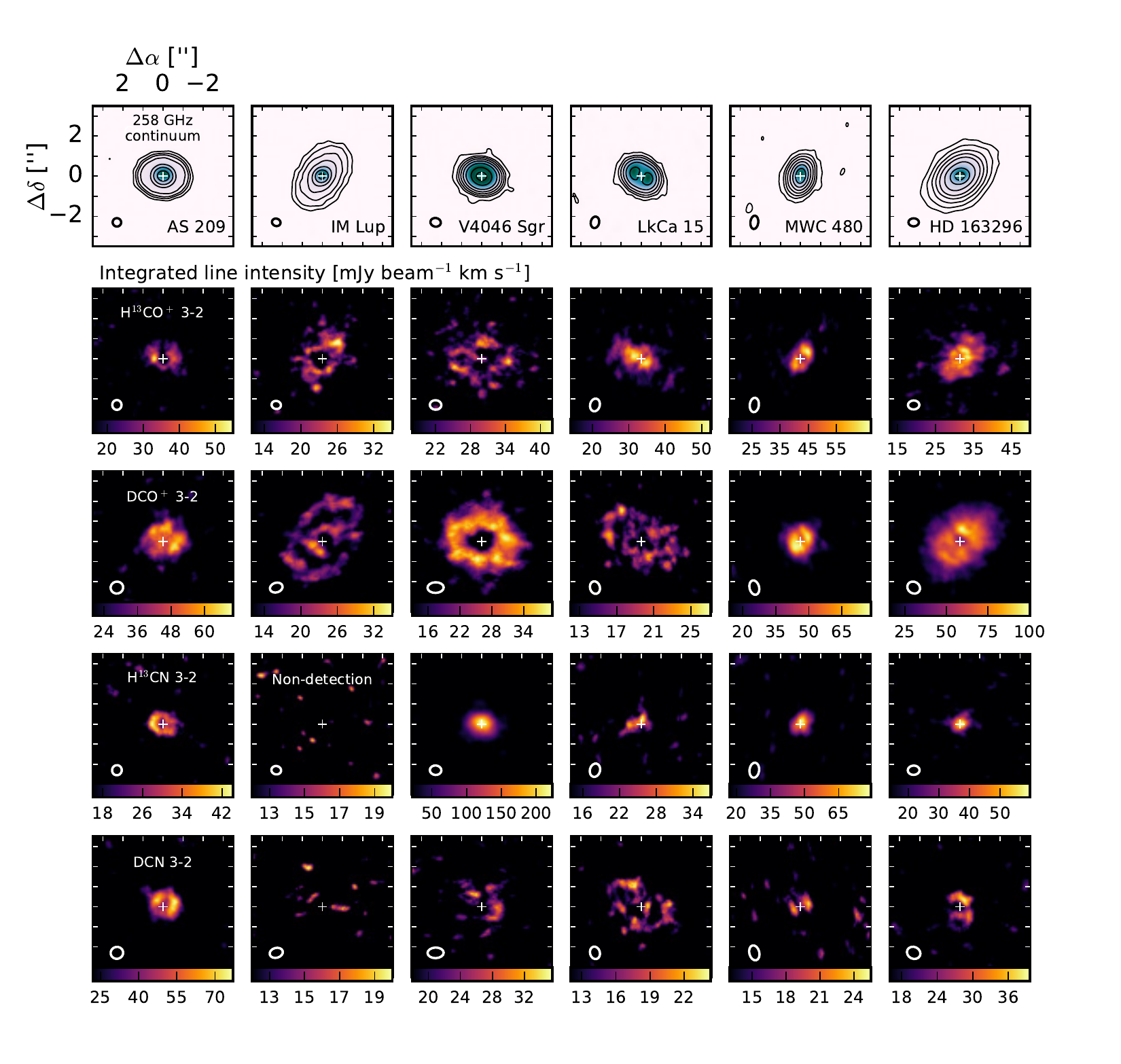}
\caption{Continuum and integrated line intensity maps for H$^{13}$CO$^+$, DCO$^+$, H$^{13}$CN, and DCN. First row: intensity maps of the 258 GHz dust continuum. Contours are drawn at [3, 10, 20, 40, 80, 160...]$\sigma$, where $\sigma$ is the rms listed in Table \ref{tab:continuum}. Rows 2 through 5: integrated intensity maps corresponding to the transition denoted in the first panel of each row, produced by clipping emission below $1\sigma$ in the channel maps. Color bars start at 2$\sigma$, where $\sigma$ is the rms of the integrated intensity map listed in Table \ref{tab:lineobservations}. Each column corresponds to the disk listed in the top panel. Synthesized beams are drawn in the lower left corners of each panel. The centroid of the continuum image is marked with white crosses. Offset from this centroid in arcseconds is marked on the axes in the upper left corner.}
\label{fig:mom0}
\end{figure*}

\begin{figure*}[htp]
\epsscale{1.1}
\plotone{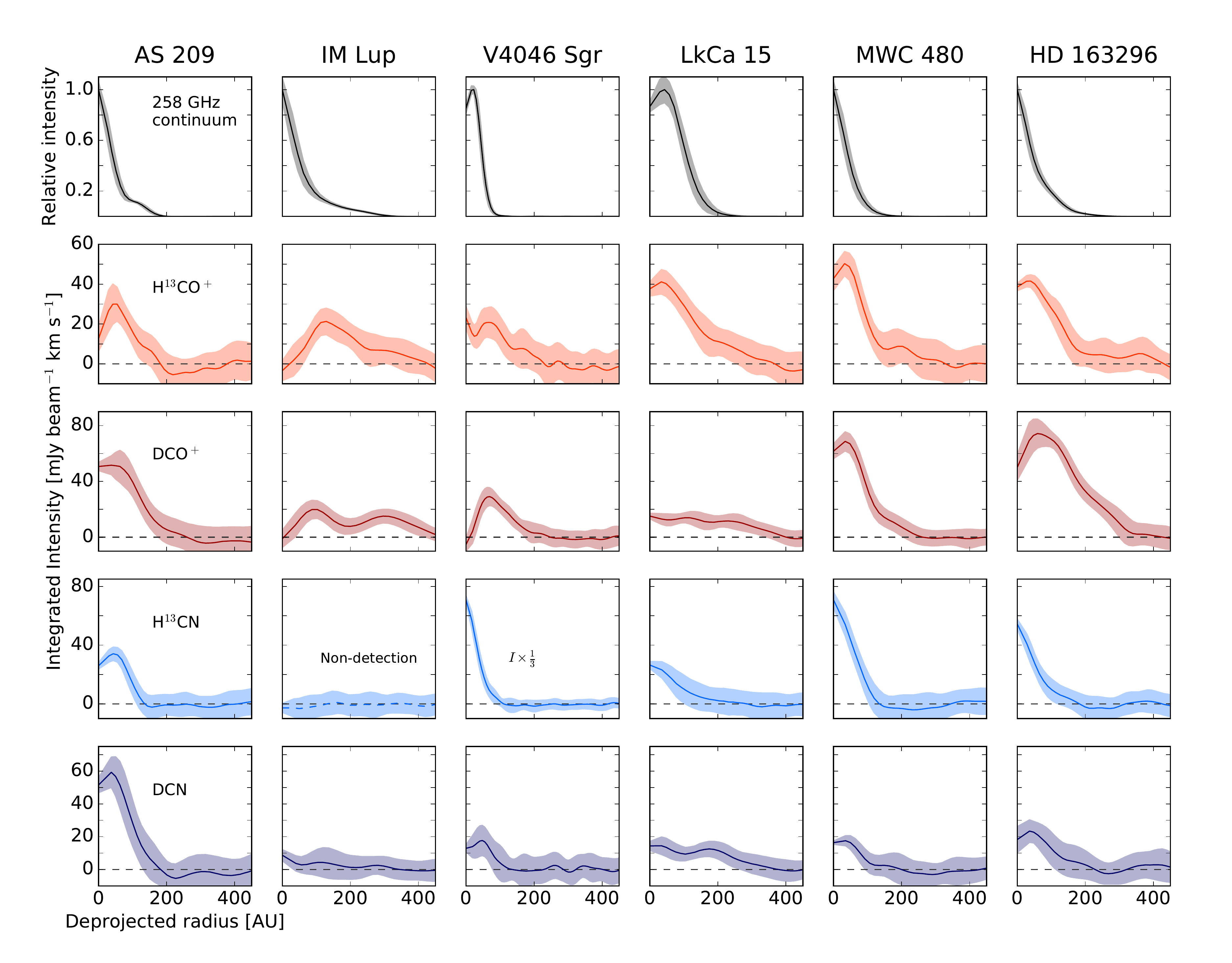}
\caption{Deprojected and azimuthally averaged 258 GHz continuum intensity profiles and integrated line intensity profiles for H$^{13}$CO$^+$, DCO$^+$, H$^{13}$CN, and DCN $J = 3-2$. Top row: 258 GHz continuum intensity profiles normalized to peak values. Rows 2\textendash5: integrated line intensity radial profiles. Each row corresponds to the molecule marked in the first column. Each column corresponds to the disk labeled above the first row. Colored ribbons show the standard deviation in pixel intensities calculated at each annulus. Adopted distances are listed in Table \ref{tab:stellarproperties}, and adopted position angles and inclinations are listed in Table \ref{tab:diskproperties}. Note that the radial profile of H$^{13}$CN $J = 3-2$ for the V4046 Sgr disk is scaled down by a factor of $1/3$  in order to be plotted on the same axes as the other disks.}
\label{fig:radialprofile}
\end{figure*}

Figure \ref{fig:mom0} shows integrated intensity maps of H$^{13}$CO$^+$, DCO$^+$, H$^{13}$CN, and DCN, with the emission in each channel clipped at the $1\sigma$ level to better isolate the signal. For most lines, the maps were produced by summing over channels corresponding to where DCO$^+$ emission was detected above the $3\sigma$ level, since DCO$^+$ was typically the strongest line. These velocity integration ranges are listed in Table \ref{tab:lineobservations}. The channel maps for the other lines were inspected to check that no emission above the 3$\sigma$ level appeared in the known disk region in channels where DCO$^+$ was not detected. For the V4046 Sgr disk, the integration ranges for H$^{13}$CN and DCN were based on the H$^{13}$CN line because the signal-to-noise ratios of the spectra are sufficiently high to confirm that H$^{13}$CN has a larger linewidth than H$^{13}$CO$^+$. Figure \ref{fig:radialprofile} shows deprojected and azimuthally averaged radial profiles, which were produced from unclipped versions of the integrated intensity maps, assuming the position angles and inclinations listed in Table \ref{tab:diskproperties}.

\begin{figure*}[htp]
\epsscale{1.1}
\plotone{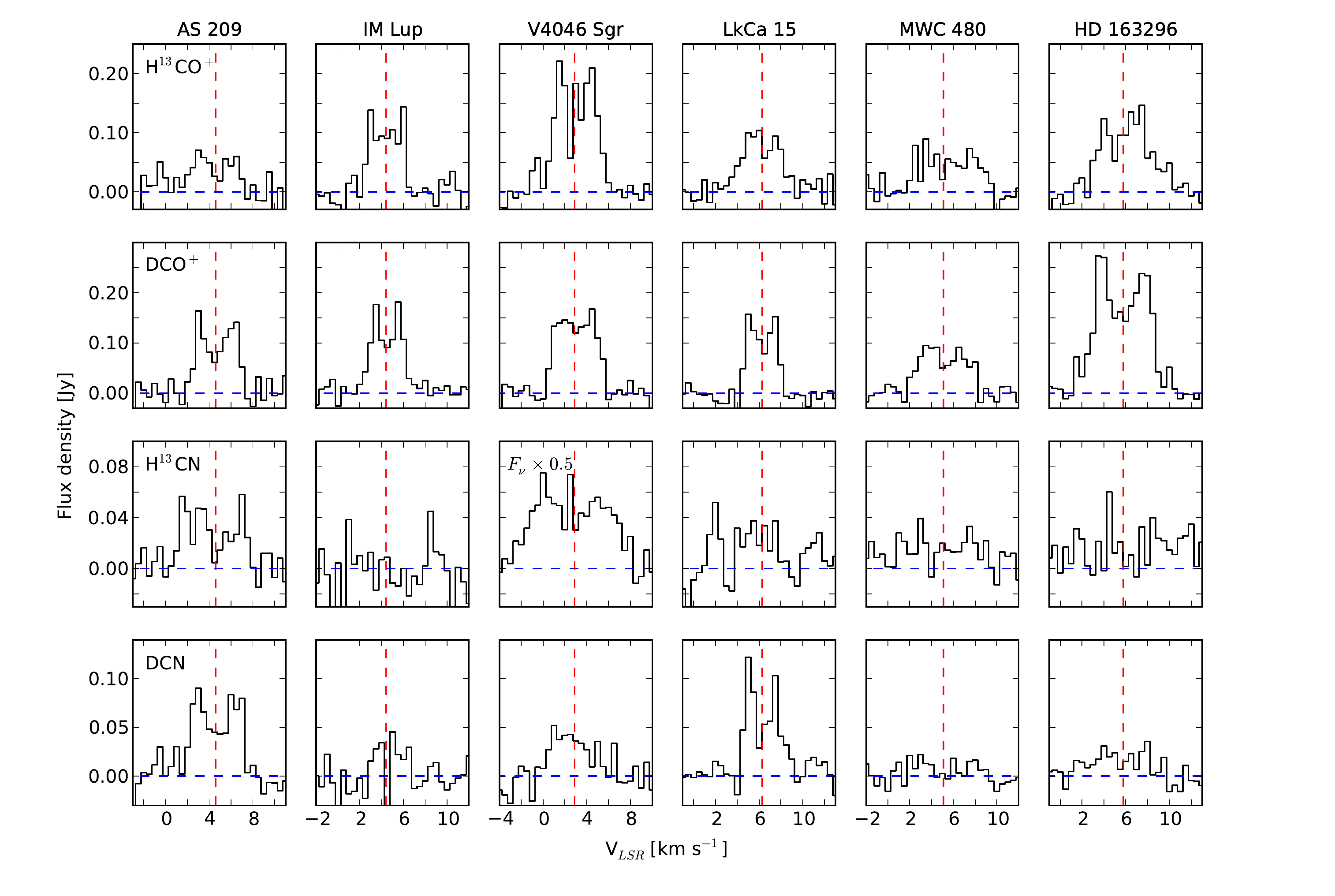}
\caption{Spectra of H$^{13}$CO$^+$, DCO$^+$, H$^{13}$CN$^+$, and DCN $J = 3-2$. The horizontal blue line marks the zero flux density level. The vertical red line marks the systemic velocity (listed in Table \ref{tab:diskproperties}). The H$^{13}$CN spectrum for the V4046 Sgr disk is scaled down by a factor of 0.5 to fit on the same axes as the other H$^{13}$CN spectra. Each row corresponds to the molecule listed in the left-most panel. Each column corresponds to the disk listed above the top panel.}
\label{fig:spectra}
\end{figure*}

Integrated line fluxes and upper limits were extracted from the image cubes using elliptical regions with centers corresponding to the centroid of the continuum image (measured in \textsc{scikit-image}) and shapes and orientations based on the inclination and position angle reported in the literature for each disk (see Table \ref{tab:diskproperties}). The major axis of the elliptical region used to measure fluxes for each pair of isotopologue lines in a given disk (see Table \ref{tab:lineobservations}) is chosen to be sufficient to cover the \textgreater$3\sigma$ line emission appearing in the image cubes. For the pairs of weaker or undetected lines (H$^{13}$CN and DCN in the MWC 480 and HD 163296 disks) with less well-defined emitting regions, the major axis is instead based on the extent of the millimeter dust disk. Spectra of the four molecules, shown in Figure \ref{fig:spectra}, were extracted from the image cubes with the same elliptical regions used to measure line fluxes. The uncertainties on the integrated line fluxes were estimated in the following manner: first, the elliptical mask used to measure line fluxes in each cube was also used to measure fluxes at random positions in randomly chosen signal-free channels (with replacement) equal in number to the channels spanning the line. The fluxes measured within the mask were summed and then multiplied by the channel width to generate a simulated integrated flux measurement. For each image cube, 500 simulated integrated flux measurements were obtained, and their standard deviation was taken to be the uncertainty in the integrated line flux. 

\begin{figure}[htp]
\epsscale{0.75}
\plotone{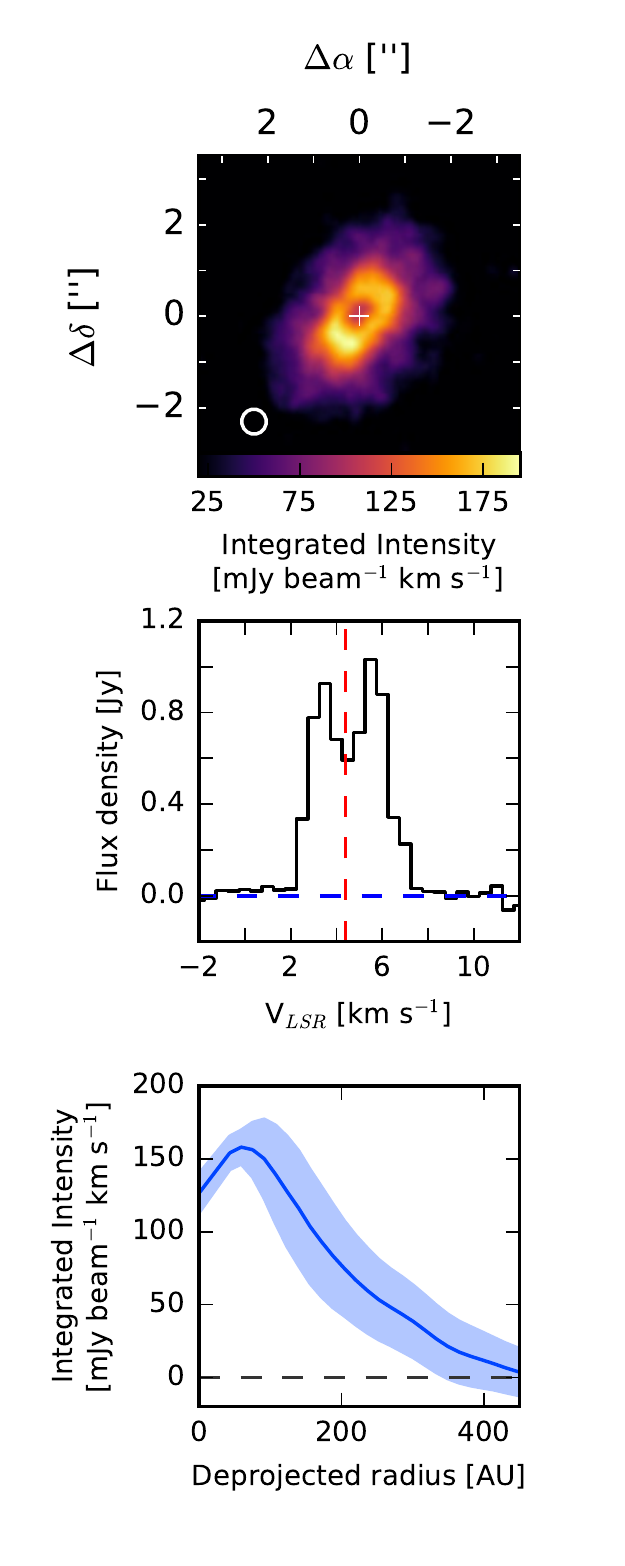}
\caption{Integrated intensity map, spectrum, and radial profile for HCN $J=3-2$ in the IM Lup disk. Top: integrated intensity map of HCN $J=3-2$ in the IM Lup disk. The color bar starts at 2$\sigma$, where $\sigma$ is the rms of the integrated intensity map listed in Table \ref{tab:lineobservations}. The synthesized beam is drawn in the lower left corner. Offset from the centroid of the continuum image in arcseconds is marked on the axes. Middle: spectrum of HCN. The horizontal blue line marks the zero flux density level. The vertical red line marks the systemic velocity (listed in Table \ref{tab:diskproperties}). Bottom: deprojected and azimuthally averaged radial profiles of integrated intensity. The colored ribbon shows the standard deviation in pixel intensities calculated at each annulus. Adopted distances are listed in Table \ref{tab:stellarproperties}, and adopted position angles and inclinations are listed in Table \ref{tab:diskproperties}. \label{fig:HCNdata}}
\end{figure}

The moment zero map, spectrum, and radial profile of HCN $J=3-2$  in the IM Lup disk are presented in Figure \ref{fig:HCNdata}. Because of the high signal-to-noise ratio of these data, we use the \textgreater3$\sigma$ emission to set integration ranges and to estimate the emitting region for measuring the fluxes of H$^{13}$CN and DCN. The uncertainties on the integrated line fluxes using the mask based on the 3$\sigma$ contours of HCN were estimated in a similar manner to that described in the preceding paragraph for the elliptical masks.

\begin{figure*}[htp]
\epsscale{1.2}
\plotone{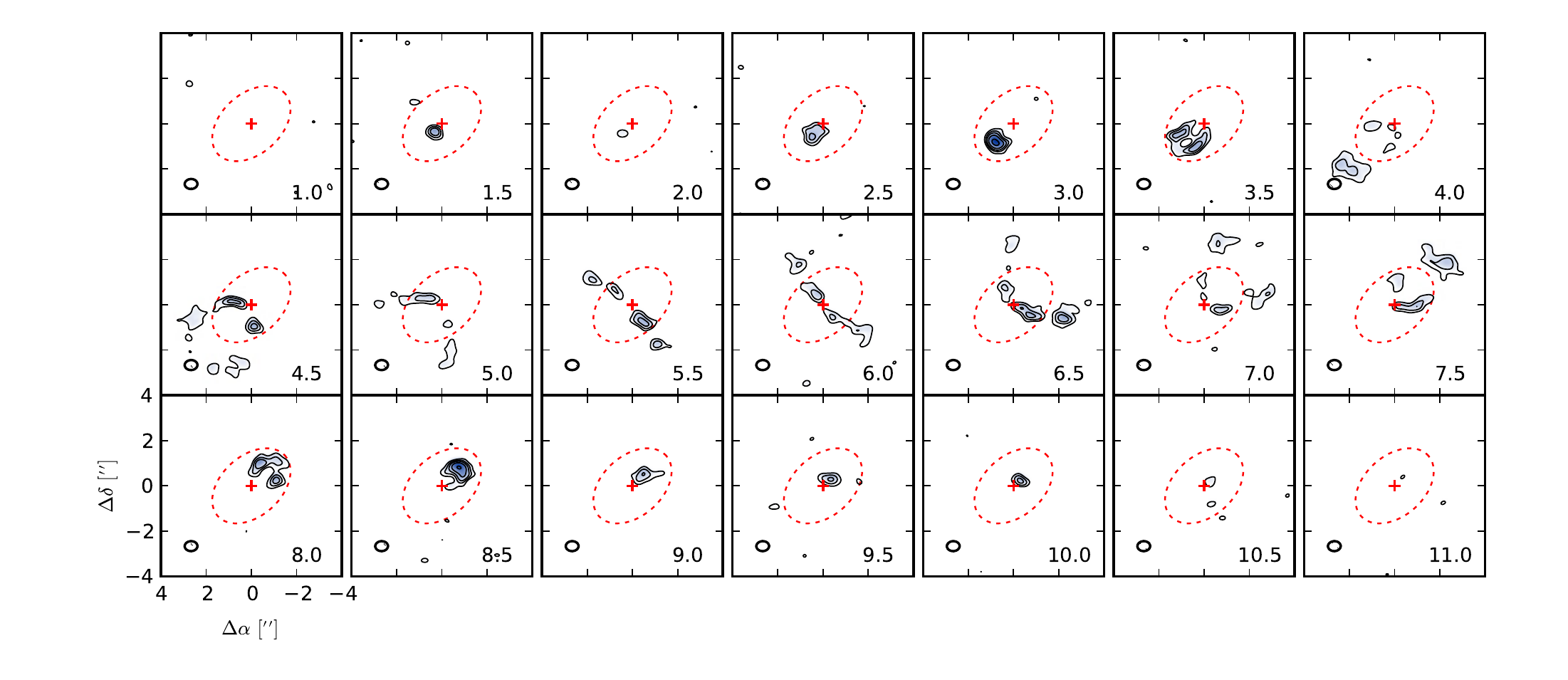}
\caption{Example channel maps of  H$^{13}$CO$^+$ $J = 3-2$ in the HD 163296 disk. Channel maps for the other detected lines are presented in Appendix \ref{sec:chanmaps}. Contours are drawn at [3, 5, 7, 10, 15, 20...]$\sigma$, where $\sigma$ is the channel rms listed in Table \ref{tab:lineobservations}. Synthesized beams are drawn in the lower left corner of each panel, and labels for the channel velocities in the LSRK frame (km s$^{-1}$) are shown in the lower right corners. Red crosses mark the position of the continuum image centroid. The offset from this centroid in arcseconds is marked on the axes in the lower left corner. The red dashed ellipse traces a projected radius of 2$''$ (assuming the P.A. and inclination listed in Table \ref{tab:diskproperties}) to highlight where the H$^{13}$CO$^+$ emission breaks (see Subsection \ref{sec:comments}).}
\label{fig:chanmapexample}
\end{figure*}

\section{Results}\label{sec:results}
In this section, we present the line detections, estimate the disk-averaged D/H ratios for each set of isotopologues, and describe the emission morphologies observed.

\subsection{Line Detections}
We classify a line as detected if emission exceeding the $3\sigma$ level is observed in at least three channels at positions consistent with the velocity field established by previous CO observations \citep[e.g.][]{2000ApJ...545.1034S, 2007AA...469..213I, 2009ApJ...700.1502A,2009AA...501..269P, 2011ApJ...734...98O}. Example channel maps for H$^{13}$CO$^+$ $J = 3-2$ in the HD 163296 disk are shown in Figure \ref{fig:chanmapexample}. The other channel maps are presented in Appendix \ref{sec:chanmaps}. 

H$^{13}$CO$^+$, DCN, and DCO$^+$ are detected in all six disks. H$^{13}$CN is detected in all but the IM Lup disk, for which we have also presented HCN observations. Most lines show the double-peaked spectra characteristic of rotating, inclined disks. In all cases of detected lines, both redshifted and blueshifted emission are observed. Integrated line fluxes and uncertainties are reported in Table \ref{tab:lineobservations}. Of the lines that have been observed previously, the fluxes measured with ALMA agree with Submillimeter Array observations of HCN $J=3-2$, DCO$^+$ $J = 3-2$, and DCN $J = 3-2$ reported in \citet{2010ApJ...720..480O} and \citet{2011ApJ...734...98O}. Separate ALMA observations of DCN and DCO+ $J=3-2$ in the HD 163296 disk at comparable spatial resolution and sensitivity have also been presented in \citet{2016ApJ...832..204Y} and show similar features.

\subsection{Estimates of Disk-averaged Deuterium Fractionation}
\subsubsection{Line Flux Ratios}
If a given pair of molecules with similar upper state energies and critical densities have co-spatial distributions, their flux ratios and column density ratios scale nearly linearly in the optically thin limit of local thermal equilibrium. The critical densities of the $J = 3-2$ lines of H$^{13}$CO$^+$ and H$^{13}$CN at 20 K in the optically thin limit are $1.3\times10^{6}$ and $6.6\times10^6$ cm$^{-3}$, respectively \citep{2015PASP..127..299S}. They decrease slightly with temperature, with the critical densities of the $J = 3-2$ lines of H$^{13}$CO$^+$ and H$^{13}$CN at 50 K in the optically thin limit being $9.5\times10^{5}$ and $4.1\times10^6$ cm$^{-3}$, respectively. Collision rates of rarer isotopologues are typically adopted from those of the main isotopologue due to lack of direct experimental constraints, and so we assume that DCO$^{+}$ and DCN $J = 3-2$ have critical densities similar to those of the non-deuterated isotopologues. Models of disk chemistry suggest that HCO$^+$, DCO$^+$, DCN, and HCN are abundant in regions of the disk where gas number densities range from $10^6$ to $10^8$ cm$^{-3}$, with HCO$^+$ and DCO$^+$ tending to populate denser and colder regions of the disk than DCN and HCN \citep{2007ApJ...660..441W,2010ApJ...722.1607W}. This indicates that LTE should be a good approximation for DCO$^+$ and H$^{13}$CO$^+$ $J = 3-2$ emission, but may be somewhat less accurate for DCN and H$^{13}$CN $J = 3-2$ emission (see also the discussion in \citealt{2001AA...377..566V,2007ApJ...669.1262P}). However, \citet{2015Natur.520..198O} tested non-LTE and LTE models for H$^{13}$CN in the MWC 480 disk and found that the derived column densities agreed within 20\%, suggesting that LTE is a reasonable first-order approximation for the targets discussed in this work. 

\begin{figure}[htp]
\epsscale{1.2}

\plotone{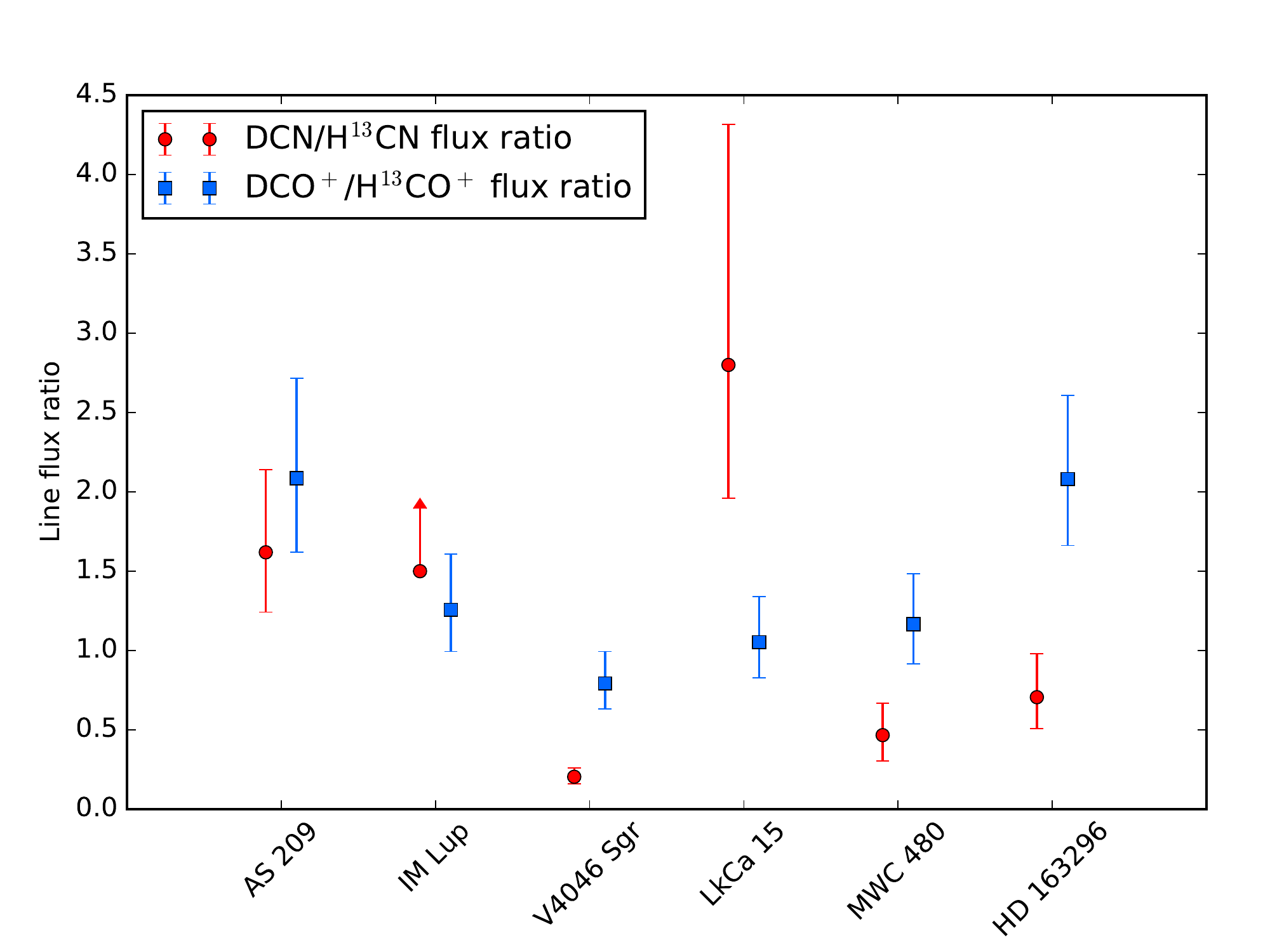}
\caption{Line flux ratios of DCN/H$^{13}$CN and DCO$^+$/H$^{13}$CO$^+$ $J = 3-2$. A lower limit is plotted for the DCN/H$^{13}$CN ratio for the IM Lup disk because H$^{13}$CN is not detected.}
\label{fig:fluxratio}
\end{figure}

The DCN/H$^{13}$CN and DCO$^+$/H$^{13}$CO$^+$ flux ratios for each disk are presented in Figure \ref{fig:fluxratio}. To account for the effects of ALMA's $\sim15\%$ systematic flux calibration uncertainty, we drew 100,000 simulated flux measurements for each line from a Gaussian distribution with mean and standard deviation given by the flux measurement and uncertainty provided in Table \ref{tab:lineobservations}. Each simulated flux measurement was multiplied by a flux rescaling factor drawn from a Gaussian centered at 1 with a standard deviation of 0.15. The plotted error bars correspond to a 68$\%$  confidence interval based on the distribution of simulated line flux ratios. 

The DCO$^+$/H$^{13}$CO$^+$ flux ratios are comparable for the whole sample (ranging from 0.8 to 2.1), while there is more scatter in the DCN/H$^{13}$CN flux ratios (spanning an order of magnitude from 0.2 to 2.8). The two transition disks represent the two extremes, with LkCa 15 having the highest DCN/H$^{13}$CN flux ratio and V4046 Sgr having the lowest. There are no clear trends with spectral type or stellar accretion rates, but the total number of sources is small. 

\subsubsection{Disk-averaged Abundance Ratios}
Detailed modeling of the radial variations in deuterium fractionation will be the subject of future papers focused on the best-resolved individual sources, but disk-averaged abundance ratios are calculated here for analysis of bulk trends and comparison with other interstellar and solar system measurements. Assuming two molecules XD and XH are co-spatial and have optically thin emission, the LTE approximation for their column density ratio is \citep[e.g.][]{2015PASP..127..266M}

\begin{equation}
\frac{N_\text{XD}}{N_\text{XH}} = \frac{(\int F \, dv \times Q_\text{rot}(T)\times \exp{\frac{E_\text{u}}{T_\text{ex}}})_\text{XD}}{(\int F \, dv \times Q_\text{rot}(T)\times \exp{\frac{E_\text{u}}{T_\text{ex}}})_\text{XH}}\times\frac{(\nu^3S\mu^2)_\text{XH}}{(\nu^3S\mu^2)_\text{XD}}
\end{equation}
where $N$ is the column density, $F$ is the flux density in Jy, $Q_\text{rot}$ is the rotational partition function as a function of gas kinetic temperature $T$, $E_u$ and $T_\text{ex}$ are the upper state energy and excitation temperature in K, $\nu$ is the transition frequency, $S$ is the line strength, and $\mu$ is the dipole moment. (Note that $T=T_\text{ex}$ since we assume LTE). 

We also neglect dust opacity when estimating the D/H ratios. Most of the flux originates from the outer disk, where previous models suggest that the millimeter dust continuum is optically thin \citep[e.g.][]{2012ApJ...760L..17P, 2016AA...588A.112G}. In addition, because the dust opacities at 1.1 mm and 1.4 mm are generally similar, the attenuation in the fluxes of the deuterated and non-deuterated isotopologues should largely cancel when taking ratios. 

The disk-averaged DCN/H$^{13}$CN and DCO$^+$/H$^{13}$CO$^+$ ratios were calculated using molecular line parameters obtained from the Cologne Database for Molecular Spectroscopy \citep{2001AA...370L..49M,2005JMoSt.742..215M}. Assumed excitation temperatures ranged from 15 to 75 degrees, which are typical of the disk regions probed by millimeter line observations \citep[e.g.][]{2007AA...467..163P, 2013ApJ...774...16R}. We note that column density ratios are less sensitive than absolute column densities to assumptions about excitation temperature, changing only by $\sim25\%$ in the temperature range evaluated, because of the similar excitation properties of the isotopologues. 

The column density ratios were then converted to DCN/HCN and DCO$^+$/HCO$^+$ abundance ratios assuming the local ISM $^{12}$C/$^{13}$C ratio of 69 \citep[e.g.][]{1999RPPh...62..143W}. Since H$^{13}$CN was not detected for IM Lup, we instead computed the DCN/HCN column density ratios directly from the DCN/HCN flux density ratio, 0.027$\substack{+0.008\\-0.006}$. The HCN flux for the IM Lup disk is about a factor of 55 larger than the H$^{13}$CN upper limit, which suggests that HCN may be optically thin over much of its emitting area. If, however, the HCN is optically thick, the resulting D/H ratios would be overestimated. The results are listed in Table \ref{tab:abundanceratios}. The DCO$^+$/HCO$^+$ ratios range from $\sim0.02$\textendash0.06, with a median of $\sim0.03$. The DCN/HCN ratios range from $\sim0.005$\textendash0.08, with a median of $\sim0.03$. Using parametric models to fit observations of the $J = 5-4$ line of DCO$^+$ and $J = 4-3$ lines of HCO$^+$ and H$^{13}$CO$^+$ in the HD 163296 disk, \citet{2013AA...557A.132M} derived a disk-averaged DCO$^+$/HCO$^+$ abundance ratio of 0.02, a factor of a few lower than our estimate. While their models confined DCO$^+$ largely within a radius of 160 AU, subsequent observations of other transitions with higher signal-to-noise ratios show emission extending out to $\sim$ 300 AU, which helps to account for our higher value \citep[][and this work]{2016ApJ...832..204Y,2015ApJ...813..128Q}.

\subsection{Emission Morphologies}
The classification of emission morphologies is dependent on the spatial resolution, since central or annular gaps may emerge at higher resolutions. However, since most of the disks are at similar distances and were observed at similar spatial resolutions, it is still informative to compare their molecular emission morphologies.  Based on the integrated intensity maps and radial profiles, the observed emission morphologies mostly fall into four categories: 
\begin{enumerate}
\item Ring-like
 \begin{enumerate}
  \item H$^{13}$CO$^+$ in the AS 209, IM Lup, LkCa 15, and MWC 480 disks
  \item DCO$^+$ in the AS 209, V4046 Sgr, MWC 480, and HD 163296 disks
  \item H$^{13}$CN in the AS 209 disk and HCN in the IM Lup disk
  \item DCN in the AS 209, V4046 Sgr, and HD 163296 disks
 \end{enumerate} 
\item Centrally peaked
 \begin{enumerate} 
  \item H$^{13}$CN in the V4046 Sgr, MWC 480, and HD 163296 disks
  \end{enumerate}
\item Profiles with multiple rings
  \begin{enumerate}
    \item H$^{13}$CO$^+$ in the HD 163296 disk
    \item DCO$^+$ in the IM Lup disk
  \end{enumerate}
\item Diffuse
  \begin{enumerate}
    \item H$^{13}$CO$^+$ in the V4046 Sgr disk
    \item DCO$^+$ in the LkCa 15 disk
  \end{enumerate}
\end{enumerate}
Like DCO$^+$ in the IM Lup disk and H$^{13}$CO$^+$ in the HD 163296 disk, the DCN emission in the LkCa 15 disk appears to feature an annular gap (i.e., a ring-like depression in the emission profile), but better spatial resolution would be necessary to confirm the morphology of the central component of the DCN emission. The emission morphologies of H$^{13}$CN in the LkCa 15 disk and DCN in the IM Lup and MWC 480 disks are not as easily categorized due to lower signal-to-noise ratios, but DCN in the IM Lup disk appears to be extended, while the DCN emission in MWC 480 and H$^{13}$CN emission in the LkCa 15 disk appear to be more compactly distributed.

No clear pattern emerges for relationships between the emission morphologies of different molecules, except that disks with more extended CO emission (see Table \ref{tab:diskproperties}) tend to feature more extended DCO$^+$ and H$^{13}$CO$^+$ emission, which is reasonable given that CO is a parent molecule. No given molecule exhibits the same emission morphology across all disks. Furthermore, for a given disk, the range of possibilities spans from similar morphologies for all molecules (AS 209) to distinct morphologies for each of the four molecules (LkCa 15). In the V4046 Sgr and HD 163296 disks, the H$^{13}$CN and DCN lines are emitted from a markedly more compact region compared to DCO$^+$ and H$^{13}$CO$^+$. In the LkCa 15 disk, the H$^{13}$CN emission originates from a more compact region than the other three molecules. Apart from the IM Lup disk, the detected emission extents for DCO$^+$ and H$^{13}$CO$^+$ in the channel maps are similar, although the radial profiles for several disks indicate that they peak in different locations. 

As a first-order approximation, the disk-averaged D/H analysis in the previous subsection assumed that the deuterated isotopologues are largely co-spatial with their non-deuterated forms. Here, we examine the effects of those assumptions in more detail. The four molecules targeted have similar excitation energies and critical densities. Thus, while the three-dimensional distributions of the molecules can only be inferred through complete structural models accounting for the gas density and temperature gradients of protoplanetary disks, molecules with very different emission patterns are unlikely to be co-spatial. Co-spatial molecules may have substantially different emission patterns if one of them is optically thick, as in the case of CO isotopologues in the cavity of transition disks \citep[e.g.][]{2016AA...585A..58V}, but the molecules targeted in this deuterium survey are orders of magnitude less abundant than CO and expected to be optically thin. 

For the IM Lup, LkCa 15, and HD 163296 disks, the radial intensity profiles of DCO$^+$ either show an enhancement beyond 200 AU or decline less steeply compared to the H$^{13}$CO$^+$ emission profiles. In the V4046 Sgr disk, there is a deep central gap in DCO$^+$ emission, whereas the H$^{13}$CO$^+$ emission is broad and diffuse. This suggests that the disk-averaged DCO$^+$/HCO$^+$ abundance ratios calculated in the previous section would likely lie between higher local values in the outer disk and lower local values in the inner disk.

\begin{deluxetable*}{ccccccc}
\tablecaption{Flux Ratios and Disk-averaged D/H Ratios \label{tab:abundanceratios}}
\tablehead{
\colhead{Source} &\multicolumn{2}{c}{Flux ratio}&\multicolumn{2}{c}{Abundance ratio assuming T$_\text{ex} = 15$ K}& \multicolumn{2}{c}{Abundance ratio assuming T$_\text{ex} = 75$ K}\\
&\colhead{$\frac{\text{DCO}^+}{\text{H$^{13}$CO}^+}$} &\colhead{$\frac{\text{DCN}}
{\text{H$^{13}$CN}}$}&\colhead{$\frac{\text{DCO}^+}{\text{HCO}^+}$} &\colhead{$\frac{\text{DCN}}
{\text{HCN}}$}&\colhead{$\frac{\text{DCO}^+}{\text{HCO}^+}$} &\colhead{$\frac{\text{DCN}}
{\text{HCN}}$}}
\vskip 5mm
\startdata
AS 209&$2.1\substack{+0.6\\-0.5}$&$1.6\substack{+0.5\\-0.4}$&$0.048\substack{+0.014 \\ -0.011}$&$0.036\substack{+0.012 \\ -0.008}$&$0.06\substack{+0.02 \\ -0.01}$&$0.045\substack{+0.014 \\ -0.010}$\\
IM Lup&$1.3\substack{+0.3\\-0.3}$&\textgreater1.5&$0.029\substack{+0.008 \\ -0.006}$&$0.044\substack{+0.013\\-0.001}$\tablenotemark{*}&$0.036\substack{+0.010 \\ -0.008}$&$0.057\substack{+0.017\\-0.013}$\tablenotemark{*}\\
V4046 Sgr&$0.8\substack{+0.2\\-0.2}$&$0.20\substack{+0.05\\-0.04}$&$0.018\substack{+0.005 \\ -0.004}$&$0.005\substack{+0.001 \\ -0.001}$&$0.023\substack{+0.006 \\ -0.005}$&$0.006\substack{+0.002 \\ -0.001}$\\
LkCa 15&$1.1\substack{+0.3\\-0.2}$&$2.8\substack{+1.5\\-0.8}$&$0.024\substack{+0.007 \\ -0.005}$&$0.06\substack{+0.03\\ -0.02}$&$0.030\substack{+0.008 \\ -0.006}$&$0.08\substack{+0.04 \\ -0.02}$\\
MWC 480&$1.2\substack{+0.3\\-0.2}$&$0.47\substack{+0.20\\-0.16}$&$0.027\substack{+0.007 \\ -0.006}$&$0.010\substack{+0.004 \\ -0.004}$&$0.034\substack{+0.009 \\ -0.007}$&$0.013\substack{+0.005 \\ -0.004}$\\
HD 163296&$2.1\substack{+0.5\\-0.4}$&$0.7\substack{+0.3\\-0.2}$&$0.048\substack{+0.012 \\ -0.009}$&$0.016\substack{+0.006 \\ -0.004}$&$0.060\substack{+0.015 \\ -0.012}$&$0.019\substack{+0.008 \\ -0.005}$
\enddata
\tablenotetext{*}{Estimated directly from DCN/HCN flux ratio}
\end{deluxetable*}

Compared to DCO$^+$ and H$^{13}$CO$^+$, DCN and H$^{13}$CN appear to have more similar distributions in this survey, except DCN features a central gap more often than H$^{13}$CN. Thus, the disk-averaged DCN/HCN ratios calculated in the previous section would likely be intermediate between higher local values in the outer disk and lower local values in the inner disk for V4046 Sgr, LkCa 15, and HD 163296. The LkCa 15 disk features the greatest discrepancy between H$^{13}$CN and DCN emission: whereas H$^{13}$CN emission is confined to a compact region at the center of the disk, DCN emission is detected out to hundreds of AU. Thus, the local DCN/HCN enhancement in the LkCa 15 relative to the other disks may be even more extreme than what is indicated by the disk-averaged estimates, for which LkCa 15 already has the largest DCN/HCN ratio.  
\subsection{Additional Comments on Individual Sources}\label{sec:comments}
In this subsection, we discuss the features observed in the integrated intensity maps ( Fig. \ref{fig:mom0}), radial profiles (Fig. \ref{fig:radialprofile}), and channel maps (Appendix \ref{sec:chanmaps}) in the context of each individual source. 
\subsubsection{AS 209}
\begin{figure}[htp]
\epsscale{1.2}
\plotone{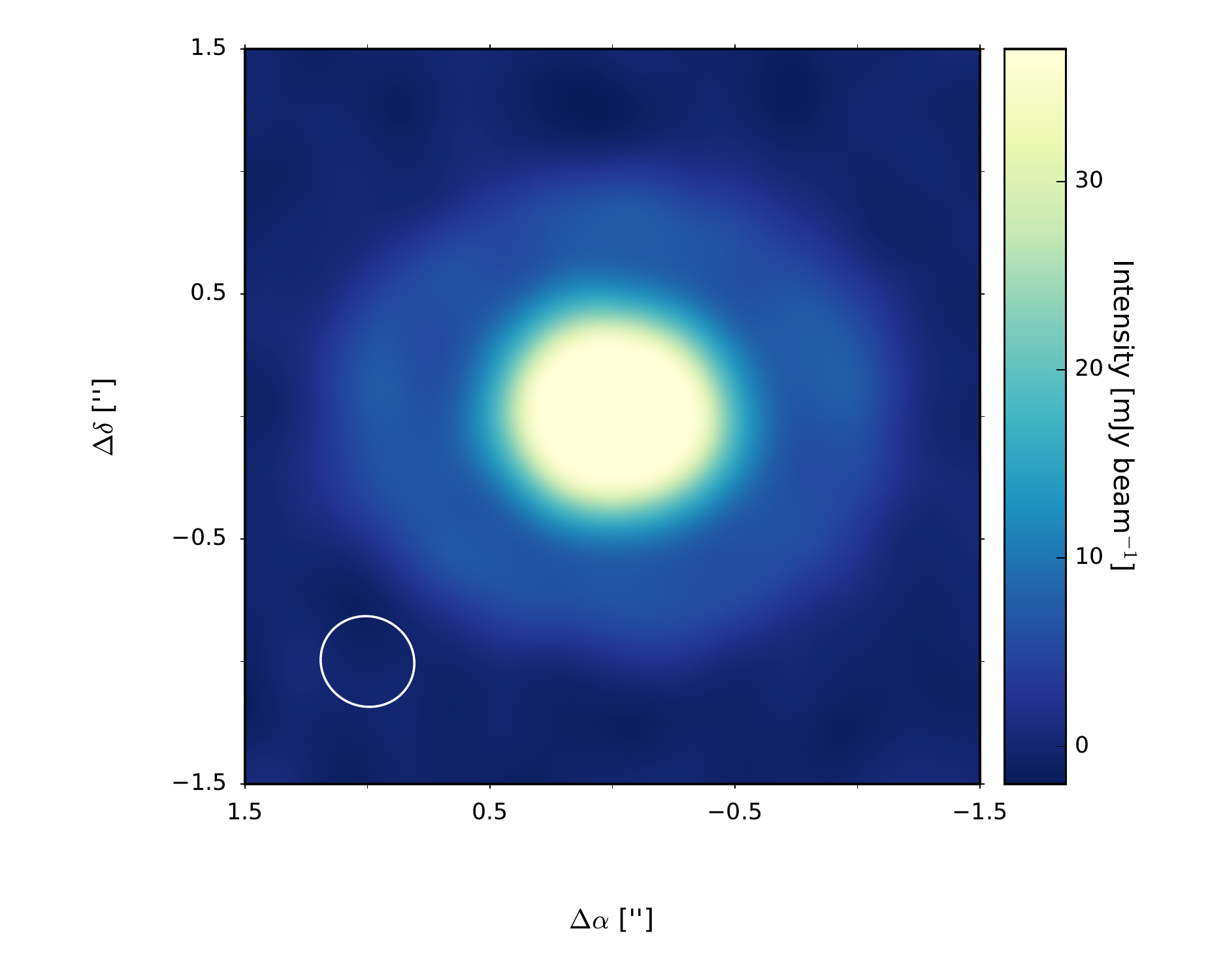}
\caption{Intensity map of the AS 209 258 GHz continuum, imaged with uniform weighting. Note the dark annulus at a radius of $\sim0.7''$ from the disk center. The color bar saturates at half the maximum intensity in order to show the outer ring more clearly. The synthesized beam is shown in the lower left corner. Axes are labeled with offsets from the centroid of the continuum image.}
\label{fig:AS209continuum}
\end{figure}

All four molecules have ring-like emission morphologies. The radial emission profiles of DCN, H$^{13}$CN, and H$^{13}$CO$^+$ peak at $R\sim50$ AU. DCO$^+$ has a shallower central depression in the emission profile compared to the other molecules, but its emission profile begins to decline rapidly past $\sim80$ AU. The radial intensity profile (see Fig. \ref{fig:radialprofile}) of the continuum shows a break at  $\sim0.7''$ ($\sim90$ AU). The 258 GHz continuum emission, re-imaged with multi-scale \textsc{CLEAN} as well as uniform weighting for better spatial resolution (synthesized beam: $0\farcs39\times0\farcs37\,(64\fdg74)$), is shown again in Figure \ref{fig:AS209continuum} and exhibits a dark annulus between the inner disk and outer emission ring. 
\subsubsection{IM Lup}
The DCO$^+$ emission features double rings. The first peak in the radial profile occurs at $R\sim110$ AU, and the second occurs at $R\sim310$ AU. H$^{13}$CO$^+$ has a ring-like morphology with the radial emission profile peaking at 130 AU, which is a wider central gap than the other H$^{13}$CO$^+$ emission rings observed in the survey, including even the transition disks. (Note that these observations were previously presented in \citet{2015ApJ...810..112O} assuming a distance of 155 pc to IM Lup). H$^{13}$CN is not detected, but HCN $J=3-2$ has a ring-like profile peaking at $\sim 60 $ AU.  DCN is only marginally detected in a few channels (see Appendix \ref{sec:chanmaps}), but the emission is spatially consistent with HCN and indicates that DCN is also extended in the disk. 

\subsubsection{V4046 Sgr}
The central dust cavity reported in \citet{2013ApJ...775..136R} is also resolved in the 258 GHz continuum ALMA data. The DCO$^+$ and DCN lines both have ring-like emission morphologies, with the averaged radial emission profiles peaking at 70 and 50 AU, respectively. \citet{2013ApJ...775..136R} determined that most of the dust mass in the V4046 Sgr disk is confined to a ring with a peak in surface density at $R = 37$ AU and a FWHM of 16 AU. Thus, the DCN emission is largely coincident with the outer regions of the millimeter dust emission. In contrast, the DCO$^+$ emission originates mostly outside the millimeter dust disk. The DCN emission appears to be brighter on the western side of the disk, but more sensitive observations would be required to confirm whether the asymmetry is real. The H$^{13}$CO$^+$ radial profile is notably flatter than those of the other lines detected in V4046 Sgr, while the H$^{13}$CN emission profile is centrally peaked. The H$^{13}$CN integrated flux is also exceptionally large compared to the other disks in the survey. Although V4046 Sgr is much closer than the other disks, its H$^{13}$CN integrated flux would still be about a factor of two higher than that of the other disks in the survey if they were at the same distance. The channel maps (Appendix \ref{sec:chanmaps}) indicate that emission above the 3$\sigma$ level for DCO$^+$ and H$^{13}$CO$^+$ is nearly absent inside the cavity. 
\subsubsection{LkCa 15}
The central dust cavity reported in \citet{2011ApJ...732...42A} is also resolved in the 258 GHz continuum ALMA data. While the DCN emission observed in the other disks is typically compact (with the possible exception of the IM Lup disk), the DCN emission in the LkCa 15 disk is clearly extended. The radial profile and integrated intensity map indicate that an annular gap in DCN emission separates a compact component near the center and an emission ring peaking at $\sim$ 180 AU near the edge of the millimeter dust emission, but higher-sensitivity observations will be needed to clarify the structure. The H$^{13}$CO$^+$ radial profile peaks at $\sim$ 40 AU, whereas the DCO$^+$ radial profile is relatively flat and extended. 
\subsubsection{MWC 480}
The H$^{13}$CO$^+$ and DCO$^+$ radial emission profiles peak at $\sim$ 40 AU. The H$^{13}$CN profile is centrally peaked. Although the DCN emission is weak and appears in only a few channels (Appendix \ref{sec:chanmaps}), it is consistent with the Keplerian rotation pattern established by the other three lines observed. The DCN emission appears to feature a central dip, but the signal-to-noise ratio is too low to be definitive. 

\subsubsection{HD 163296}
The H$^{13}$CN profile is centrally peaked, while the DCN radial profile peaks at $\sim50$ AU. Our DCN data do not show clear evidence for the offset from center noted by \citet{2016ApJ...832..204Y} for  separate HD 163296 observations. However, the average DCN radial profiles appear similar for both observations. The DCO$^+$ emission is ring-like, with the radial profile peaking at $\sim70$  AU. Like \citet{2016ApJ...832..204Y}, we find that the DCO$^+$ $J=3-2$ emission peaks in brightness northwest of the disk center. A similar feature was also reported for the DCO$^+$ $J=5-4$ line by \citet{2013AA...557A.132M}. The H$^{13}$CO$^+$ line features a compact  emission ring peaking at $\sim$ 50 AU, but its most striking feature is the emission break near $R\sim200$ AU, as seen most clearly in the channel maps in Figure \ref{fig:chanmapexample} and faintly in the integrated intensity map in Figure \ref{fig:mom0}. A similarly positioned break is hinted at in a few channels in the DCO$^+$ $J  = 3-2$ channel maps in Appendix \ref{sec:chanmaps} as well as the channel maps presented for the DCO$^+$ $J = 4-3$ line in \citet{2015ApJ...813..128Q}, but in neither case is the break as unambiguous as for H$^{13}$CO$^+$. 
\section{Discussion} \label{sec:discussion}
\subsection{Comparison of D/H ratios}

\begin{figure}[htp]
\begin{center}
\subfloat{\includegraphics[scale = 0.5]{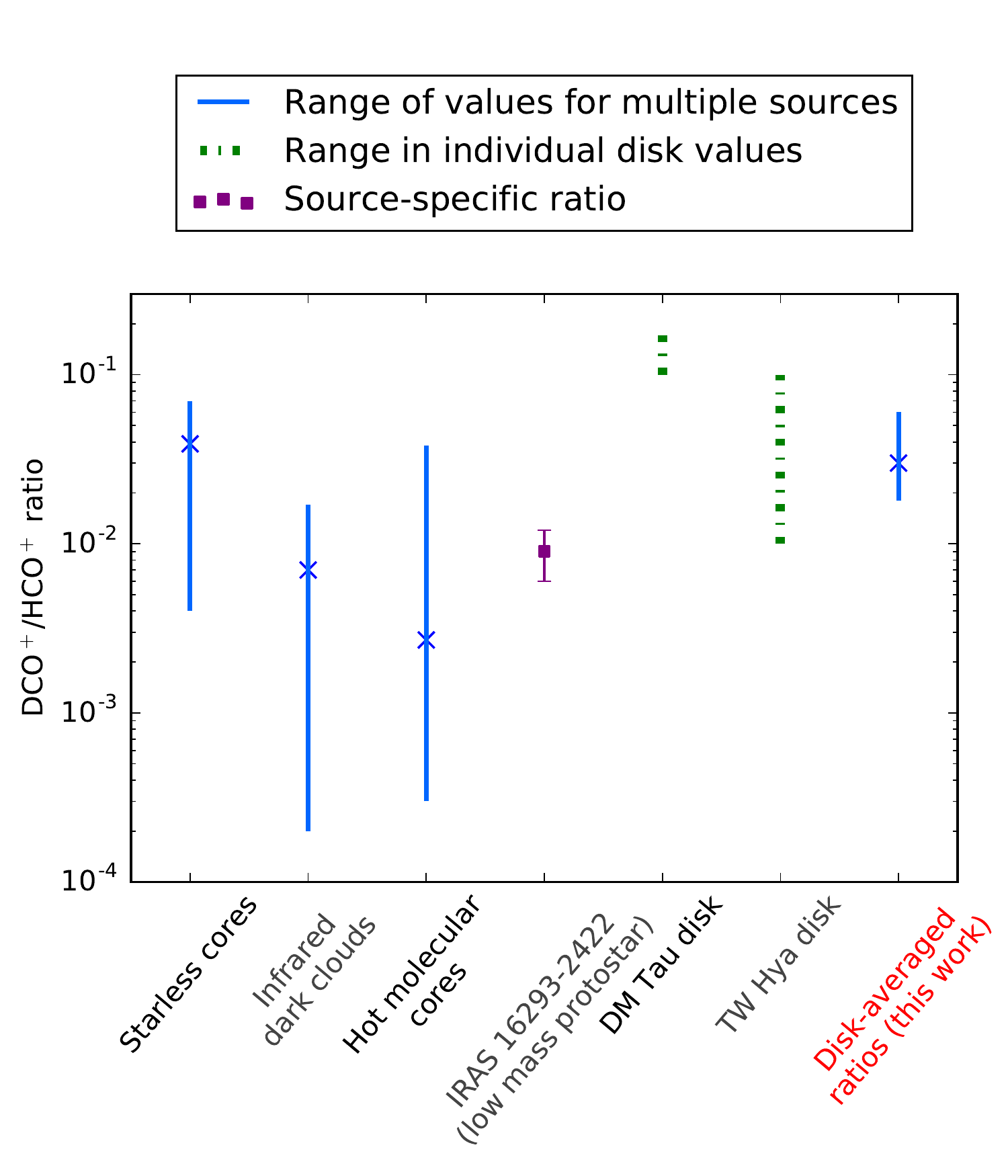}}
\end{center}
\begin{center}
\subfloat{\includegraphics[scale = 0.5]{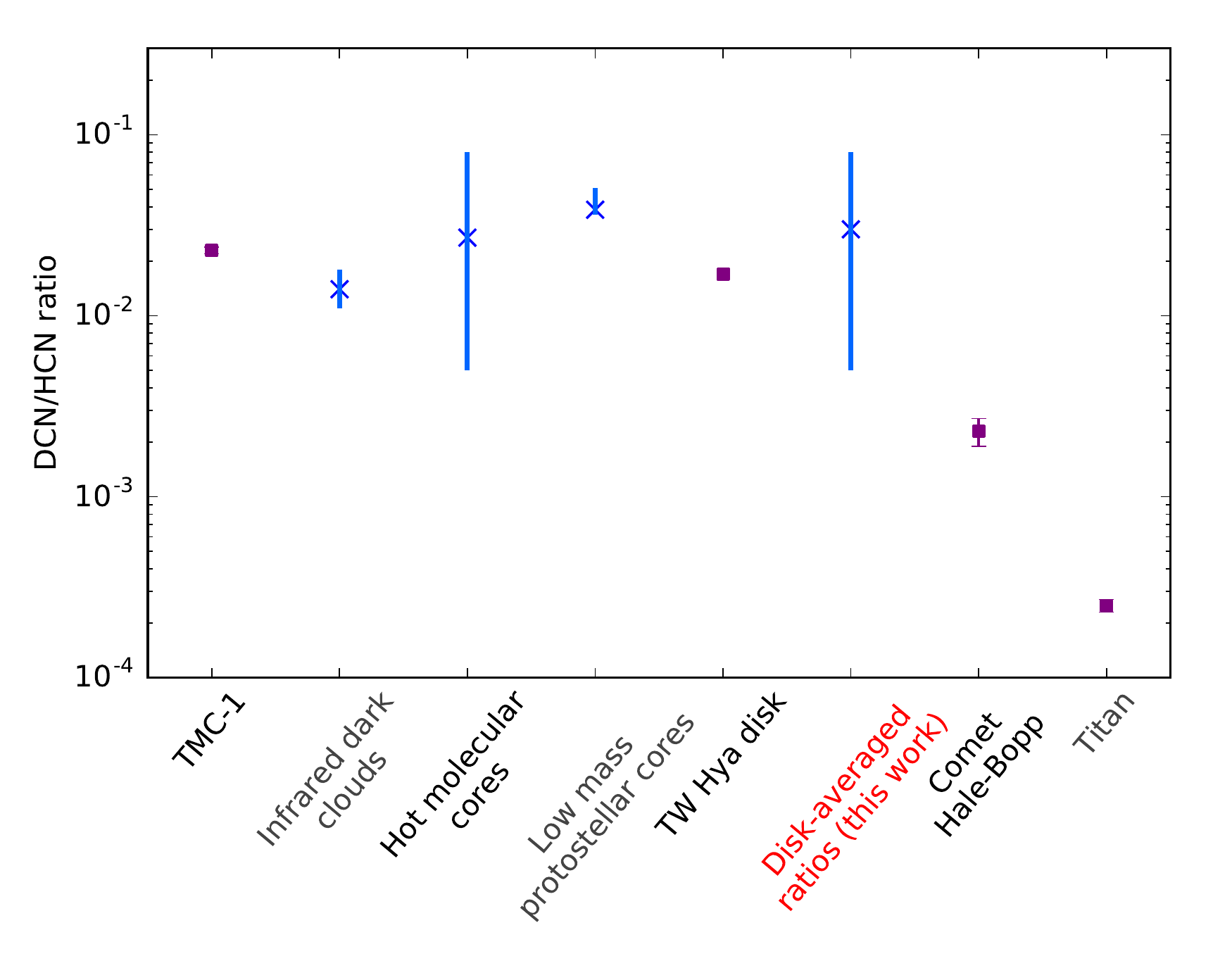}}

\end{center}
\caption{Comparison of literature D/H ratios to disk-averaged D/H ratios estimated in this work. \textbf{Top:} DCO$^+$/HCO$^+$ abundance ratios. \textbf{Bottom:} DCN/HCN abundance ratios. Blue markers plotted for starless cores \protect\citep{1995ApJ...448..207B, 2002ApJ...565..344C, 2006AA...455..577T}, infrared dark clouds \protect\citep{2011AA...534A.134M,2015AA...579A..80G}, hot molecular cores \protect\citep{2015AA...579A..80G},  low-mass protostellar cores \protect\citep{2002AA...381.1026R}, and protoplanetary disks (this work) represent the range of values observed for multiple objects in each category.  Crosses mark the median values. Dashed green lines plotted for the DM Tau \protect\citep{2015AA...574A.137T} and TW Hya \protect\citep{2008ApJ...681.1396Q} disks represent the range of values derived for various radii. Purple squares represent values for the following individual sources: TMC-1 \protect\citep{1987IAUS..120..311W}, low-mass protostar IRAS 16293-2422 \protect\citep{2002AA...390.1001S}, the TW Hya disk \protect\citep{2012ApJ...749..162O}, comet Hale-Bopp \protect\citep{1998Sci...279.1707M}, and Titan \protect\citep{2016AJ....152...42M}. }
\label{fig:fractionationplots}
\end{figure}

The range of disk-averaged DCN/HCN and DCO$^+$/HCO$^+$ abundance ratios estimated for the survey are plotted in Figure \ref{fig:fractionationplots}, along with measurements reported in the literature for other interstellar environments and solar system objects. 
\subsubsection{Comparison to Other Sources}
As with previous findings for the DM Tau \citep{2015AA...574A.137T}, TW Hya \citep{2003AA...400L...1V, 2008ApJ...681.1396Q, 2012ApJ...749..162O}, and HD 163296 \citep{2013AA...557A.132M} disks, our survey indicates that the high deuterium fractionation in earlier star formation stages can also persist during the disk stage for a diverse set of sources. The DCO$^+$/HCO$^+$ estimates for the six disks are comparable to values measured for starless cores \citep{1995ApJ...448..207B, 2002ApJ...565..344C, 2006AA...455..577T}, and slightly higher than the ratio derived for low-mass protostar IRAS 16293-2422 \citep{2002AA...390.1001S}. They are also a factor of a few higher than typical values measured for infrared dark clouds \citep{2011AA...534A.134M,2015AA...579A..80G} and hot molecular cores \citep{2015AA...579A..80G}, both of which are associated with high-mass star formation rather than the low- and intermediate-mass stars from this work. 

Likewise, the disk-averaged DCN/HCN abundance ratios for our survey are comparable to the values obtained for infrared dark clouds \citep{2015AA...579A..80G}, hot molecular cores \citep{2015AA...579A..80G}, and low-mass protostellar cores \citep{2000AA...361..388R}. \citet{2012ApJ...749..162O} derived a disk-averaged value of 0.017 for the TW Hya disk, which lies within the range of values estimated in this work. DCN is a particularly useful tracer of deuterium chemistry because its fractionation has been measured in a range of sources across the stages of star and planetary system formation, including within our own solar system. Most of the disk-averaged abundance ratios are higher than that of comet Hale-Bopp, which has a DCN/HCN ratio of 2.3$\pm0.4\times10^{-3}$ \citep{1998Sci...279.1707M}. However, the DCN/HCN ratio of the V4046 Sgr disk is within a factor of a few of that of Hale-Bopp. Since V4046 Sgr is much older than the other disks in the survey, it would be interesting to observe other old disks to investigate whether DCN/HCN fractionation depends on disk age, which would in turn help to constrain the origin of cometary ices. (TW Hya is also advanced in age, but thought to be younger than V4046 Sgr). The disk-averaged DCN/HCN ratios are also typically several orders of magnitude larger than that of Titan \citep{2016AJ....152...42M}. While Titan features very active photochemistry, its current deuterium fractionation may still partially reflect primordial fractionation \citep{2008ApJ...689L..61C}.

\subsubsection{Comparison to Disk Models}
For a general T Tauri disk model, \citet{2002AA...386..622A} predict that both DCN/HCN and DCO$^+$/HCO$^+$ vary between 0.01 to 0.1 from 50 to 400 AU.  The T Tauri disk model from \citet{2007ApJ...660..441W} predicts that DCN/HCN varies between 0.01 to 0.1 from 50 to 400 AU (similar to Aikawa et al.), and DCO$^+$/HCO$^+$ varies from 0.1 to 2. Both the Aikawa and Willacy models evolve from initial abundances mirroring those of dark clouds and produce DCO$^+$ through the H$_2$D$^+$ pathway (Eq. \ref{dcop_1}). DCN is primarily produced through the CH$_2$D$^+$ pathway (Eq. \ref{DCN_warm_1} and \ref{DCN_warm_2}) in the models from \citet{1999ApJ...526..314A}, which were extended into two dimensions in \citet{2002AA...386..622A}. The Willacy models, in contrast, produce a substantial portion of DCN through the H$_2$D$^+$ pathway (Eq. \ref{DCN_cold_1} and \ref{DCN_cold_2}). The Willacy models also incorporate grain-surface reactions and multiply-deuterated species in their chemical network. 

While we calculate disk-averaged abundance ratios rather than D/H ratios as a function of radius, we find that our estimated disk-averaged ratios are for the most part comparable to the values quoted above for disk models. V4046 Sgr does feature a disk-averaged DCN/HCN ratio that is a factor of a few lower than the values in the Aikawa and Willacy models, but  it is a binary star system that is much older than the disk models (typically $\leq$ 1 Myr), so its disk properties may have evolved substantially. 

Generally, the abundance ratios derived for the disk survey are compatible with scenarios of \textit{in situ} deuterium fractionation. A more compelling case for \textit{in situ} fractionation may be the diverse and highly structured emission patterns observed for the same molecules across different disks, which would suggest that their distributions are strongly responsive to the particular disk environment. Chemical modeling of each disk would be necessary to draw firmer conclusions regarding the extent to which $\textit{in situ}$ deuteration processes are required to explain the observed chemistry. 
\subsubsection{Considerations on $^{12}$C/$^{13}$C Fractionation}
H$^{13}$CN and H$^{13}$CO$^+$ were targeted because line emission from the main isotopologues was assumed to be too optically thick to trace their distributions. Of the disks in the survey with previously reported detections of HCO$^+$ $J = 3-2$ \citep{2010ApJ...720..480O, 2011ApJ...734...98O}, the HCO$^+$ to  H$^{13}$CO$^+$ $J = 3-2$ line flux ratios range from $\sim$ 13 to 25. The flux ratios of the HCN to H$^{13}$CN $J = 3-2$ lines (based on observations of HCN $J = 3-2$ reported in  \citealt{2010ApJ...720..480O, 2011ApJ...734...98O,2015ApJ...807L..15G}) range from $\sim$10 to 60, with AS 209 and V4046 Sgr on the low end and LkCa 15 on the high end. Assuming that the $^{12}$C/$^{13}$C ratio is  the ISM value of 69 \citep[e.g.][]{1999RPPh...62..143W}, the flux ratios indicate that HCO$^+$ $J = 3-2$ is optically thick for all the targets, while HCN $J = 3-2$ is likely optically thick for at least part of the sample. 

In addition to the adoption of a $^{12}$C/$^{13}$C ratio, inferences about deuterium fractionation based on optically thin isotopologues require the assumption that the $^{12}$C and $^{13}$C isotopologues are co-spatial. The appropriateness of these two assumptions depends on the extent of $^{12}$C/$^{13}$C fractionation in disks, which has not yet been constrained observationally. Obtaining such constraints will improve interpretations of disk chemistry, but the ALMA systematic flux calibration uncertainty and uncertainties in disk temperature structures currently pose some challenges in measuring $^{12}$C/$^{13}$C fractionation.

Models offer some insight into the potential impact of carbon fractionation on the conclusions. H$^{13}$CO$^+$ may be somewhat depleted in the outer disk compared to HCO$^+$ due to the less-abundant $^{13}$CO being photodissociated more easily than $^{12}$CO, but the full T Tauri and Herbig Ae disk models from \citet{2014AA...572A..96M} indicate that its effects should be modest interior to a couple hundred AU, the region that our observations trace. In contrast, full T Tauri disk models also indicate that  the H$^{13}$CN/HCN ratios may vary by up to factors of a few in the inner disk \citep{2009ApJ...693.1360W}. (Similar calculations have not been performed yet for transition disks and for Herbig Ae disks, but it is possible that fractionation could vary from the full T Tauri disk models due to the different radiation fields.) Thus, the use of H$^{13}$CN to trace the distribution of HCN ought to be treated with more caution than the use of H$^{13}$CO$^+$ to trace HCO$^+$, particularly if HCN may be optically thin enough in certain disks to trace distributions directly. Fitting models to observations of multiple transitions of $^{12}$C and $^{13}$C isotopologues could provide better insight into the robustness of assumed isotope ratios. However, even if the $^{12}$C/$^{13}$C ratios are overestimated by a factor of a few, the D/H levels derived for this survey would still be indicative of efficient fractionation.

Greater uncertainty in the H$^{13}$CN/HCN ratios may partially account for the greater variability of DCN/HCN abundance ratios derived in this work compared to DCO$^+$/HCO$^+$. Nevertheless, we note that the unusually bright H$^{13}$CN $J = 3-2$ detection in the V4046 Sgr disk relative to the rest of the sample is not likely to be due primarily to carbon fractionation, since HC$^{15}$N emission is also exceedingly strong in the V4046 Sgr disk compared to the other disks \citet{Guzmaninpress}. 

\subsubsection{Potential Effects of Selection Bias}
The disks discussed in this work are often considered to be benchmark systems for studies of disk structure and chemistry due to their large size, bright molecular lines, and relative isolation \citep[e.g.][]{2007AA...467..163P, 2007ApJ...660..441W, 2010ApJ...720..480O, 2011ApJ...734...98O, 2013ApJ...774...16R, 2013ApJ...775..136R}. The diversity of molecular emission patterns observed in this survey of six disks demonstrates the value in observing multiple systems with varying physical properties in order to explore the range of chemical outcomes. 

On the other hand, the characteristics that have often made these disks favored for chemical studies may also lead to some bias, leaving some facets of deuterium chemistry not fully explored.  Because the targets were primarily selected on the basis of previous DCO$^+$ detections, the similar DCO$^+$/H$^{13}$CO$^+$ flux ratios measured with ALMA may be a result of  the selection process favoring disks with higher levels of deuterium enrichment. Furthermore, even though MWC 480 and HD 163296 are chemically rich, observations of other Herbig Ae stars suggest that their disks tend to be more chemically poor than those around T Tauri stars, perhaps because the higher UV luminosities of Herbig Ae stars lead to more photodissociation \citep[e.g.][]{2008AA...491..821S, 2010ApJ...720..480O, 2011ApJ...734...98O, 2016AA...592A.124G}. In addition, because the survey covers only relatively massive and isolated disks (in part a legacy of which sources were most readily observable on earlier mm/sub-mm instruments), the most extreme effects of the interstellar radiation field are not well constrained. Finally, because the disks in the survey have large radial extents, they have very cold outer regions that would be expected to be most hospitable to deuterium fractionation. Thus, extension of the survey toward lower-mass or more compact disks would provide additional insight into what aspects of deuterium chemistry observed in this work are most common in disks. Recent ALMA continuum and CO surveys of disks in various star-forming regions have now created a rich selection of potential chemistry follow-up targets spanning a range of stellar and disk properties \citep[e.g.][]{2016ApJ...828...46A, 2016ApJ...827..142B, 2016AA...593A.111T, 2016ApJ...831..125P}. 
\subsection{Chemical Implications}
Prior to this work, the LkCa 15, TW Hya, and HD 163296 disks were the only disks in which both DCN and DCO$^+$ had been detected \citep{2001PhDT.........1Q, 2003AA...400L...1V, 2012ApJ...749..162O, 2016ApJ...832..204Y}. With the additional detections from this survey, the number of sources in which DCN has been observed has now doubled. 
\subsubsection{Formation of DCN}
As noted in the introduction, the DCN distribution was reported to be much more radially compact than that of DCO$^+$ in the TW Hya and HD 163296 disks. The previous LkCa 15 observations were not sufficiently resolved to compare DCN and DCO$^+$. For TW Hya, the spatial differentiation has been attributed to production of DCN through CH$_2$D$^+$ (e.g., Eq. \ref{DCN_warm_1} and \ref{DCN_warm_2}) in the warm inner disk and production of DCO$^+$ through H$_2$D$^+$ (Eq. \ref{dcop_1}) in the cold outer disk. 

In our survey, we observe the following relationships between DCN and DCO$^+$ emission:
\begin{enumerate}
  \item Substantial spatial differentiation between DCN and DCO$^+$ in the V4046 Sgr and HD 163296 disks. In these sources, DCN is much more compact and peaks in intensity within the central gap of DCO$^+$ emission. This is similar to the TW Hya disk, which suggests that DCN production in these disks is also dominated by the CH$_2$D$^+$ pathway.
  \item Similar radial extents for DCN and DCO$^+$ emission in the AS 209 and LkCa 15 disks. This may be indicative of more overlap in the formation pathways for DCN and DCO$^+$ in these disks than in the V4046 Sgr, HD 163296, or TW Hya disks. 
  \item Ambiguous relationship due to the DCN detections with low signal-to-noise ratios in the IM Lup and MWC 480 disks.
\end{enumerate}

DCN and DCO$^+$ are not necessarily co-spatial in the disks in which they have similar emission profiles, since it is possible that they are confined to different heights. Even so, the results indicate that the inner disk/outer disk tracer dichotomy that was previously suggested for DCN and DCO$^+$ by the TW Hya and HD 163296 data does not apply to all disks. 

The DCN emission profile in the LkCa 15 disk is particularly interesting, since it appears to peak once near the central star and again at $\sim$ 180 AU. Fitting a vertically isothermal model to CO isotopologue observations,   \citet{2007AA...467..163P} derived a gas temperature of $\sim22$ K at 100 AU from the central star. While this method may lead to imprecise temperature values due to real disks having vertical temperature gradients, the effect would likely be to overestimate rather than underestimate the temperature at which DCN resides, since $^{12}$CO typically probes the upper regions of the warm molecular layer. Therefore, much of the DCN emission in the LkCa 15 disk originates from disk regions expected to be too cold for CH$_2$D$^+$ to be the dominant deuterated ion, which suggests that an H$_2$D$^+$ formation pathway is a significant contributor in the outer LkCa 15 disk. The marginal detection of DCN in the IM Lup disk appears to be similarly extended into the cold outer disk, but more sensitive observations would be necessary to confirm this. 

Unlike LkCa 15, which has a gas disk estimated to extend to $\sim$900 AU \citep{2012ApJ...747..136I}, the TW Hya gas disk only extends to $\sim$215 AU \citep{2012ApJ...744..162A}. The TW Hya disk may simply not have enough material in its outer regions for a cold DCN formation pathway to be important. An alternative scenario is that radial drift in the LkCa 15 disk creates a radial thermal inversion allowing for the warm CH$_2$D$^+$ pathway to produce DCN just outside the millimeter dust disk \citep{2016ApJ...816L..21C}. Then, other disks exhibiting evidence of radial drift, such as TW Hya and AS 209 \citep{2014AA...564A..93M, 2012ApJ...760L..17P}, might be expected to show similar DCN emission, but they do not at the current sensitivity.   
 \subsubsection{Formation of DCO$^+$}
 Chemical models generally predict two types of radial profiles for DCO$^+$: 
 \begin{enumerate}
 \item Ring-like, if forming predominantly from H$_2$D$^+$ and CO. This is a consequence of H$_2$D$^+$ being readily destroyed by the back-reaction of \ref{eq:H2D+} as well as by gas-phase CO in the warmer inner disk \citep[e.g.][]{1999AA...351..233A, 2001AA...371.1107A, 2007ApJ...660..441W}.
 \item Radially decreasing (or with a very small central gap), if forming predominantly from CH$_2$D$^+$ and CO, since CH$_2$D$^+$ is abundant in warmer gas compared to H$_2$D$^+$ \citep{2015ApJ...802L..23F}.  
 \end{enumerate}
Ring-like distributions have previously been inferred for DCO$^+$ in the TW Hya, HD 163296, and DM Tau disks \citep{2008ApJ...681.1396Q, 2013AA...557A.132M, 2015AA...574A.137T}, and ring-like emission patterns are observed for AS 209, V4046 Sgr, MWC 480, and HD 163296 in this work. These results suggest that ring-like DCO$^+$ distributions are common, and support a cold H$_2$D$^+$ formation scenario. The double DCO$^+$ rings observed in the IM Lup disk deviate from this picture somewhat, but \citet{2015ApJ...810..112O} found that the presence of multiple rings could be explained with models of DCO$^+$ abundances first decreasing in the outer disk due to CO freezeout, then rising again as CO non-thermally desorbs in regions of the disk with lower surface density where UV radiation penetrates closer to the midplane. 

In addition, while the LkCa 15 DCO$^+$ emission is not as distinctly ring-like as that of the other disks, the radially averaged integrated intensity of H$^{13}$CO$^+$ drops by about a factor of 3 from 50 to 200 AU, whereas the DCO$^+$ radial profile remains relatively flat in the same interval. Likewise, the radially averaged integrated intensity of H$^{13}$CO$^+$ in the HD 163296 disk drops by about a factor of 5 from 50 to 200 AU, whereas that of DCO$^+$ drops only by about a factor of 2. This behavior suggests that either the column densities or average excitation temperatures of H$^{13}$CO$^+$ are decreasing more rapidly with radius than those of DCO$^+$. Since disks are expected to have radially decreasing temperature gradients, either scenario would require the DCO$^+$/HCO$^+$ ratio to increase at lower temperatures.

We do not observe radially decreasing emission profiles for DCO$^+$. In addition, given that the dominant formation pathway of DCN is expected to begin with CH$_2$D$^+$ \citep[e.g.][]{1989ApJ...340..906M, 1999ApJ...526..314A}, one would expect the DCO$^+$ emission to be bright in the same regions if its formation followed the models of \citet{2015ApJ...802L..23F}. On the contrary, for at least a few disks (V4046 Sgr and HD 163296 from this work, as well as TW Hya from  \citealt{2008ApJ...681.1396Q} and \citealt{2012ApJ...749..162O}), the DCO$^+$ emission features prominent central gaps where DCN is brightest. Our observations therefore do not provide evidence that the CH$_2$D$^+$ pathway dominates DCO$^+$ production. 

It is interesting to note, though, that the two Herbig Ae sources in the survey, MWC 480 and HD 163296, have disk-averaged DCO$^+$/HCO$^+$ ratios comparable to those of the T Tauri sources. This result may seem counter-intuitive, since the Herbig Ae disks are expected to be warmer and therefore less favorable toward high deuterium fractionation, although no deuterium chemistry models have been published for Herbig Ae disks yet. The observations may hint that CH$_2$D$^+$ is a secondary contributor to fractionation. In addition, based on DCO$^+$ $J = 4-3$ observations in the HD 163296 disk, \citet{2015ApJ...813..128Q} concluded that DCO$^+$ remains abundant in regions of the disk warmer than 30 K, which suggests that a warm pathway is partially responsible for DCO$^+$ production. A larger sample of both T Tauri and Herbig Ae disks would be needed to determine whether they indeed typically have similar levels of DCO$^+$/HCO$^+$ fractionation.

\subsection{Relationship Between Molecular Distributions and Disk Dust Structures}
\subsubsection{Transition Disk Cavities}
The V4046 Sgr and LkCa 15 disks both feature prominent cavities, but markedly different line emission patterns, implying that the presence of a cavity does not lead to predictable molecular distributions. The DCN and DCO$^+$ in the V4046 Sgr disk are clearly not co-spatial, while they have more similar radial extents in the LkCa 15 disk.  The disk gaps in the millimeter dust continuum do not have an obvious relationship to the observed line morphologies. In the V4046 Sgr disk, H$^{13}$CN line emission appears at the 5$\sigma$ level up to $\pm$5.5 km s$^{-1}$ offset from the systemic velocity. This suggests that H$^{13}$CN is present at least within 15 AU from the central binary and well within the millimeter dust cavity ($R = 29$ AU), assuming a Keplerian velocity field (i.e. $v_{\text{projected}} = \sqrt{\frac{GM_*}{r}}\sin{(\text{incl})}$) and the stellar mass and disk inclination given in Tables \ref{tab:stellarproperties} and \ref{tab:diskproperties}. On the other hand, almost all the H$^{13}$CO$^+$ and DCO$^+$ emission in V4046 Sgr at the \textgreater$3\sigma$ level in the channel maps are detected well outside the cavity. 

\subsubsection{The Edge of the Millimeter Dust Disk}
Recent observations have indicated that molecular emission rings arising at the edge of the millimeter dust disk are a recurring feature \citep[e.g.][]{2015ApJ...814...53G, 2015ApJ...810..112O, 2016ApJ...831..101B}. Indeed, in the deuterium survey, this feature is observed for DCN in LkCa 15, H$^{13}$CO$^+$ in HD 163296, and DCO$^+$ in V4046 Sgr and IM Lup (as first pointed out by \citet{2015ApJ...810..112O}), providing further evidence that molecular emission rings encircling millimeter dust disks are common. Models suggest that such rings could be due to dust evolution facilitating either thermal or non-thermal desorption in the outer disk \citep[e.g.][]{2015ApJ...810..112O, 2016ApJ...831..101B, 2016ApJ...816L..21C}. However, since this feature seems to arise in different molecules for different disks, a consistent connection is not yet established. Given that outer emission rings are sometimes very faint, a systematic study of molecular emission at very high sensitivity in conjunction with imaging of dust at multiple wavelengths would be needed to investigate whether the evolution of grain size distributions leads to predictable locations for molecular emission rings. 

\subsubsection{The AS 209 disk}
The outer ring in the 258 GHz continuum emission in the AS 209 disk (Fig. \ref{fig:AS209continuum}) appears to coincide with a similar outer ring in the C$^{18}$O $J = 2-1$ emission and with a sharp change in slope of the 1.4 mm continuum emission observed at slightly lower spatial resolution in \citet{2016ApJ...823L..18H}. (Note that \citet{2016ApJ...823L..18H} used an older distance estimate of 120 pc rather than 126 pc in this work). \citet{2016ApJ...823L..18H} proposed that the outer C$^{18}$O emission ring was the result of desorption in the outer disk outside the CO snowline. This would appear to be consistent with predictions that continuum emission rings can arise near volatile snowlines, perhaps due to changes in dust opacity from mechanisms such as pebble growth \citep{2015ApJ...806L...7Z} or sintering \citep{2016ApJ...821...82O}. 

The H$^{13}$CO$^+$, DCO$^+$, H$^{13}$CN, and DCN emission morphologies of the AS 209 disk are quite similar to one another, unlike those of the other disks in the survey. All four are distributed in emission rings peaking near the dark ring in the 258 GHz dust continuum. Given that H$^{13}$CN and H$^{13}$CO$^+$ do not have similar formation pathways, the similar emission patterns in the four molecules observed for AS 209 may instead be related to the interplay between the dust and gas structures near the continuum emission gap. 

However, for AS 209, higher-resolution continuum studies at multiple wavelengths to constrain spectral indices and surface density profiles would be needed to distinguish between the numerous mechanisms for producing dust emission rings that have been proposed for protoplanetary disks. Ring-like structures in millimeter dust continuum emission have also been directly imaged in several other protoplanetary disks, but their origins are still debated \citep{2015ApJ...808L...3A, 2016ApJ...820L..40A, 2016Natur.535..258C, PhysRevLett.117.251101}. In addition to the mechanisms discussed in the previous paragraph, other hypotheses for the origins of rings often invoke the presence of a forming planet \citep[e.g.][]{2013AA...549A..97R, 2015ApJ...809...93D, 2015AA...573A...9P, 2015MNRAS.454L..36G} or magnetorotational instability \citep[e.g.][]{2012AA...538A.114P, 2013AA...549A..97R, 2015AA...574A..68F}. 
\subsection{Assessing the Viability of DCO$^{+}$ as a Midplane CO Snowline Tracer}
Because temperatures in a protoplanetary disk generally decrease with radius and increase with height above the midplane, CO is present in the gas phase in the inner regions of the disk and in warm layers above the midplane, but it is mostly frozen out onto dust grains in the outer midplane. Measuring the location of the CO midplane snowline, which marks the radius where this freezeout starts to occur, is important for constraining planet formation theories because changes in the abundance of CO gas and ice are thought to influence the composition of planetary atmospheres as well as the movement and growth of dust grains \citep[e.g.][]{2011ApJ...743L..16O, 2014ApJ...794L..12M, 2014ApJ...793....9A, 2015ApJ...815..109P,2016ApJ...821...82O}. Because CO is almost always optically thick in protoplanetary disks, the CO snowline is usually traced instead through interferometric imaging of optically thin molecules that have distributions dependent on gas-phase CO \citep{2013AA...557A.132M,2013Sci...341..630Q,2015ApJ...813..128Q}. \citet{2013AA...557A.132M} suggested that DCO$^+$ traces the CO snowline because the formation of DCO$^+$ requires both H$_2$D$^+$ and gas-phase CO, which coexist only in a narrow temperature range. N$_2$H$^+$ is also used as a tracer because it is quickly destroyed by gas-phase CO, so it is expected to become abundant when CO starts freezing out \citep{2013Sci...341..630Q, 2015ApJ...813..128Q, 2015ApJ...807..120A}.

Although the DCO$^+$ emission morphologies and high fractionation levels observed in this work are broadly consistent with predictions from disk models invoking the standard H$_2$D$^+$ pathway, observations of the other molecular lines in this survey provide evidence that DCO$^+$ is not a robust tracer of the CO snowline. \citet{2015ApJ...813..128Q} make a similar argument based on the relationship between the inferred DCO$^+$ and N$_2$H$^+$ distributions, while our line of reasoning is based on the relationship between DCO$^+$ and H$^{13}$CO$^+$. Whereas \citet{2013AA...557A.132M} suggested that the inner radius of the DCO$^+$ distribution is determined by where H$_2$D$^+$ becomes abundant, the AS 209, IM Lup, LkCa 15, and MWC 480 disks feature central dips in their H$^{13}$CO$^+$ radial emission profiles that are comparable to or even larger than the central  dips in the corresponding DCO$^+$ profiles. The H$^{13}$CO$^+$ dip in the LkCa 15 disk can be plausibly explained by a drop in gas density, which has already been inferred for several transition disks based on CO isotopologue observations \citep{2016AA...585A..58V}, but no models or observations so far have indicated that a similar drop in central gas density is likely for the full disks. Instead, the observations of AS 209, IM Lup, and MWC 480 raise the possibility that the inner radius of the DCO$^+$ ring is not always controlled by the availability of H$_2$D$^+$, but that instead, the inner edges of the DCO$^+$ and H$^{13}$CO$^+$ rings may be set by common chemical or physical factors, such as the ionizing environment due to cosmic rays or X-rays \citep[e.g.][]{2014ApJ...794..123C}.

Alternatively, high dust opacity at the disk center could create the appearance of a hole in the H$^{13}$CO$^+$ distribution \citep[e.g.][]{2015ApJ...810..112O, 2016ApJ...832..110C}, but then optically thin DCO$^+$ emission would likely suffer similar effects, which would also increase the difficulty of using it to trace the CO snowline. Given that optically thin line transitions can now be imaged by ALMA with much greater ease than by previous instruments, the question of how dust opacity affects the interpretation of their emission patterns will warrant more rigorous examination through continuum and line observations in multiple bands (e.g., Band 3 in conjunction with Band 6 or 7) as well as simultaneous modeling of gas and dust structure.

The DCO$^+$ observations in the IM Lup and LkCa 15 disks, as well as the H$^{13}$CO$^+$ emission in the HD 163296 disk, also indicate that the outer edge of the DCO$^+$ emission profile does not necessarily correspond closely to the location of the classical midplane CO snowline. \citet{2015ApJ...810..112O} proposed that the second DCO$^+$ emission ring in the IM Lup disk arises from CO ice desorption outside the CO snowline, whereas the inner DCO$^+$ emission ring was likely located near the CO snowline. At lower spatial resolutions, multiple DCO$^+$ emission rings could appear to be a single ring, for which neither the inner nor outer edge would mark the location of the CO snowline. No estimates of CO snowline locations are available in the literature for the LkCA 15 disk, but estimates for other disks range from 17 to 90 AU \citep{2013Sci...341..630Q,2015ApJ...813..128Q,2016ApJ...823...91S}. Rather than being confined to the single narrow ring predicted by disk chemistry models, the LkCa 15 DCO$^+$ radial profile remains relatively flat out to hundreds of AU, well past likely locations of the midplane CO snowline. Finally, the annular gap in H$^{13}$CO$^+$ emission in the HD 163296 disk at $\sim$ 200 AU is well beyond the 90 AU CO snowline that \citet{2015ApJ...813..128Q} estimated from N$_2$H$^+$ data, which could indicate that gas-phase CO is depleted through some other mechanism besides CO freezeout. Using millimeter observations at $\sim0.2''$, \citet{PhysRevLett.117.251101} identified a number of rings in the HD 163296 disk where both the CO and dust are depleted. While the most distant ring identified is at 160 AU, which does not appear to be an obvious match with the H$^{13}$CO$^+$ gap, our observations are also at coarser resolution, so forward modeling will be necessary to determine more precisely the nature of the relationship between the H$^{13}$CO$^+$ and CO distributions in this source.

\section{Summary}\label{sec:summary}
We present sensitive, high-resolution ALMA observations of DCO$^+$, DCN, H$^{13}$CO$^+$, and H$^{13}$CN $J=3-2$ lines, as well as HCN $J=3-2$ in the IM Lup disk, that indicate that deuterium fractionation of these molecules is efficient in disks and manifests in diverse forms. The main findings are summarized below. 
\begin{enumerate}
 \item DCO$^+$, H$^{13}$CO$^+$, and DCN are detected in all disks, and H$^{13}$CN is detected in all but the IM Lup disk. 
  \item Disk-averaged DCO$^+$/HCO$^+$ and DCN/HCN abundance ratios derived for the survey are on the order of $\sim0.02$\textendash0.06 and $\sim0.005$\textendash0.08, respectively. These high values are comparable to previous disk observations and disk chemical models, as well as the D/H ratios measured for earlier stages of star formation.
 \item Given the high rate of DCN detections (four of which are new) with relatively short integration times, DCN is especially valuable as a tracer of disk chemistry in the context of chemical evolution because it has also been observed in cold interstellar environments and within the solar system. 
\item The diverse emission morphologies observed for DCN suggest that both warm and cold pathways can contribute significantly to its production in disks. 
 \item The observations point to a cold formation pathway dominating DCO$^+$ production in disks. However, even though DCO$^+$ has been proposed as a CO snowline tracer on the basis of DCO$^+$ formation being most favored in gas near the CO freezeout temperature, observations of similar central and annular gaps in H$^{13}$CO$^+$ and DCO$^+$ emission suggest that the ring-like emission profiles of DCO$^+$ may not reliably constrain the location of the CO snowline. 
\end{enumerate}

\acknowledgments
We thank the referee for comments improving this paper. We also thank M. A. MacGregor, I. Czekala, L. I. Cleeves, X. Bai, A. Tripathi, M. Holman, D. Sasselov, and E. Fayolle for helpful discussions, and A. Moullet, B. Mason, and the ALMA helpdesk staff for their advice on data reduction. This paper makes use of ALMA data \dataset[ADS/JAO.ALMA\#2013.1.00226.S]{https://almascience.nrao.edu/aq/?project\_code=2013.1.00226.S} and \dataset[ADS/JAO.ALMA\#2015.1.00964.S]{https://almascience.nrao.edu/aq/?project\_code=2015.1.00964.S}. ALMA is a partnership of ESO (representing its member states), NSF (USA) and NINS (Japan), together with NRC (Canada) and NSC and ASIAA (Taiwan), in cooperation with the Republic of Chile. The Joint ALMA Observatory is operated by ESO, AUI/NRAO and NAOJ. The National Radio Astronomy Observatory is a facility of the National Science Foundation operated under cooperative agreement by Associated Universities, Inc. This work has made use of data from the European Space Agency (ESA) mission {\it Gaia} (\url{http://www.cosmos.esa.int/gaia}), processed by the {\it Gaia} Data Processing and Analysis Consortium (DPAC, \url{http://www.cosmos.esa.int/web/gaia/dpac/consortium}). Funding for the DPAC has been provided by national institutions, in particular the institutions participating in the {\it Gaia} Multilateral Agreement. This material is based upon work supported by the National Science Foundation Graduate Research Fellowship under Grant No. DGE-1144152. K. I. O. acknowledges funding from the Simons Collaboration on the Origins of Life (SCOL) and the David and Lucile Packard Foundation. Astrochemistry in Leiden is supported by the EU A-ERC grant 291141 CHEMPLAN. 
\software{\textsc{CASA}, \texttt{numpy}, \texttt{scipy}, \texttt{matplotlib}, \texttt{scikit-image}, \texttt{pwkit} (\url{https://github.com/pkgw/pwkit}), \texttt{analysisUtils} (\url{https://casaguides.nrao.edu/index.php?title=Analysis\_Utilities})}

\appendix 
\section{Spectral Setups}\label{sec:setups}
Details of the spectral setups and flux calibration are given in Tables \ref{tab:1.1 mm}\textendash\ref{tab:fluxcal}.
\begin{deluxetable}{ccc}
\tablecaption{Cycle 2 1.1 mm setting spectral setup \label{tab:1.1 mm}}
\tablehead{
\colhead{Frequency} &\colhead{Bandwidth} &\colhead{Resolution}\\
\colhead{(GHz)}&\colhead{(MHz)}&\colhead{(MHz)}}
\startdata
\hline
\multicolumn{3}{c}{Lower sideband} \\
\hline
241.56155&58.6&0.061\\
241.700219&58.6&0.061\\
241.767224&58.6&0.061\\
241.791431&58.6&0.061\\
245.53288 & 117.2 &0.061\\
245.5634219 & 117.2 &0.061\\
\hline
\multicolumn{3}{c}{Upper sideband} \\
\hline
256.3366289&58.6&0.061\\
257.527444&58.6&0.061\\
258.156996&58.6&0.061\\
258.2558259&58.6&0.061\\
258.9421992&58.6&0.061\\
259.011787&58.6&0.061\\
260.255342&58.6&0.061\\
260.51802&58.6&0.061\\
\enddata
\end{deluxetable}

\begin{deluxetable}{ccc}
\tablecaption{Spectral Setup of the Cycle 2 1.4 mm setting \label{tab:1.4 mm}}
\tablehead{
\colhead{Frequency} &\colhead{Bandwidth} &\colhead{Resolution}\\
\colhead{(GHz)}&\colhead{(MHz)}&\colhead{(MHz)}}
\startdata
\hline
\multicolumn{3}{c}{Lower Sideband} \\
\hline
216.37332&58.6&0.061\\
216.11258&58.6&0.061\\
217.10498&58.6&0.061\\
217.23853&58.6&0.061\\
218.475632&58.6&0.061\\
218.760066&58.6&0.061\\
219.560358&58.6&0.061\\
220.3986842&58.6&0.061\\
\hline
\multicolumn{3}{c}{Upper Sideband} \\
\hline
231.41027&58.6&0.061\\
231.3218283&58.6&0.061\\
231.220686&58.6&0.061\\
230.53800&58.6&0.061\\
234.935663&468.8 &0.122
\enddata

\end{deluxetable}

\begin{deluxetable*}{ccc}
\tablecaption{Spectral Setup of the Cycle 3 1.1 mm setting for the IM Lup disk\label{tab:HCNsetting}}
\tablehead{
\colhead{Frequency} &\colhead{Bandwidth} &\colhead{Resolution}\\
\colhead{(GHz)}&\colhead{(MHz)}&\colhead{(MHz)}}
\startdata
\hline
\multicolumn{3}{c}{Lower Sideband} \\
\hline
249.054368&468.8&0.244\\
251.3143364&117.2&0.122\\
251.5273023&117.2&0.122\\
\hline
\multicolumn{3}{c}{Upper Sideband} \\
\hline
262.00426&234.4&0.122\\
265.886431&234.4 &0.122
\enddata

\end{deluxetable*}

\begin{deluxetable*}{ccccc}
\tablecaption{Flux Calibration Models for Quasars \label{tab:fluxcal}}
\tablehead{
\colhead{Flux Calibrator} &\colhead{Date} &\colhead{Reference Frequency}&\colhead{Flux Density}&\colhead{Spectral Index}\\
&&\colhead{(GHz)}&\colhead{(Jy)}}
\startdata
J1924-2914&2014 June 9& 223.34197 & 3.90764 & -0.63934\\
J0150+1800 & 2014 June 15\tablenotemark{a}&251.92326 & 1.22655& -0.35351\\
& 2014 July 29 &223.35276 & 1.173892 & -0.54114 \\
& 2015 June 6 & 223.33933 & 2.25581 & -0.32657\\
J1733-1304 & 2014 July 2\tablenotemark{b} & 223.33105& 1.28031 & -0.71777\\
 & 2014 July 16 & 251.90750 & 1.130841 & -0.70073
\enddata
\tablenotetext{a}{For the 1.1 mm observations of MWC 480/LkCa 15, we found that the original calibrated data delivered by ALMA specified the flux density model for J0510+1800 to be 1.03 Jy for all SPWs, which is about 20\% lower than the result from querying the ALMA calibrator catalogue for the date of observation. An inquiry to the ALMA helpdesk indicated that this discrepancy was likely due to an inaccurate retrieval of the flux densities when generating the original reduction script. We therefore used NRAO's \texttt{analysisUtils} package to update the flux density model to the one listed in this table.}
\tablenotetext{b}{The calibration for the original delivered 1.4 mm HD 163296 data included a final flux rescaling step that used a model flux density of 0.99 Jy for J1733-1304 at 1.4 mm, which is about 20$\%$ lower than the model used in the initial flux calibration step as well as the value retrieved by querying the ALMA calibrator catalogue. After consultation with the ALMA helpdesk, we edited the reduction scripts to remove this final flux rescaling step. We verified that all other calibration scripts provided by ALMA (and available in the Archive) specify quasar flux models consistent with the ALMA calibrator catalogue. }
\end{deluxetable*}

\section{Channel Maps} \label{sec:chanmaps}

Channel maps are shown in Figures \ref{as209chanmaps}\textendash\ref{hd163296chanmaps}.

\begin{figure*}[htp]
\begin{center}
\subfloat[Channel maps of H$^{13}$CO$^+$ $J = 3-2$.]{\includegraphics[scale = 0.85]{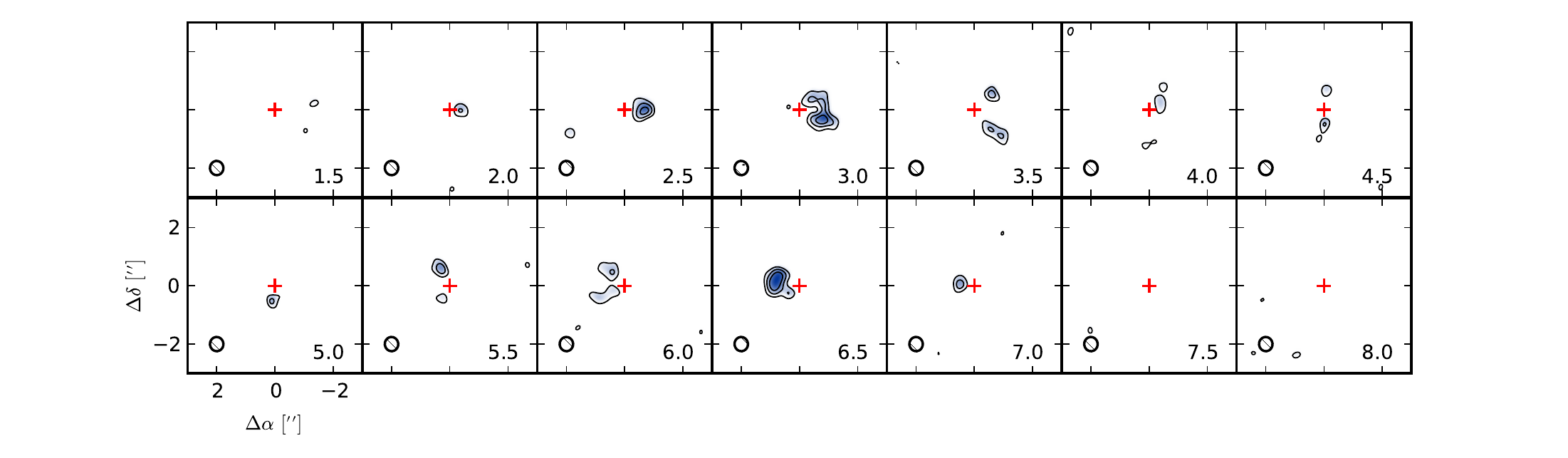}}
\end{center}
\begin{center}
\subfloat[Channel maps of DCO$^+$ $J = 3-2$.]{\includegraphics[scale = 0.85]{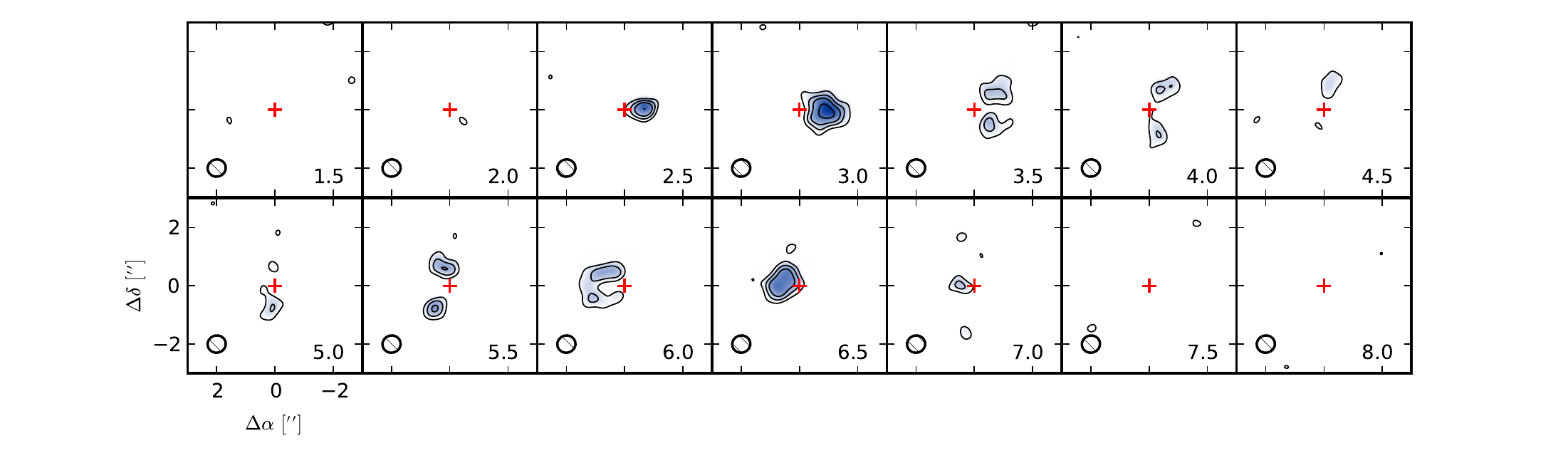}}

\end{center}
\begin{center}
\subfloat[Channel maps of H$^{13}$CN $J = 3-2$.]{\includegraphics[scale = 0.85]{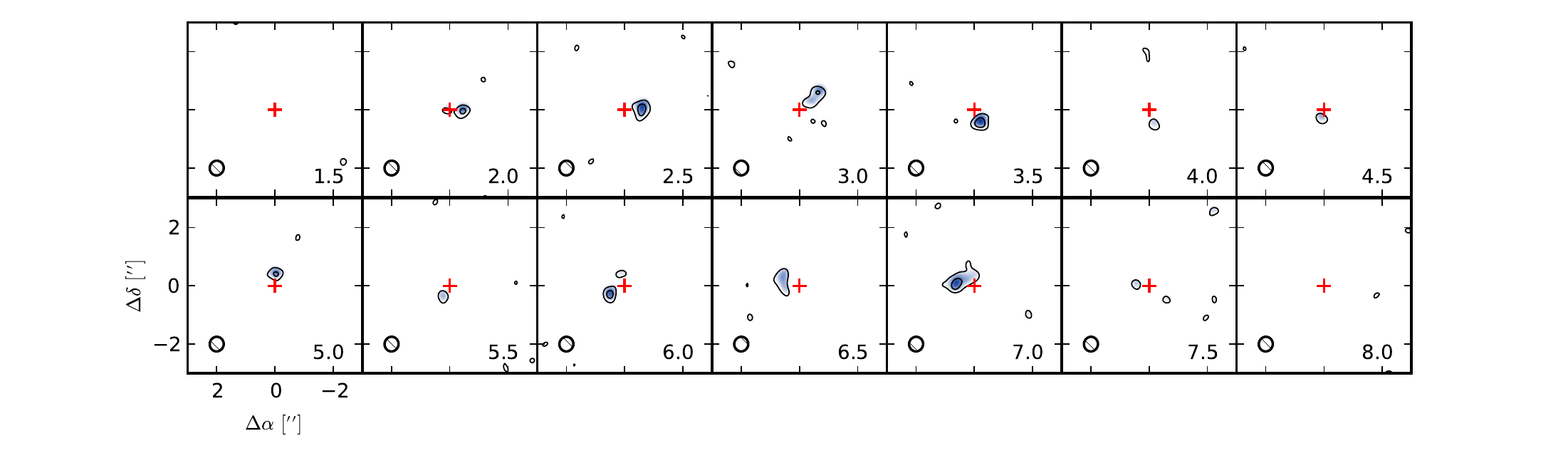}}
\end{center}
\caption{Channel maps of the AS 209 disk. Contours are drawn at [3, 5, 7, 10, 15, 20...]$\sigma$, where $\sigma$ is the channel rms listed in Table \ref{tab:lineobservations}. Red crosses mark the position of the continuum image centroid. Synthesized beams are drawn in the lower left corner of each panel, and labels for the channel velocities in the LSRK frame (km s$^{-1}$) appear in the lower right corners. Offset from the continuum image centroid in arcseconds is marked on the axes in the lower left corner.}
\label{as209chanmaps}
\end{figure*}

\begin{figure*}[htp]

\begin{center}
\ContinuedFloat
\subfloat[Channel maps of DCN $J = 3-2$.]{\includegraphics[scale = 0.85]{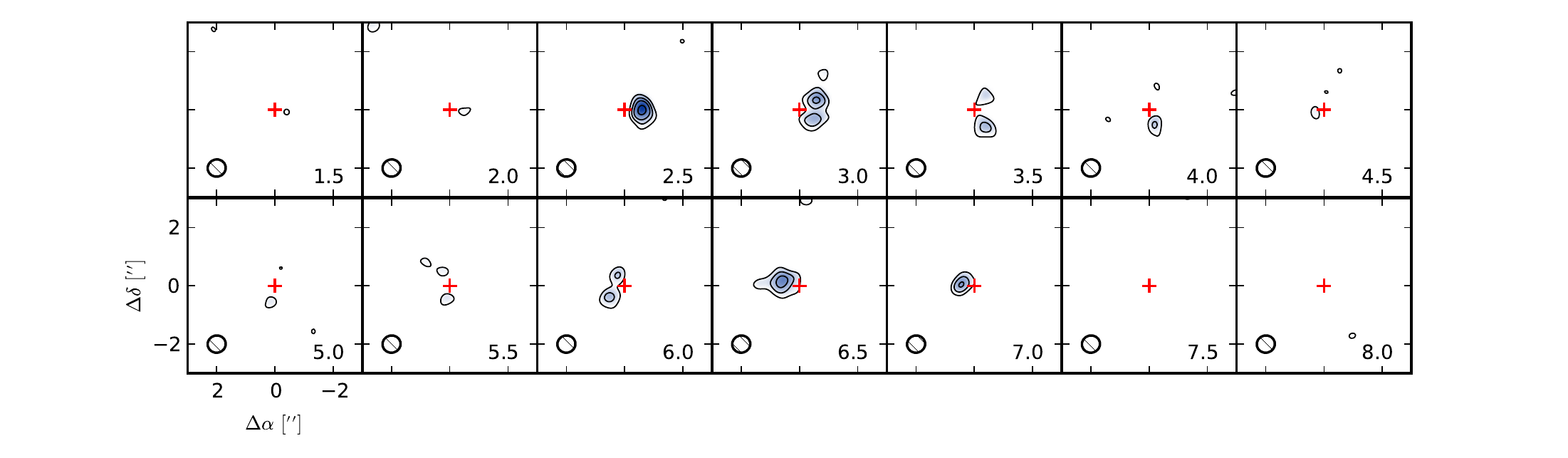}}
\end{center}
\caption{Continued}
\end{figure*}

\begin{figure*}[htp]
\begin{center}
\subfloat[Channel maps of H$^{13}$CO$^+$ $J = 3-2$.]{\includegraphics[scale = 0.85]{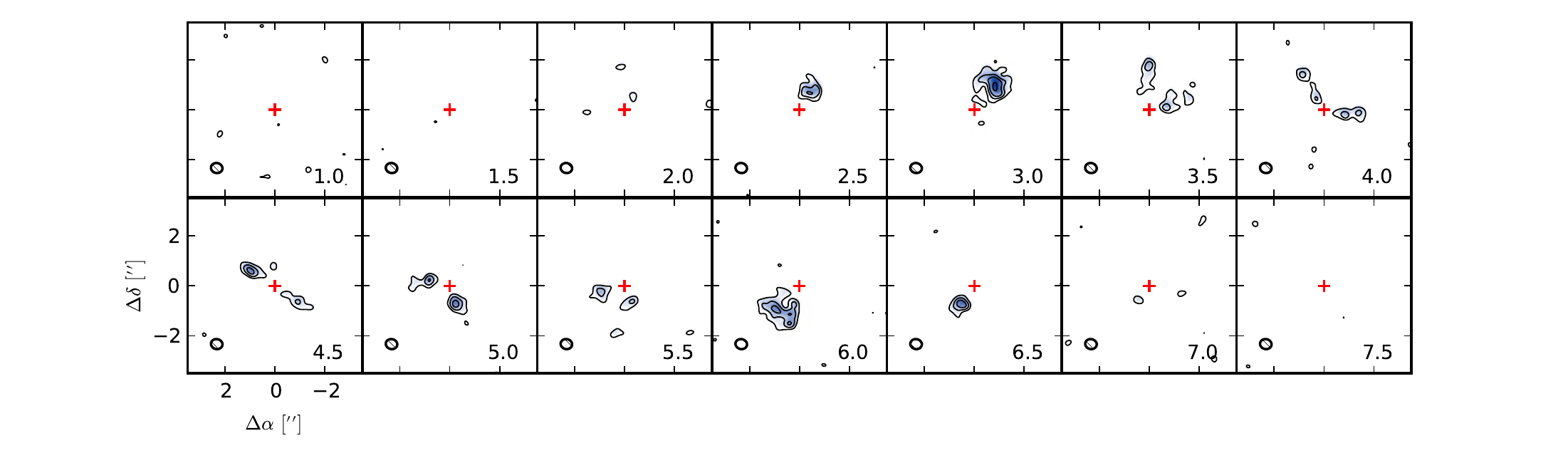}}
\end{center}
\begin{center}
\subfloat[Channel maps of DCO$^+$ $J = 3-2$.]{\includegraphics[scale = 0.85]{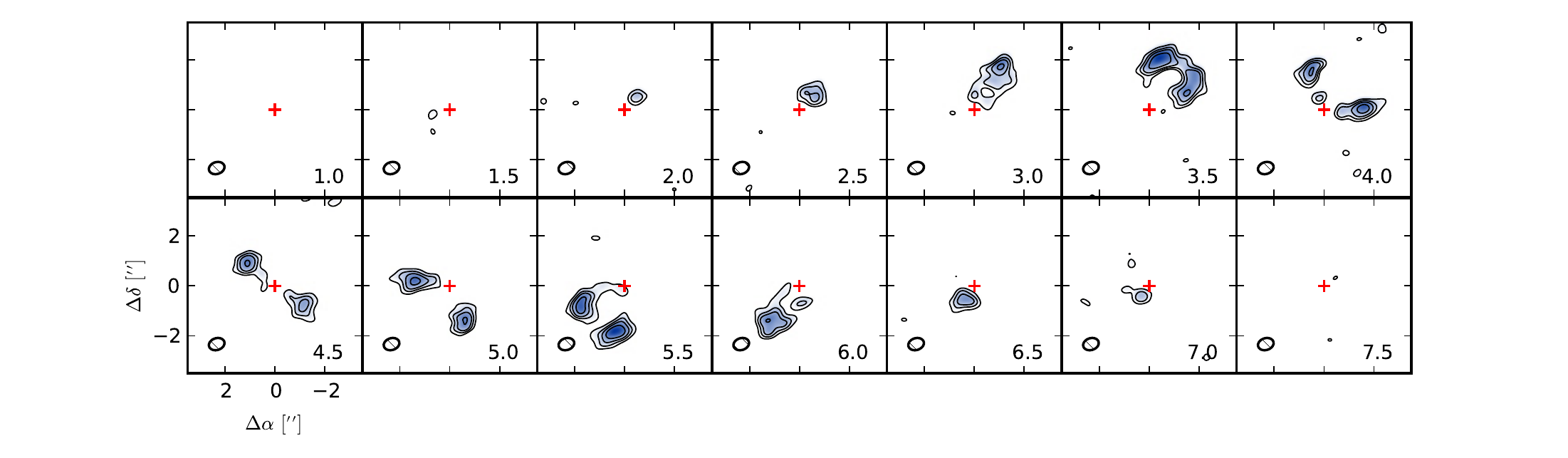}}

\end{center}
\caption{Similar to Fig. \ref{as209chanmaps}, but for the IM Lup disk. }
\label{imlupchanmaps}
\end{figure*}

\begin{figure*}[htp]
\begin{center}
\ContinuedFloat
\subfloat[Channel maps of HCN $J = 3-2$ (H$^{13}$CN $J=3-2$ was not detected in IM Lup).]{\includegraphics[scale = 0.85]{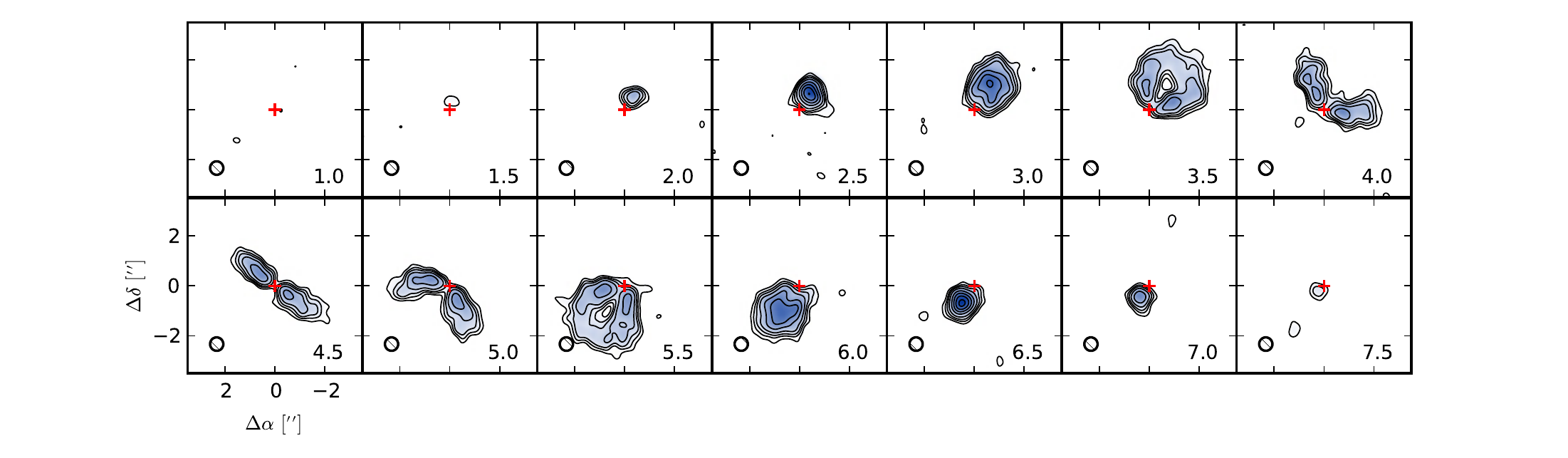}}
\end{center}

\begin{center}

\subfloat[Channel maps of DCN $J = 3-2$. Gray dashed contours mark the region where HCN emission exceeds $3\sigma$, used as an estimate for the DCN emitting region.]{\includegraphics[scale = 0.85]{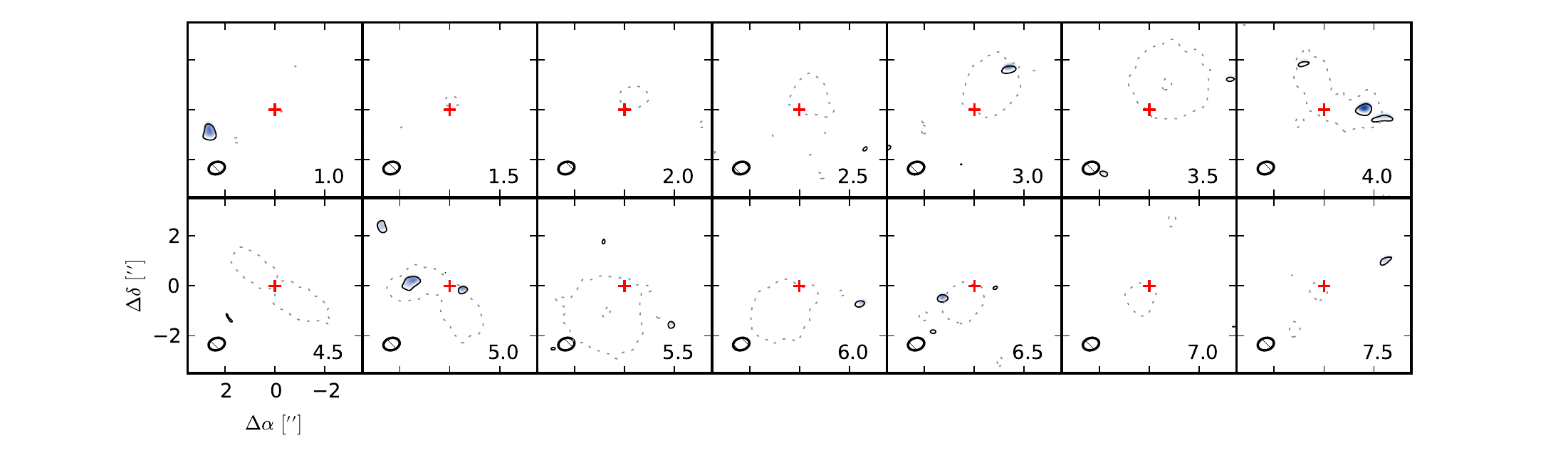}}

\end{center}
\caption{Continued}
\end{figure*}

\begin{figure*}[htp]
\begin{center}
\subfloat[Channel maps of H$^{13}$CO$^+$ $J = 3-2$.]{\includegraphics[scale = 0.65]{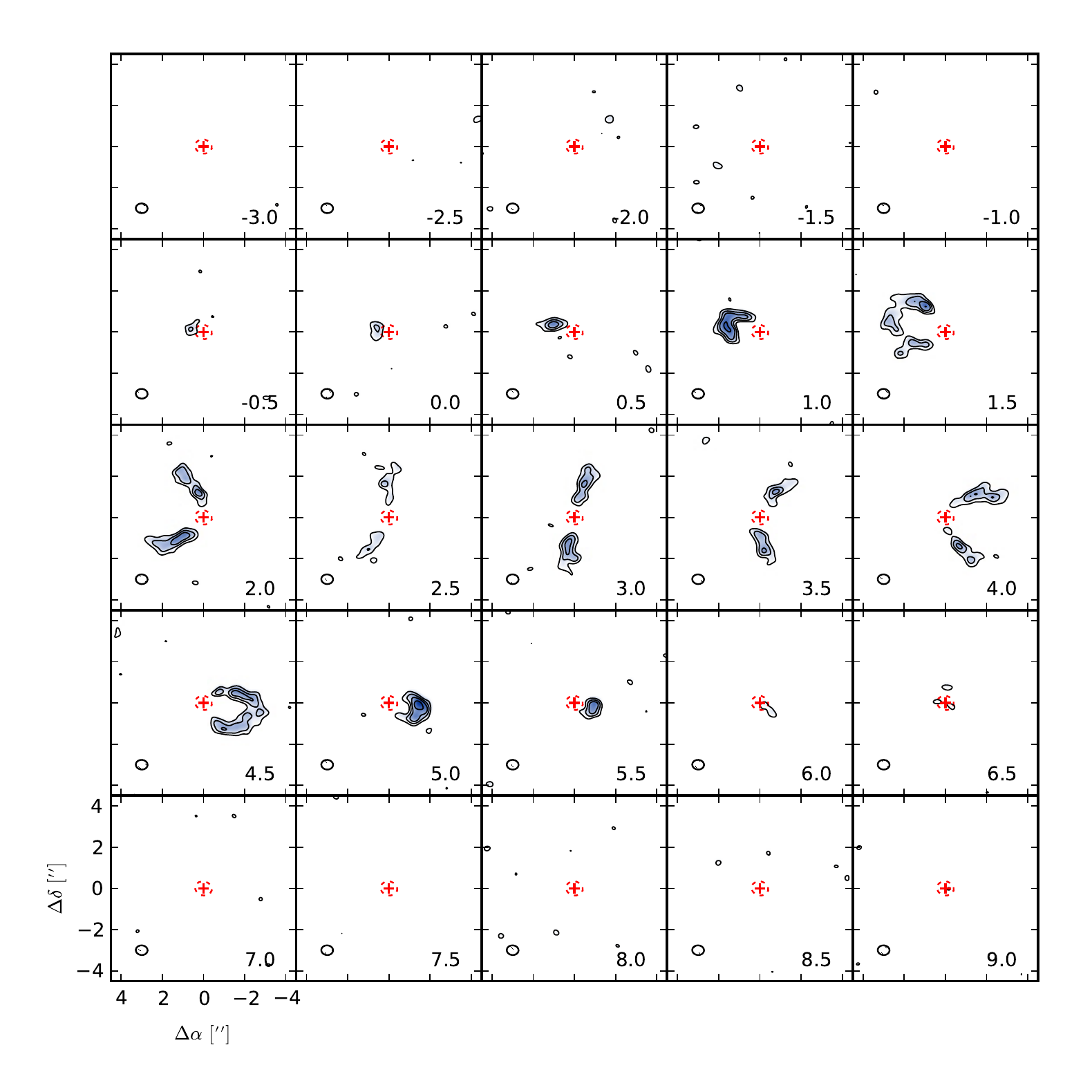}}
\end{center}
\begin{center}
\subfloat[Channel maps of DCO$^+$ $J = 3-2$.]{\includegraphics[scale = 0.65]{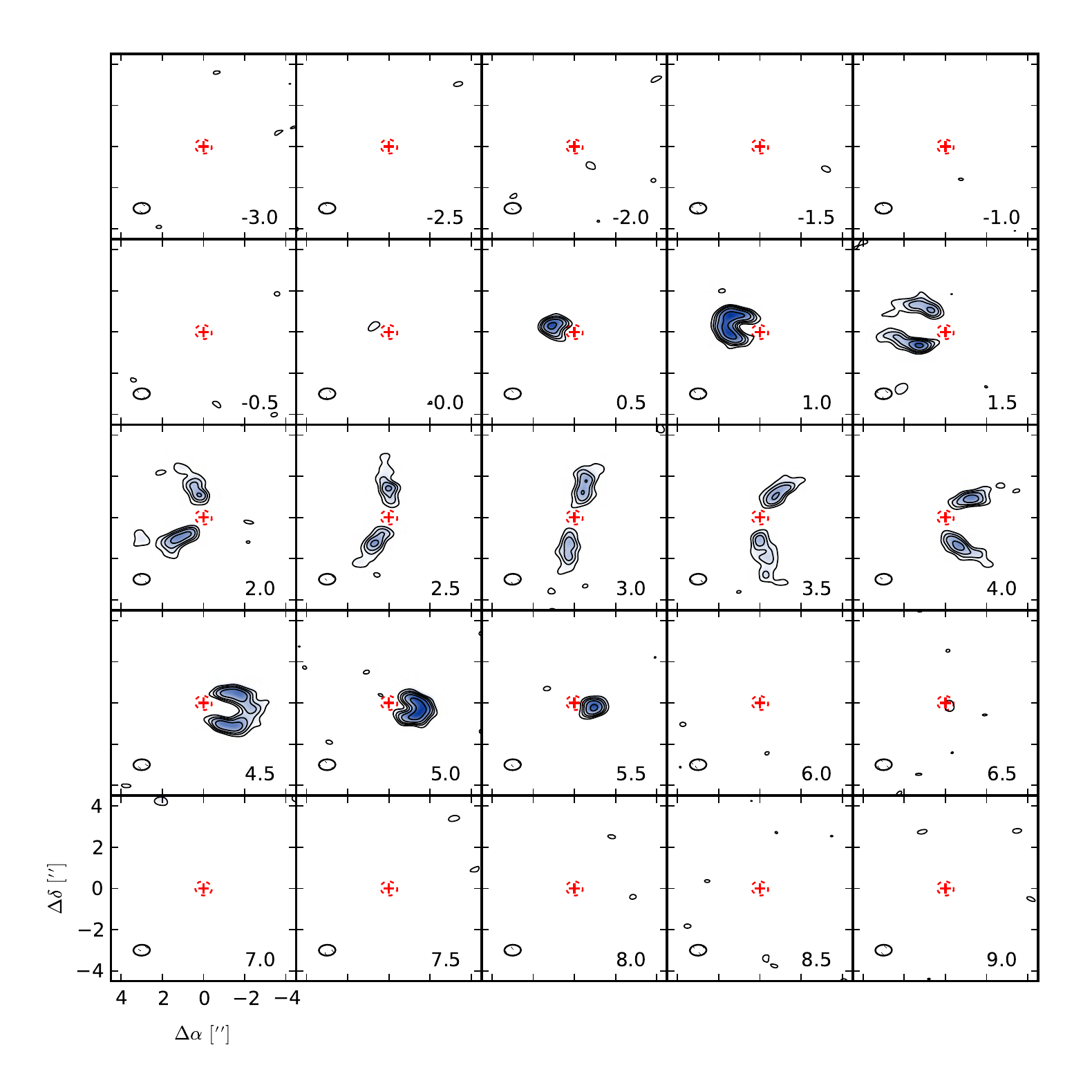}}

\end{center}
\caption{Similar to Fig. \ref{as209chanmaps}, but for the V4046 Sgr disk. The red dashed ellipse traces a projected radius of 29 AU, the size of the 1.3 mm dust cavity reported in  \citet{2013ApJ...775..136R}.}
\label{v4046sgrchanmaps}
\end{figure*}

\begin{figure*}[htp]
\begin{center}
\ContinuedFloat
\subfloat[Channel maps of H$^{13}$CN $J = 3-2$.]{\includegraphics[scale = 0.65]{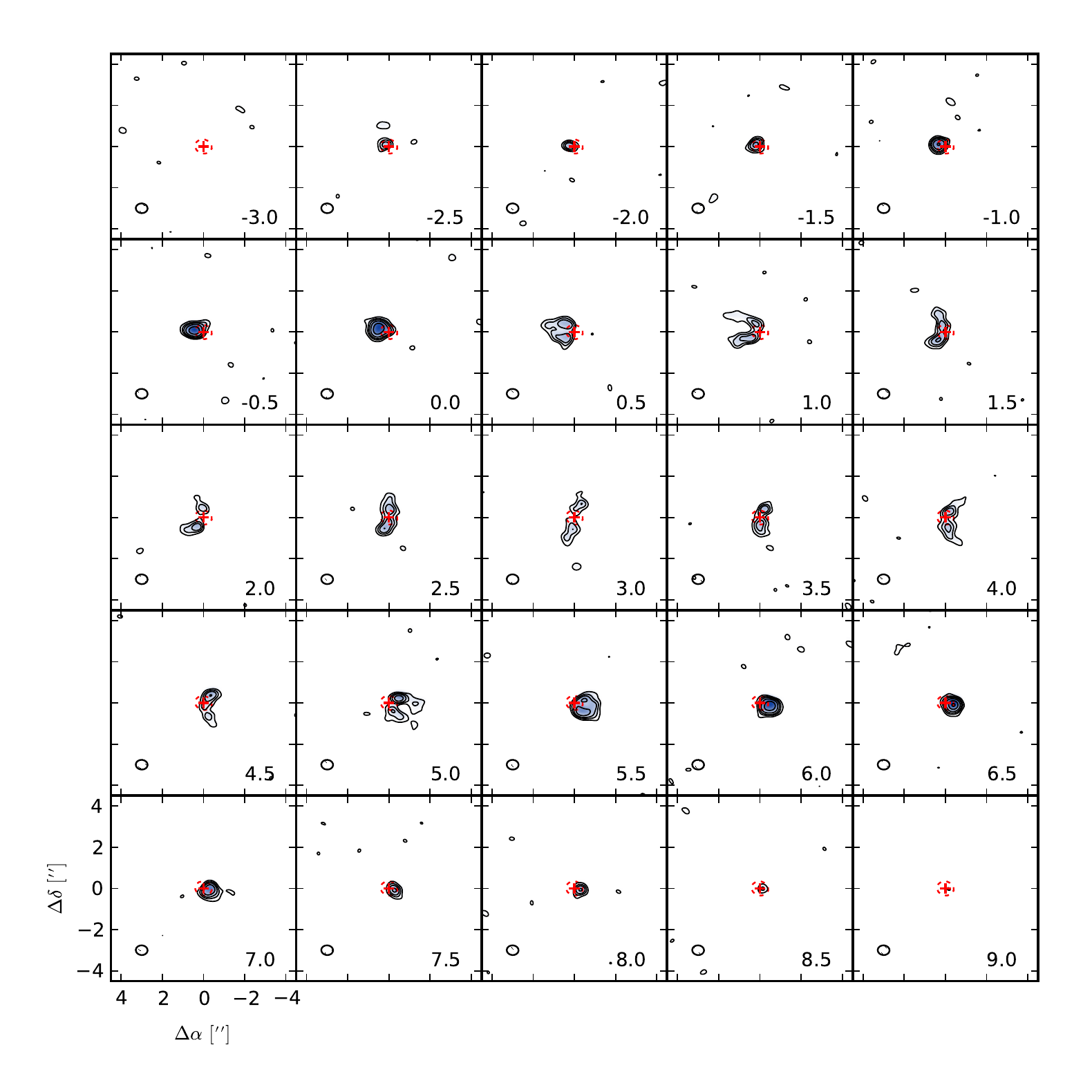}}
\end{center}

\begin{center}

\subfloat[Channel maps of DCN $J = 3-2$.]{\includegraphics[scale = 0.65]{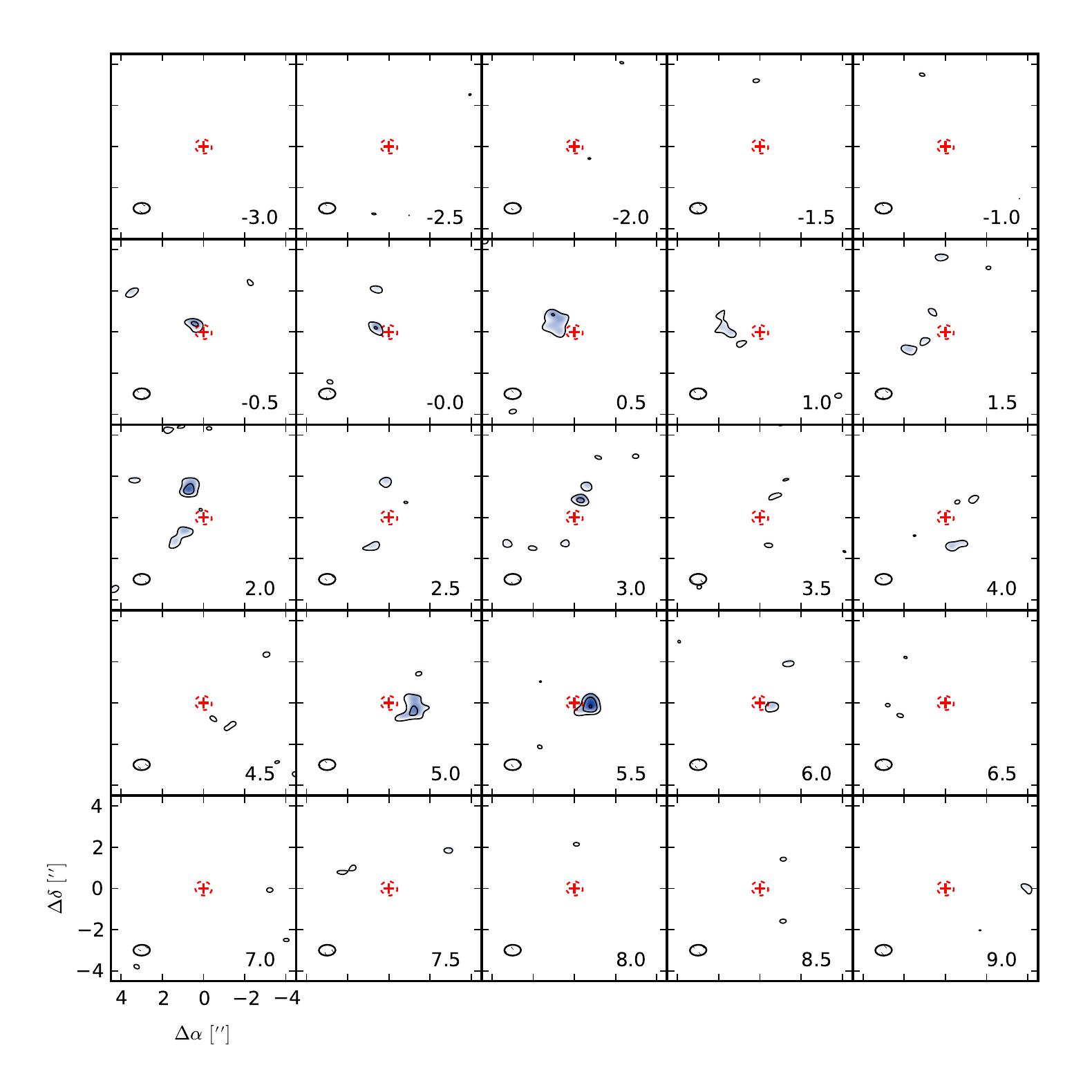}}

\end{center}
\caption{Continued}
\end{figure*}

\begin{figure*}[htp]
\begin{center}
\subfloat[Channel maps of H$^{13}$CO$^+$ $J = 3-2$.]{\includegraphics[scale = 1.0]{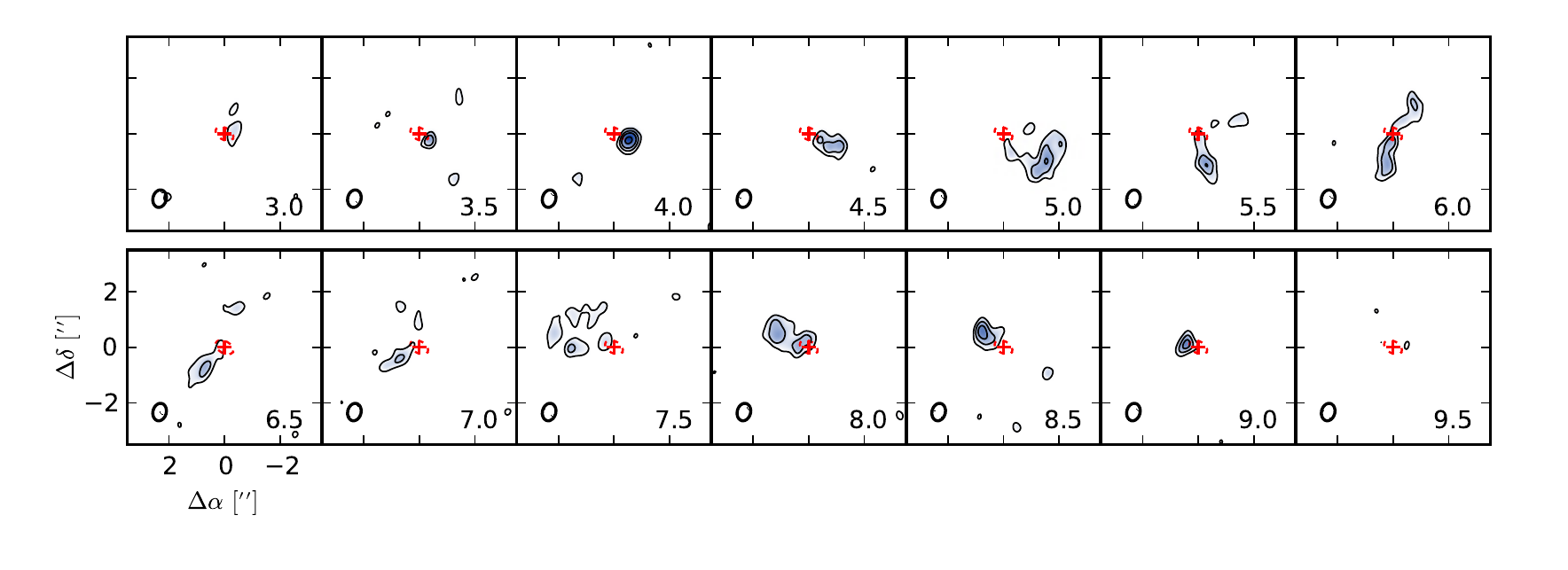}}
\end{center}
\begin{center}
\subfloat[Channel maps of DCO$^+$ $J = 3-2$.]{\includegraphics[scale = 1.0]{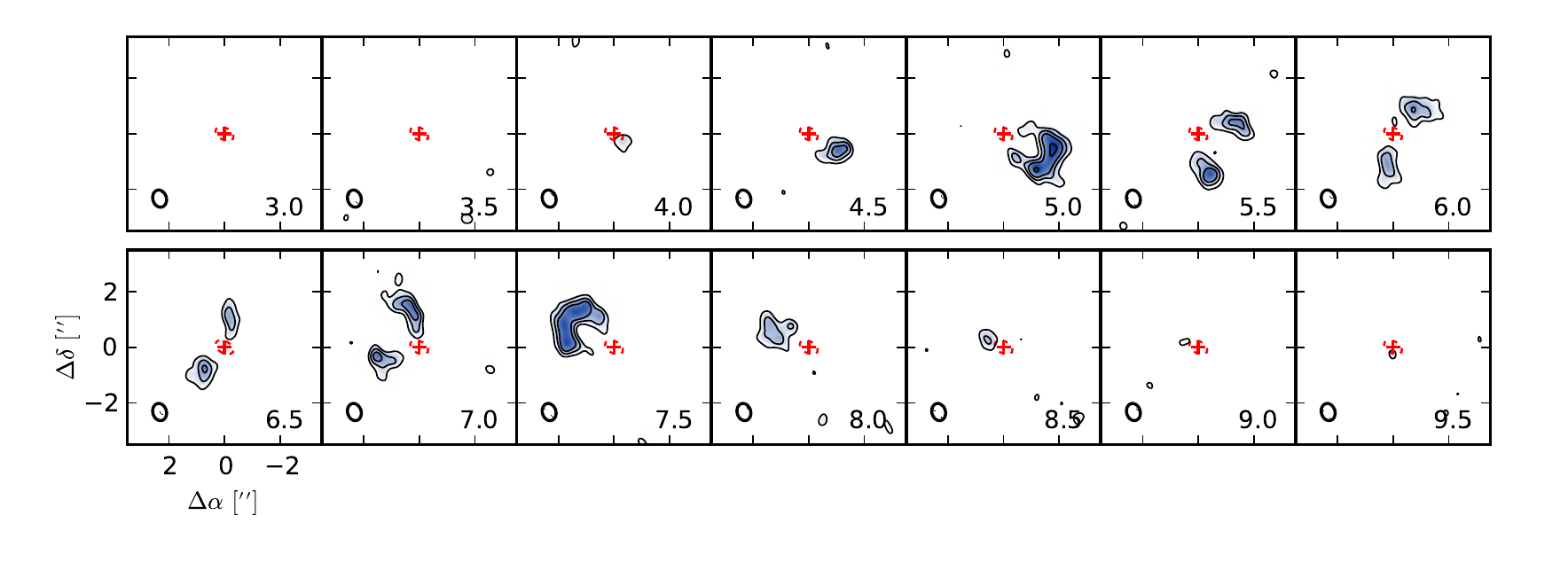}}

\end{center}

\caption{Similar to Fig. \ref{as209chanmaps}, but for the LkCa 15 disk. The red dashed ellipse traces a projected radius of 50 AU, roughly the size of the dust cavity reported in  \citet{2011ApJ...742L...5A}.}
\label{lkca15chanmaps}
\end{figure*}

\begin{figure*}[htp]

\begin{center}
\ContinuedFloat
\subfloat[Channel maps of H$^{13}$CN $J = 3-2$.]{\includegraphics[scale = 1.0]{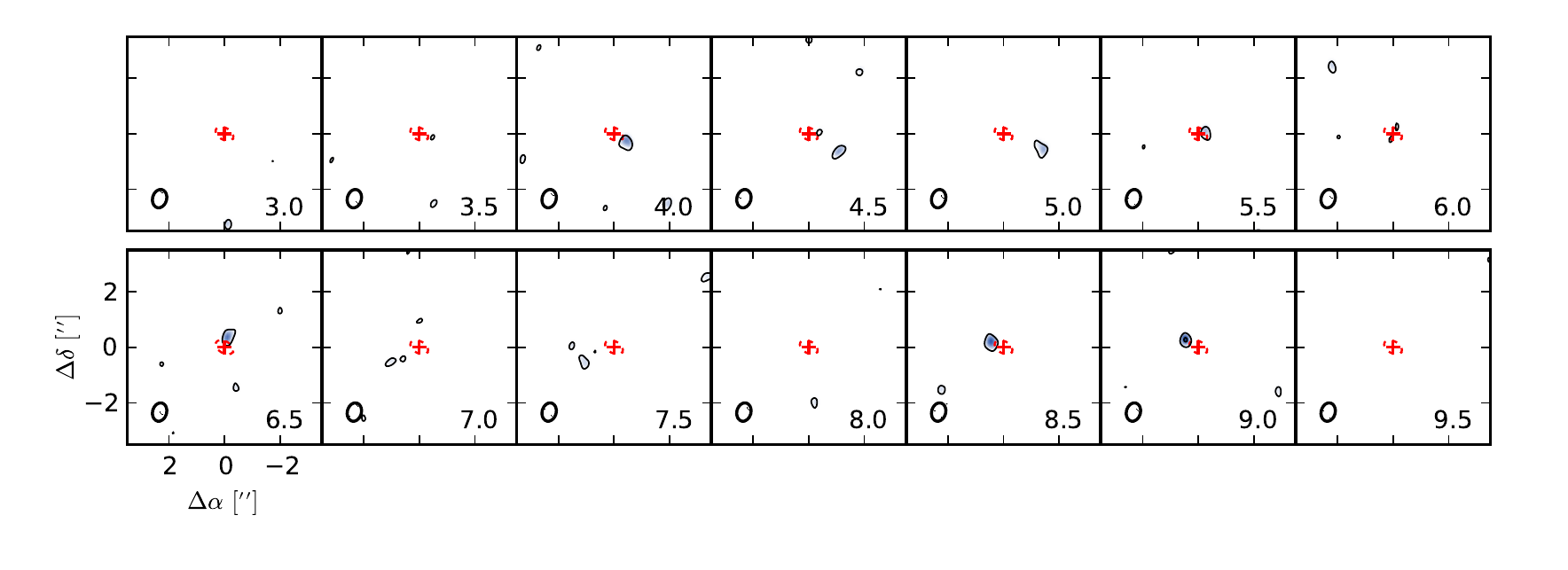}}
\end{center}
\begin{center}
\subfloat[Channel maps of DCN $J = 3-2$.]{\includegraphics[scale = 1.0]
{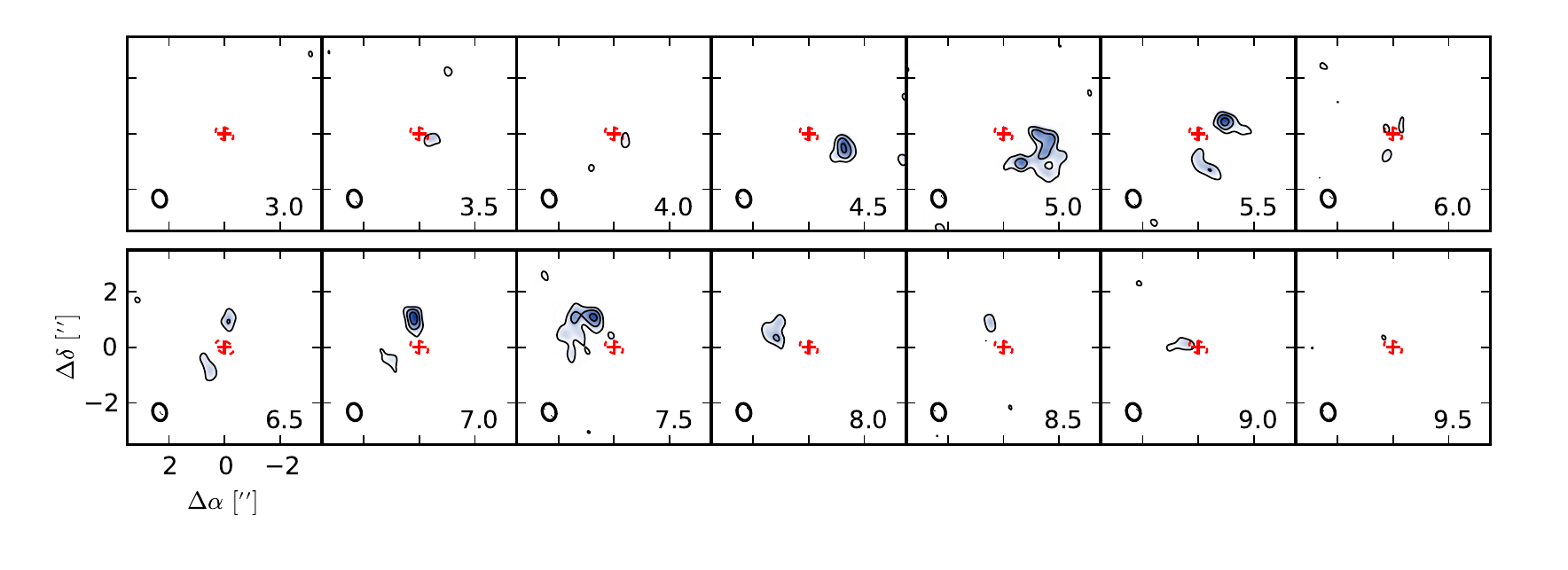}}

\end{center}
\caption{Continued}
\end{figure*}

\begin{figure*}[htp]
\begin{center}
\subfloat[Channel maps of H$^{13}$CO$^+$ $J = 3-2$.]{\includegraphics[scale = .9]{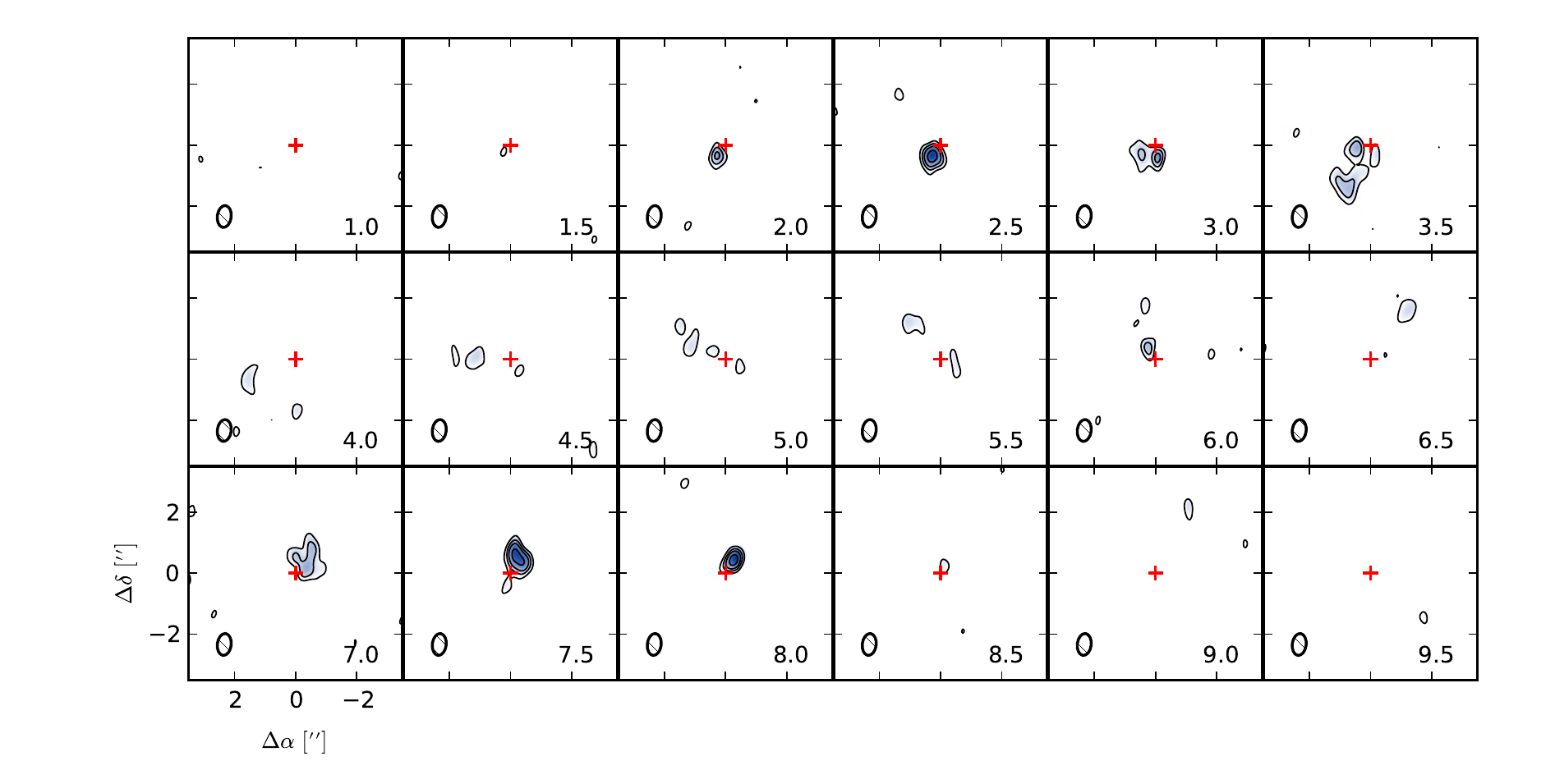}}
\end{center}
\begin{center}
\subfloat[Channel maps of DCO$^+$ $J = 3-2$.]{\includegraphics[scale = .9]{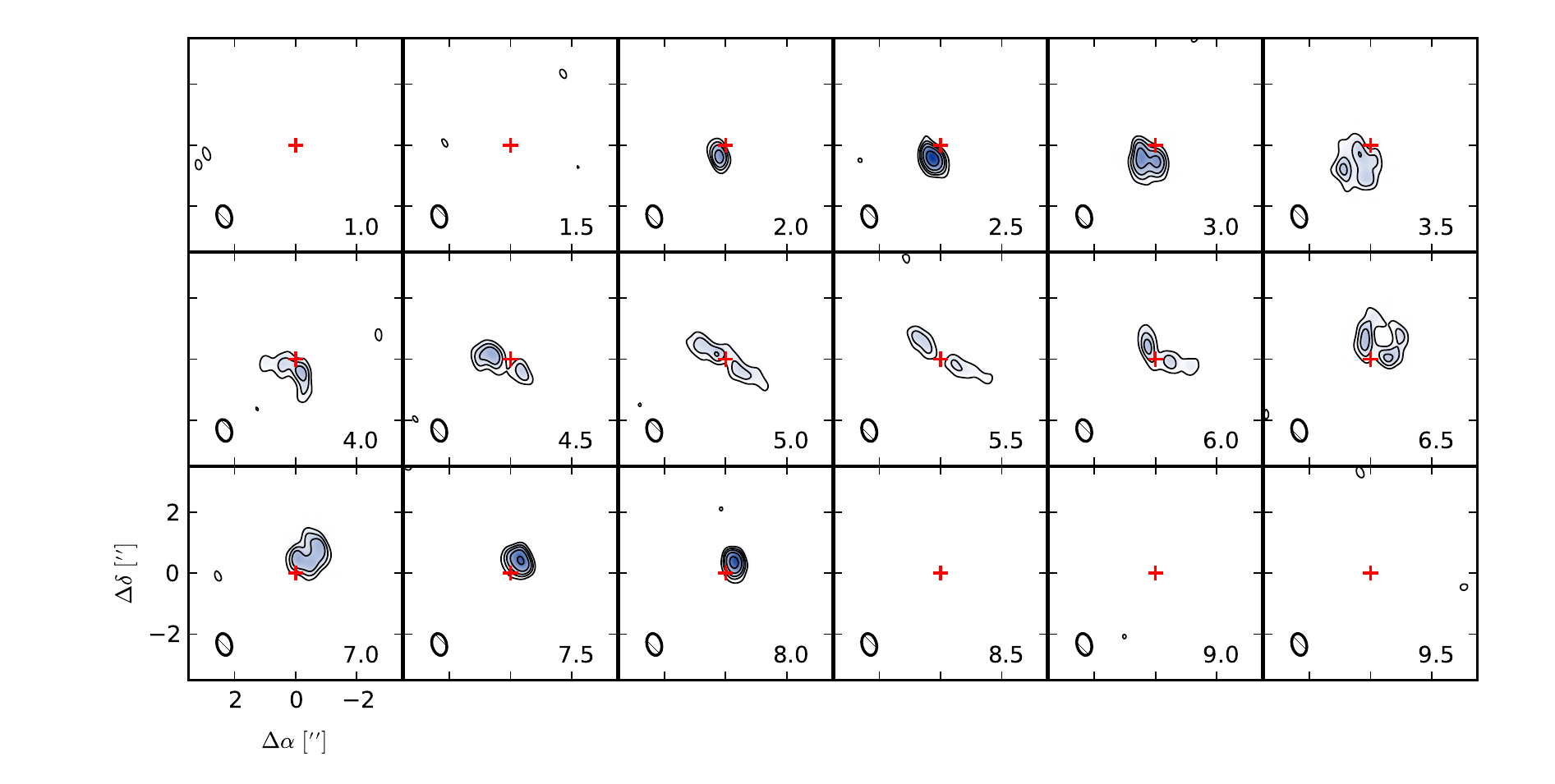}}

\end{center}
\caption{Similar to Fig. \ref{as209chanmaps}, but for the MWC 480 disk.}
\label{mwc480chanmaps}
\end{figure*}

\begin{figure*}[htp]
\begin{center}
\ContinuedFloat
\subfloat[Channel maps of H$^{13}$CN $J = 3-2$.]{\includegraphics[scale = .9]{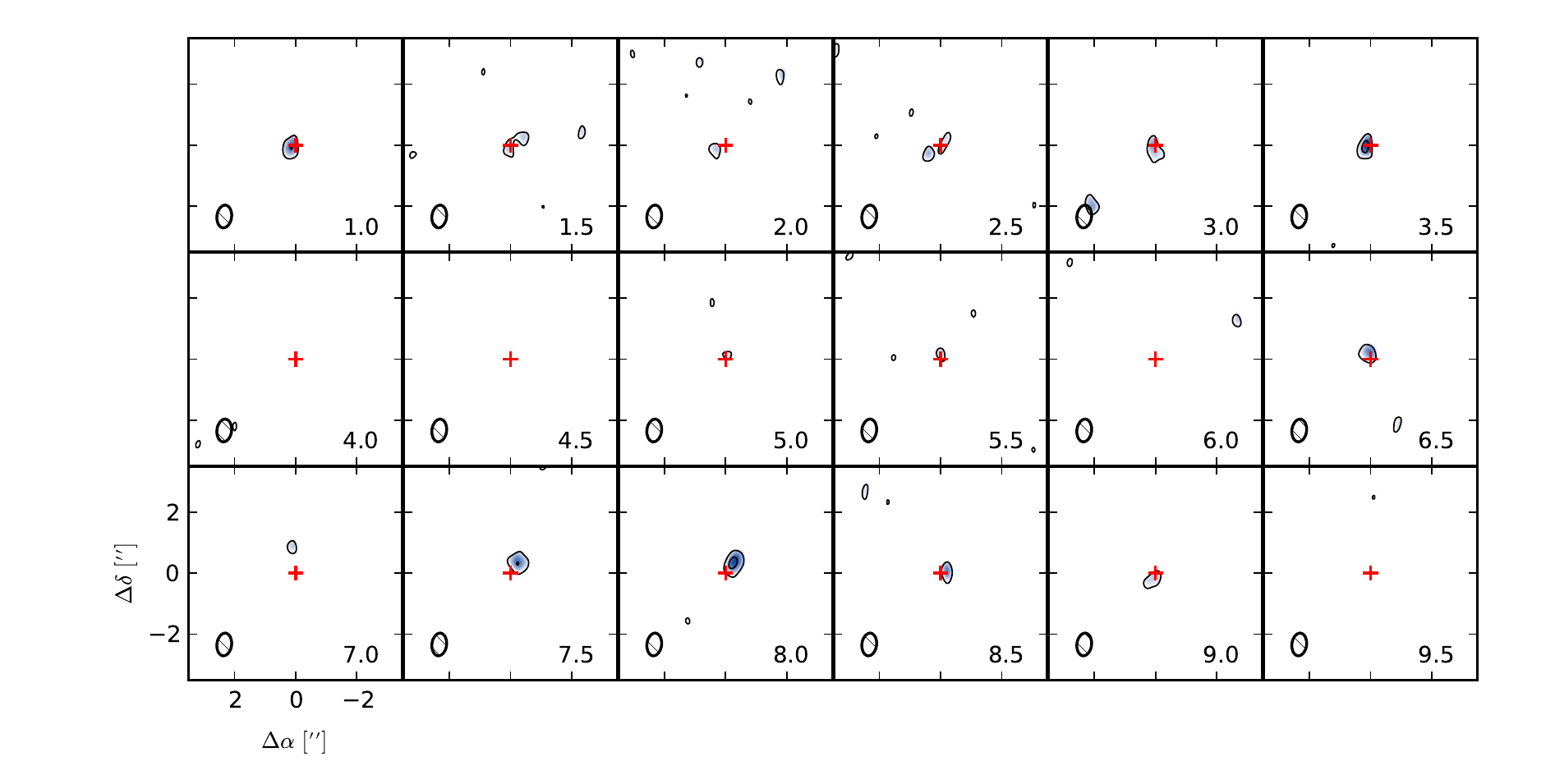}}
\end{center}

\begin{center}

\subfloat[Channel maps of DCN $J = 3-2$.]{\includegraphics[scale = .9]{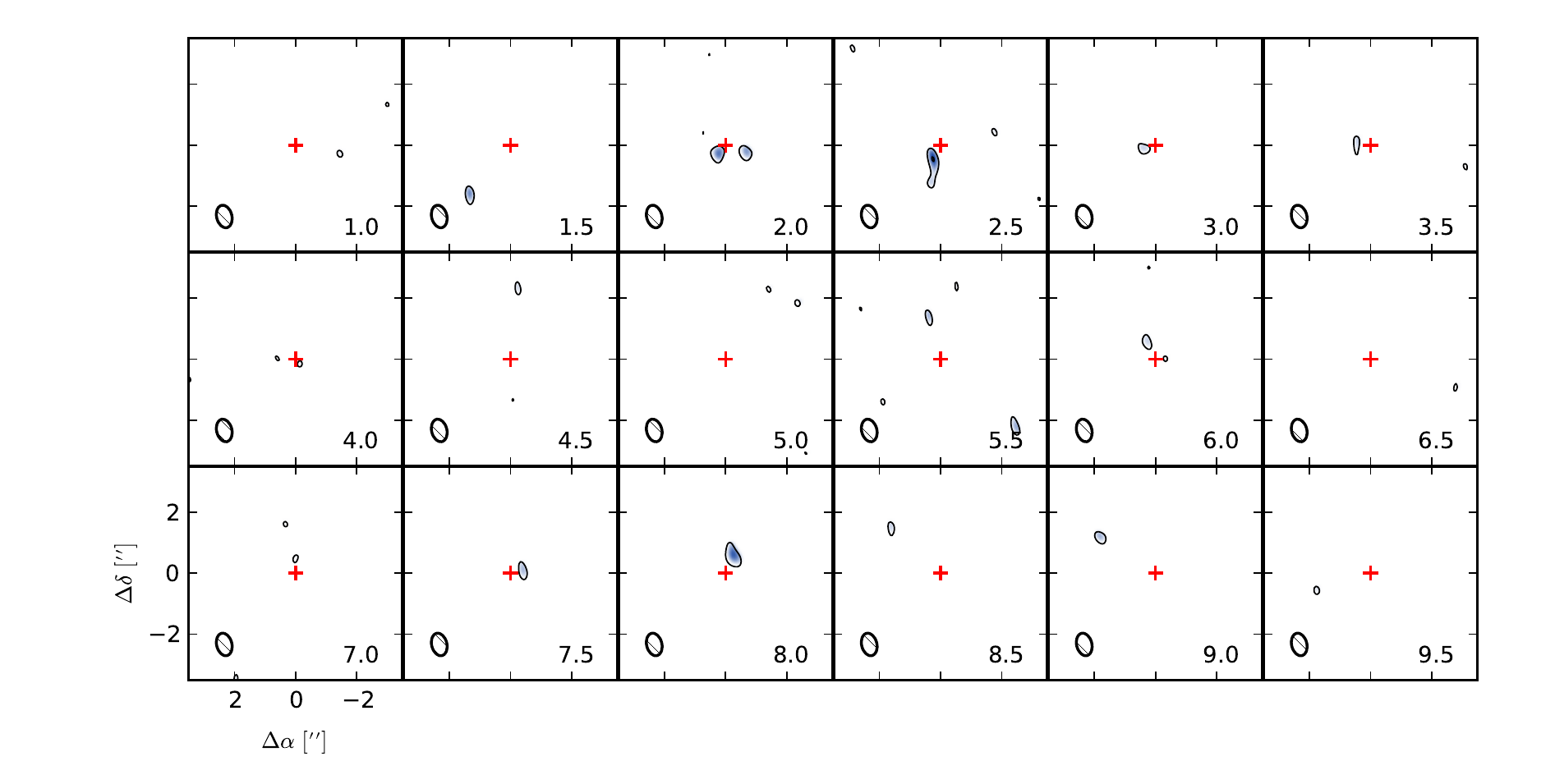}}

\end{center}
\caption{Continued}
\end{figure*}

\begin{figure*}[htp]
\begin{center}
\subfloat[Channel maps of DCO$^+$ $J = 3-2$.]{\includegraphics[scale = 0.8]{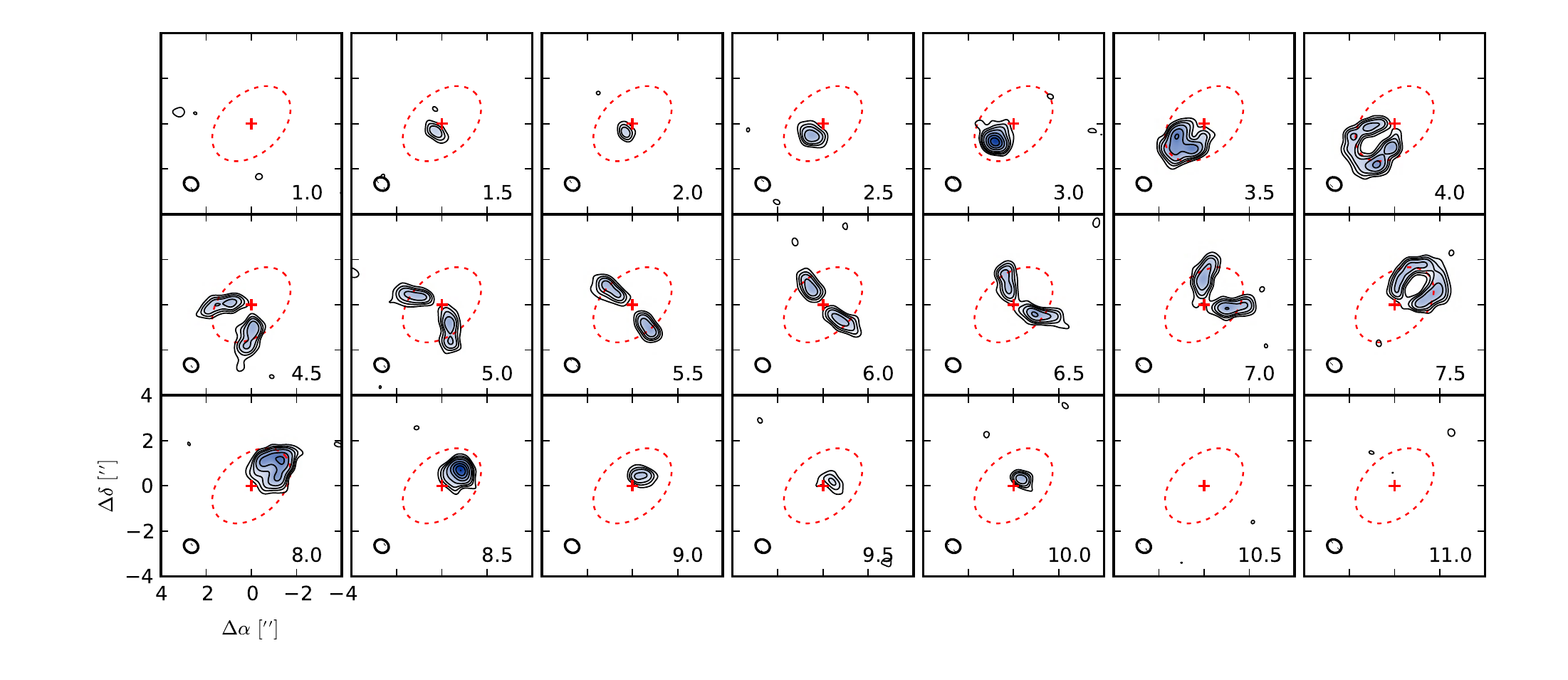}}

\end{center}

\begin{center}
\subfloat[Channel maps of H$^{13}$CN $J = 3-2$.]{\includegraphics[scale = 0.8]{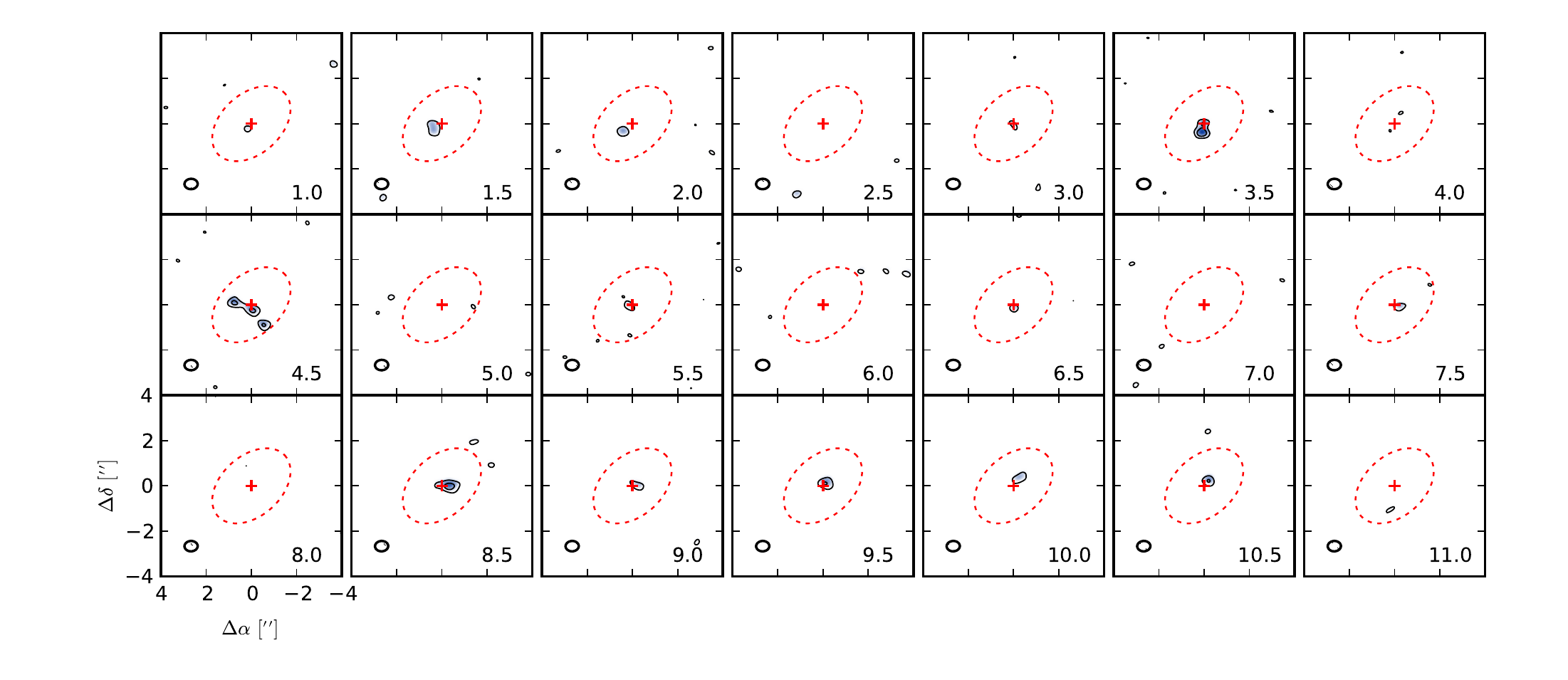}}
\end{center}
\caption{Similar to Fig. \ref{as209chanmaps}, but for the HD 163296 disk. (The H$^{13}$CO$^+$ $J = 3-2$ channel map appears in the main text in Figure \ref{fig:chanmapexample}). The red dashed ellipse traces a projected radius of 2$''$ to demonstrate where the H$^{13}$CO$^+$ emission breaks.}
\label{hd163296chanmaps}
\end{figure*}

\begin{figure*}[htp]

\begin{center}
\ContinuedFloat
\subfloat[Channel maps of DCN $J = 3-2$.]{\includegraphics[scale = 0.8]{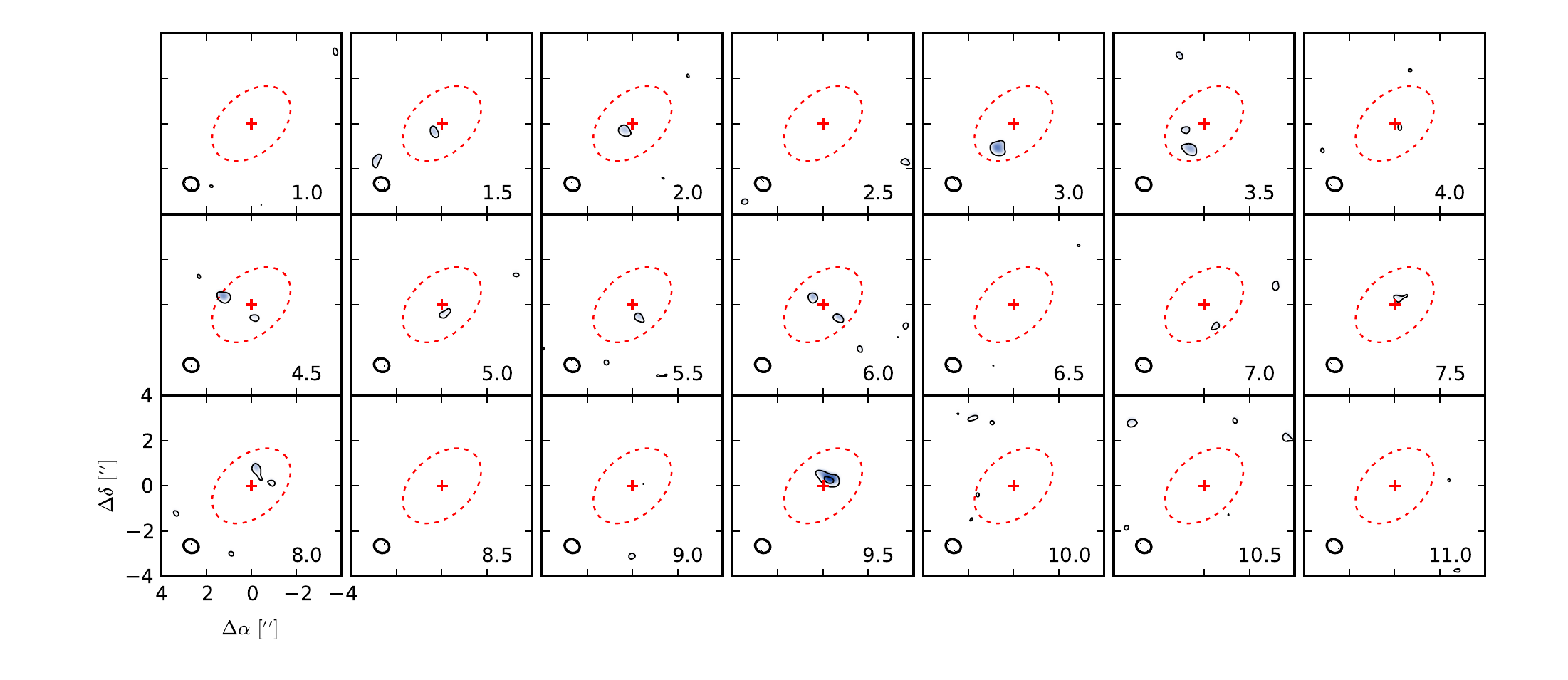}}

\end{center}
\caption{Continued}
\end{figure*}

\end{document}